\documentclass[journal,twocolumn,romanappendices]{IEEEtran}
\IEEEoverridecommandlockouts
\usepackage{cite}
\usepackage{amsmath,amssymb,amsfonts,amsthm}
\usepackage{mathtools,bbm}  
\usepackage{algorithmic}
\usepackage{graphicx}
\usepackage{textcomp}
\usepackage{xcolor}
\usepackage[thinc]{esdiff}
\usepackage[section]{placeins}
\usepackage{float}
\usepackage{mathabx}
\usepackage{physics}
\usepackage{tikz} 
\usepackage{enumitem} 
\usepackage{stmaryrd} 
\usepackage{subcaption} 
\usepackage{ragged2e} 
\usepackage{svg} 
\usepackage{caption} 
\usepackage{booktabs} 
\usepackage{multicol} 
\usepackage{hyperref} 
\usepackage{wrapfig} 
\usepackage{quantikz}
\usepackage[T1]{fontenc}
\usepackage{soul}
\soulregister\cite7 
\allowdisplaybreaks
\newtheorem{theorem}{Theorem}

\newtheorem{example}{Example}
\newenvironment{example*}
  {\addtocounter{example}{-1}\example}
  {\endexample}
\newtheorem{observation}[theorem]{\textcolor{black}{Observation}}
\newcommand{\newtext}[1]{{\color{black} #1}}
\newcommand{\newtextnew}[1]{{\color{black} #1}}

\def\BibTeX{{\rm B\kern-.05em{\sc i\kern-.025em b}\kern-.08em
    T\kern-.1667em\lower.7ex\hbox{E}\kern-.125emX}}
\begin{document}
\bstctlcite{IEEEexample:BSTcontrol}

\title{Adaptive Entanglement Distillation}

\author{\IEEEauthorblockN{Sijie Cheng and Narayanan Rengaswamy} \\
\thanks{The authors are with the School of Electrical, Computing and Software Engineering, University of Arizona, Tucson AZ 85721. E-mail: \{ sijiecheng , narayananr\}@arizona.edu}
}

\maketitle

\begin{abstract}
Quantum network applications impose a variety of requirements on entanglement resources in terms of rate, fidelity, latency, and more.
The repeaters in the quantum network must combine good methods for entanglement generation, effective entanglement distillation, and smart routing protocols to satisfy these application requirements.
In this work, we focus on entanglement distillation in a linear chain of quantum repeaters.
While conventional approaches reuse the same distillation scheme over multiple hop lengths after entanglement swaps, we propose a novel adaptive quantum error correction (QEC) scheme that boosts end-to-end metrics.
Specifically, depending on the network operating point, we adapt the code used in distillation over successive rounds to monotonically increase the rate while also improving fidelity.
We demonstrate the effectiveness of this strategy using three codes, with parameters $\llbracket 9,1,3\rrbracket$, $\llbracket 9,2,3\rrbracket$, $\llbracket9,3,3\rrbracket$, and a new performance metric, \emph{efficiency}, that incorporates both overall rate and fidelity.
Since the minimum input fidelity for QEC-based distillation is high, we then extend our study to include non-QEC-based purification protocols, specifically DEJMPS since it outperforms others.
We compare the performance of end-to-end DEJMPS against adapting from DEJMPS to QEC once DEJMPS improves the initial fidelity to the threshold for QEC.
Through a refined efficiency metric, we illuminate the regime where QEC is beneficial.
These results provide a detailed outlook for entanglement purification and distillation in first and second generation quantum repeaters.
\end{abstract}

\begin{IEEEkeywords}
Quantum repeaters, entanglement distillation, stabilizer code, depolarizing channel, Werner state, DEJMPS
\end{IEEEkeywords}

\section{Introduction}

\IEEEPARstart{Q}{uantum} networks fundamentally differ from classical networks since they rely upon entanglement distribution across the network nodes.
Since direct transmission of quantum information across long distances suffers from an exponential scaling of loss~\cite{Pirandola-natcomm17}, quantum repeaters are critical for wide area quantum networks.
A repeater generates entanglement across single hops -- short links with its neighboring nodes, which could be end-users or other repeaters -- and performs entanglement swaps to create longer-range entanglement across longer hops.
Since each link introduces loss, and swaps involve local operations that are noisy, \emph{entanglement distillation} is used to distill fewer high fidelity entangled states from many noisy entangled states.
Such distillation can be performed on single hops as well as longer hops, as long as classical communication is available across those hops.
When end-users place requests for entanglement, the network must execute a mechanism to deliver entanglement at the necessary rate and fidelity, besides other metrics such as latency.
This must be done through a combination of good entanglement generation, efficient distillation and smart routing protocols.
Hence, it is essential to develop scalable strategies for these processes that can meet the needs of different applications.

In this paper, we focus specifically on entanglement distillation across single and multiple hops.
Entanglement can be distilled using local operations and classical communication (LOCC), which can be one- or two-way~\cite{Bennett-pra96}.
Classic $2$-to-$1$ protocols require the nodes to perform two-way LOCC whereas protocols based on quantum error correction (QEC) only require one-way LOCC, significantly improving efficiency and reducing latency at the cost of increased complexity of LO.
We consider QEC-based distillation protocols for second-generation quantum repeaters, where heralded entanglement generation addresses photon loss and QEC is used to correct errors introduced during LO~\cite{Muralidharan-scirep16}.
Since recent years have witnessed an exciting amount of experimental demonstrations of QEC in different hardware technologies~\cite{Berthusen-arxiv24,Bluvstein-nature24,daSilva-arxiv24,Gupta-nature24,Lescanne-natphys20,Reichardt-arxiv24,SalesRodriguez-arxiv24}, these protocols could indeed be implemented in the next few years.
Our goal here is to demonstrate a novel approach to \emph{adaptively} improve end-to-end metrics that combine both rate and fidelity.
Hence, to highlight the key insights and keep the setup simple, we incorporate noise in the initial generation of Bell pairs but assume that all other components are noiseless.
In principle, memory decoherence or noise in LO can be mapped back to the initial fidelity of the Bell pairs, potentially with some Pauli twirling.
\newtext{We emphasize that the proposed framework is not limited to the simplified noise model considered here and can be extended to more realistic settings by incorporating decoherence, imperfect operations, and fault-tolerance constraints.}
Since these considerations will distract from the central focus of this work, and warrant an independent investigation on their own, we leave them for subsequent follow-up work.

We consider a linear chain of odd number of repeaters where each repeater generates entanglement with its neighboring nodes, distills, and then swaps them.
\newtextnew{The restriction to an odd number of repeaters is not fundamental to the proposed protocol. It is introduced only to simplify the description of the recursive construction. For a chain with an even number of repeaters, the same recursive procedure can be applied to the largest subchain satisfying the construction, while the remaining repeater can be appended at the end and connected in the final stage.}
The process continues until the end nodes have shared Bell pairs that they have distilled themselves.
But independent of the length of the chain, we set the maximum number of distillation rounds to $3$, as explained later.
The entanglement generation is deterministic but the created state is a \emph{Werner state}, where the Werner parameter could be interpreted as incorporating decoherence as well as noise in LO.
These are the minimal assumptions we need to demonstrate our adaptive protocol.
Conventional schemes apply the same distillation, or purification, procedure in each round to generate end-to-end entanglement of desired fidelity.
However, we start with an $\llbracket n,k,d\rrbracket$ stabilizer code, $C_1$, with low rate, $R = \frac{k}{n}$, to correct many errors and improve the fidelity after the first round of distillation at single hops.
After swapping the resulting $k$ Bell pairs, we adaptively switch to a different $\llbracket n,k'>k,d'\rrbracket$ code, $C_2$, that is able to further improve the fidelity while having a higher rate $R' = \frac{k'}{n} > R$.
This is done carefully only when the input fidelity to the first round is just large enough to ensure that the resulting (output) fidelity is above the input fidelity threshold for the second code.
For the third round of distillation, we adapt once again by picking yet another $\llbracket n,k''>k'>k,d''\rrbracket$ code, $C_3$, if the channel operating point, i.e., depolarizing rate, is suitable.
\newtext{We fix $n$ throughout so that the circuits for LO have a fixed size and it is easier to study the tradeoff between rate and fidelity when adapting the code.
But the technique can be generalized appropriately for varying $n$ as well.}

For illustration of the idea, our simulations are conducted using the following codes: $C_1: \llbracket 9,1,3\rrbracket, C_2: \llbracket 9,2,3\rrbracket, C_3: \llbracket 9,3,3\rrbracket$.
\newtext{These were the smallest codes from the \texttt{CodeTables} database~\cite{Grassl:codetables} that satisfy the above requirements while correcting at least one error.}
We compare four protocols, $C_1-C_1-C_1$, $C_1-C_2-C_2$, $C_1-C_2-C_3$, and $C_2-C_2-C_2$, and show the regions where each one is advantageous.
Our metric for comparison is a combination of overall rate (not just the code rate) and fidelity, which we call \emph{efficiency}.
Such adaptive QEC for entanglement distillation in quantum repeaters is unprecedented and opens up exciting new ways to achieve end-to-end metrics of quantum network applications.
These can be adapted for more realistic noise settings and the advantages will still qualitatively hold, but for a more sophisticated definition of network operating point.
We leave those investigations, e.g., a general construction of code sequences as a function of target metrics, their fault tolerance requirements, and resource estimates, for future work.

In the above exploration of adaptive QEC, the minimum required input fidelities for the codes are high, $\gtrapprox 90\%$.
For near-term implementation, it is important to keep the codes small but plan for low input fidelities, $\approx 70-80\%$, especially while accounting for decoherence and noise in LO.
Even in the long-term, the fault-tolerant thresholds for large quantum low-density parity-check (QLDPC) codes are also expected to be high.
Hence, we consider the use of classic non-QEC-based entanglement purification protocols such as BBPSSW~\cite{PhysRevLett.76.722_BBPSSW} and DEJMPS~\cite{PhysRevLett.77.2818_DEJMPS} to improve initial fidelities to the threshold of the above QEC codes%
\footnote{Note that in the QEC literature this is called a `pseudo-threshold' since it is for a specific code and not a property of a family of codes. A `threshold' usually refers to a family. Here, we will use `threshold' for convenience.}.
We first show that DEJMPS with no twirling comprehensively outperforms BBPSSW (with/without twirling), so we restrict our consideration to only DEJMPS for the rest of the paper.
We illuminate the asymmetric nature of the Pauli error probabilities for DEJMPS and show that it enables the protocol to rapidly improve fidelities in each round.
Then we use DEJMPS to improve initial fidelities to the threshold of the $\llbracket 9,3,3 \rrbracket$ code, chosen for its high rate compared to the other two codes.
We highlight the emergence of ``checkpoints'' in the input-output fidelity plot as the minimum required number of DEJMPS rounds decreases with increasing input fidelity.
After DEJMPS improves the input fidelity above the code threshold, we consider switching to one round of the QEC protocol and compare it to continuing with further DEJMPS rounds to achieve the same performance.
We refine our previously defined efficiency metric to account for the subtleties of this setting.
Using this refined metric, we highlight the regions where pure DEJMPS outperforms the switch to QEC and vice-versa.
Importantly, we observe that it is the high rate of the QEC code that provides an edge at very high input fidelities, rather than the output fidelity.

This suggests that the design of second generation repeaters must carefully inspect the choice and value of a QEC code to replace or co-exist with simpler protocols such as DEJMPS.
We expect that QEC would become a preferable choice under a fault-tolerant setting, but the details of the code and protocol will be critical to ensure superior performance to DEJMPS.
We are not aware of prior works that have closely examined the evolution from first- to second-generation repeaters, so we think that our initial investigation will lead to many more careful studies of this important technological aspect.
\newtext{Since the writing of this paper, there is at least one other work exploring adaptive distillation approaches through a code-puncturing framework~\cite{Abouamer-arxiv26}.}

As a final note, adaptivity in quantum repeater protocols can involve a broader range of design choices. 
In this work, we focus specifically on adaptive QEC code selection while keeping the distillation schedule fixed so that the benefit of code-level adaptation can be clearly demonstrated. 
However, one could combine this with adaptivity in the distillation schedule or other network degrees of freedom as appropriate.

Recent work has begun to study entanglement purification as an adaptive network control problem rather than as a fixed local operation. In particular, the Adaptive Purification Controller (APC) proposed in \cite{APC} focuses on 1G recurrence purification and dynamically selects the purification depth, as well as the protocol family such as BBPSSW or DEJMPS, along a routed path. Its main adaptive degree of freedom is the number of recurrence rounds. This choice is optimized by balancing the fidelity gain from additional purification rounds against the associated costs in success probability, raw Bell pair consumption, latency, memory decoherence, classical communication delay, Bell state measurement success probability, and local gate and measurement errors. APC therefore provides a network-level control-plane framework for deciding how much 1G purification should be applied under realistic timing and hardware constraints.

Our work addresses a different aspect of adaptive entanglement distillation. We do not aim to build a full network controller for optimizing 1G recurrence rounds under all realistic hardware parameters. Instead, in our work, the role of 1G purification is mainly to assist low fidelity input states. More specifically, the pseudo-threshold of a chosen 2G distillation  code can be used as an effective target fidelity for the preceding 1G purification stage. In this way, recurrence purification is used only when the input fidelity is too low for the 2G step to be beneficial, and its purpose is to raise the state into the region where 2G distillation becomes effective.

The main adaptive mechanism studied in this paper is therefore not the selection of the 1G purification depth itself, but the switching of QEC codes in the 2G distillation stage. Once the input fidelity has been raised above the relevant pseudo-threshold, the distillation code can be changed across successive rounds.
Use a lower-rate code in the earlier stage, where stronger error suppression is needed, and then switch to higher-rate codes once the fidelity has improved. This allows the protocol to exploit the improved input quality by increasing the output rate while potentially improving the output fidelity.


The paper is organized as follows.
Section~\ref{sec:background} introduces the necessary technical background, \newtext{Section~\ref{sec:1G_protocols} analyzes BBPSSW and DEJMPS,} Section~\ref{sec:protocol} discusses our adaptive \newtext{QEC} protocol, Section~\ref{sec:results} demonstrates the simulation results for our protocol, Section~\ref{sec:extension_to_1G} explores the extension to \newtext{DEJMPS+QEC} protocols, and Section~\ref{sec:conclusion} concludes the paper by exploring avenues for future work. 
\newtextnew{Appendix~\ref{app:SFESWS} introduces essential background concepts such as the stabilizer formalism.
Appendix~\ref{app:BBPSSW_DEJMPS_comparison} provides a detailed comparison between the BBPSSW and DEJMPS protocols.
Appendix~\ref{app:BBPSSW_DEJMPS} provides the proofs for the results in Section~\ref{sec:extension_to_1G}.
Appendix~\ref{app:1G Simulator} describes our simulator for first generation purification protocols.} 
\newtext{Appendix~\ref{app:3 rounds} presents a quantitative analysis to decide the number of rounds of distillation in our adaptive protocol. Appendix~\ref{app:sim_setup} details the simulation setup of our adaptive protocol. Finally, Appendix~\ref{app:multiple_repeaters} provides additional details for our results in multiple-repeater settings.}

\section{Background}
\label{sec:background}

To better understand the ideas we propose, let us first recall the fundamental concepts and standardize the terminology used in this paper. 
\newtextnew{
For details on the stabilizer formalism, error syndromes, and Werner states, please refer to Appendix~\ref{app:SFESWS}.
}

\subsection{Distillable Entanglement}

The efficiency of one-way hashing method~\cite{Bennett-pra96} is determined by the number of distilled ideal Bell pairs, \( m \), from $n$ pairs of qubits each in a Werner state, given by:
\begin{align}
D_H = \frac{m}{n} = 1 - S(W),
\end{align}
where \( S(W) \) is the von Neumann entropy of \( \rho_W \):
\begin{align}
    S(W) &= -F \log_2 F - (1-F) \log_2 \left(\frac{1-F}{3}\right).
\end{align}
This method is effective when \( F > 0.81071 \), i.e., $D_H > 0$.

When we have three nodes, let $W_1$ be the parameter between nodes 1 and 2, and $W_2$ be the parameter between nodes 2 and 3. 
Then, after a swap at node 2, we have the effective Werner parameter between nodes 1 and 3,
\begin{align}
W_{\rm eff} &= W_1 W_2 \nonumber \\
          &= \left(\frac{4F_1-1}{3}\right) \left(\frac{4F_2-1}{3}\right), \\
F_{\rm eff} &= \frac{1}{4}+\frac{3}{4}W_{\rm eff} \nonumber \\
          &= \frac{1}{4}+\frac{3}{4}\left(\frac{4F_1-1}{3}\right) \left(\frac{4F_2-1}{3}\right).\label{eq:quantum swap}
\end{align}
When we have $n$ nodes, we will have $W_1, \ldots, W_{n+1}$, so
\begin{align}
F_{\text{eff}} = \frac{1}{4} + \frac{3}{4} \prod_{i=1}^{n+1} \left(\frac{4F_i-1}{3}\right).
\end{align}
For successful distillation, the threshold condition is:
\begin{align}
\prod_{i=1}^{n+1} \left(\frac{4F_i-1}{3}\right) > 0.747613.
\end{align}

\newtext{%
\subsection{Terminology Standardization}

To avoid any misunderstanding regarding the terminology used in this paper, let us provide standardized terminology.
\begin{itemize}
    
\item 1G purification refers to the use of circuits without error correction, where some Bell pairs are discarded and the remaining pairs are kept to improve the output fidelity.

\item 2G distillation refers to a process in which Bell pairs are not discarded. Instead, an encoding procedure is used to convert the physical Bell pairs into logical Bell pairs. Error correction, such as a simple lookup table procedure, is then applied to improve the output fidelity.

\item A 1G protocol refers to the use of several rounds of 1G purification to improve the output fidelity. 

\item A 2G protocol refers to the use of three rounds of adaptive distillation to improve the output fidelity.

\item A 1G+2G protocol refers to a protocol in which the 1G protocol is first used to improve the output fidelity to, or above, the pseudo-threshold of the 2G protocol. Then, 2G distillation is applied to further improve the fidelity.
\end{itemize}
}

\section{Classic 1G Entanglement Purification}
\label{sec:1G_protocols}

\subsection{BBPSSW and DEJMPS}

\newtext{There are many types of 1G protocols, but the most well-known ones are BBPSSW~\cite{PhysRevLett.76.722_BBPSSW} and DEJMPS~\cite{PhysRevLett.77.2818_DEJMPS}. }

\begin{figure}[h]  
    \centering  
    \includegraphics[width=0.3\textwidth]{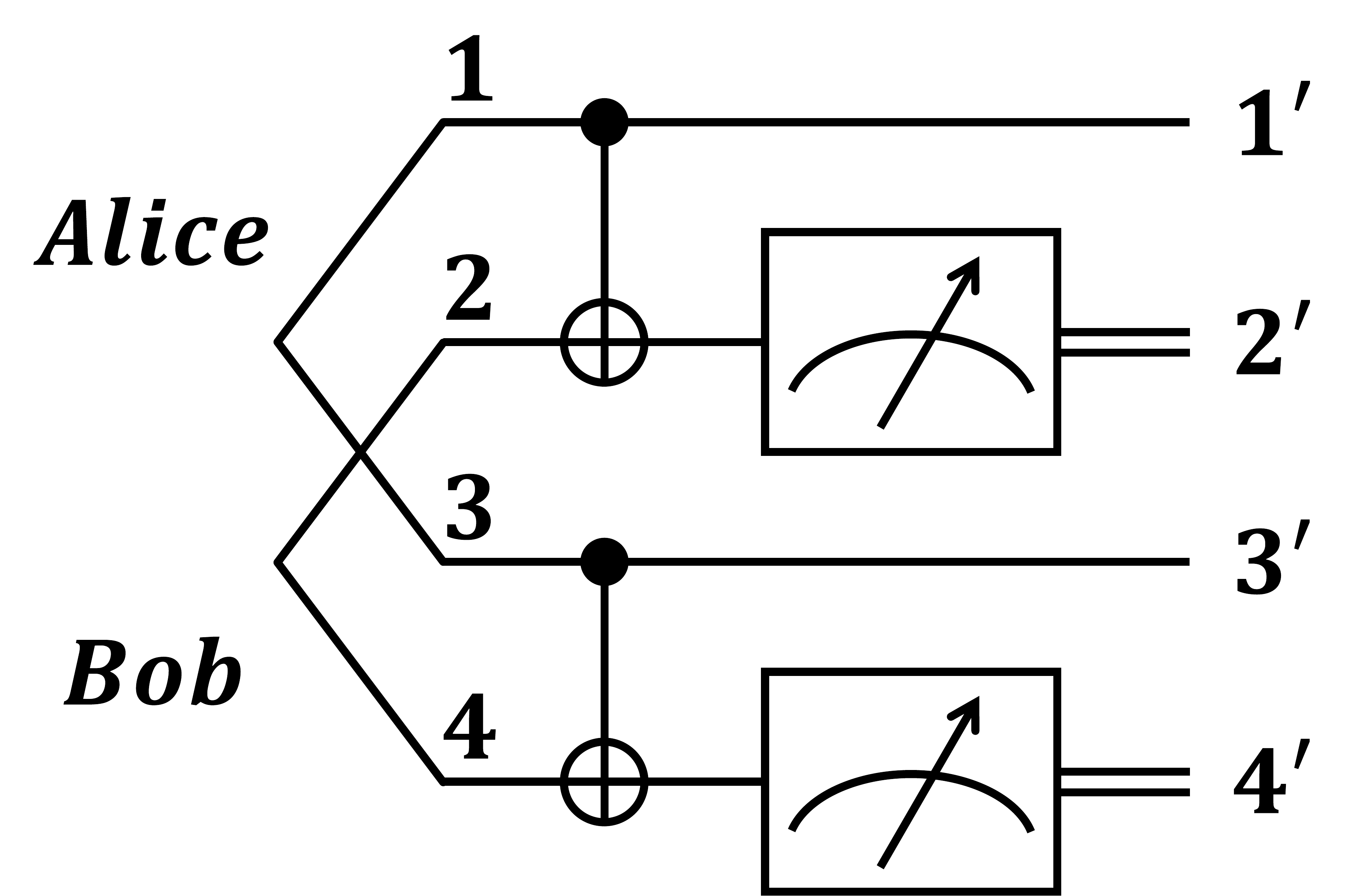}
    \caption{BBPSSW circuit diagram} 
    \label{fig:Cir_BBPSSW}  
    \vspace*{-5pt}
\end{figure}

The BBPSSW protocol operates as follows: after generating two Bell pairs (1 and 3, 2 and 4), Alice and Bob apply a CNOT gate, using one pair as the control and the other as the target, and then perform measurements on the target pair. If 2' and 4' match, the Bell pair 1' and 3' is kept. Otherwise, the Bell pair is discarded. After each round, BBPSSW requires a twirling operation to ensure that the protocol converges efficiently.

\begin{figure}[h]  
    \centering   
    \includegraphics[width=0.4\textwidth]{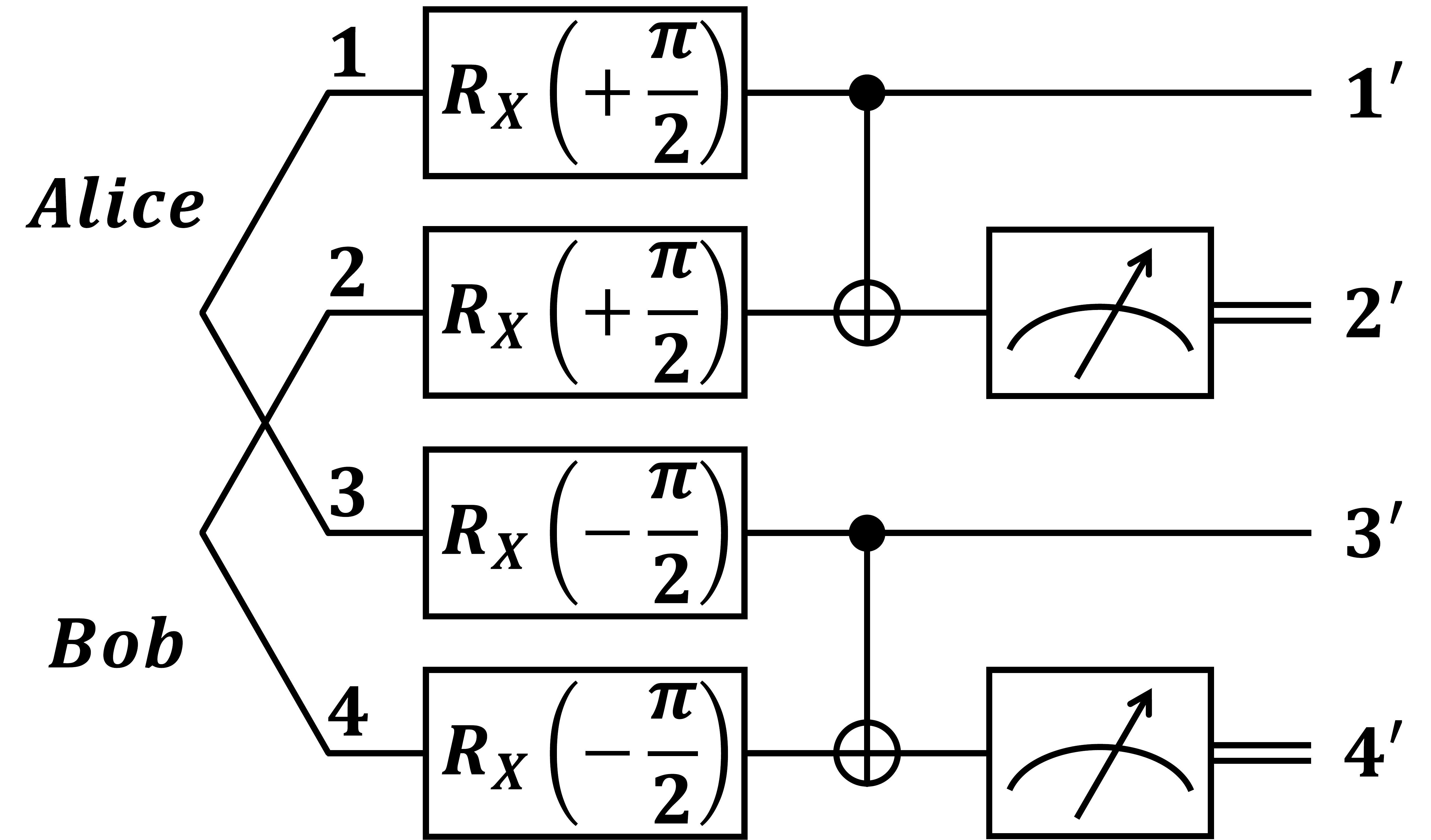}
    \caption{DEJMPS circuit diagram} 
    \label{fig:Cir_DEJMPS}  
    \vspace*{-5pt}
\end{figure}

The DEJMPS protocol is a modified version of BBPSSW that adds an X-axis rotation to the circuit. Interestingly, if DEJMPS also performs twirling, its performance becomes nearly identical to that of BBPSSW. However, DEJMPS typically omits the twirling step because the asymmetry of Z errors relative to X and Y introduces a bias that allows DEJMPS to converge much faster than BBPSSW.

In the DEJMPS circuit, the $R_X(\pi/2)$ gate transforms the stabilizer group $\{X,Y,Z\}$ into $\{X,Z,-Y\}$, while $R_X(-\pi/2)$ gate transforms $\{X,Y,Z\}$ into $\{X,-Z,Y\}$, according to the rotation operator $R_X(\theta)=e^{-\frac{i\theta X}{2}}$.

\newtextnew{
In Appendix~\ref{app:BBPSSW_DEJMPS_comparison} and~\ref{app:BBPSSW_DEJMPS}, we provide a detailed comparison of BBPSSW (with and without twirling) and DEJMPS (with and without twirling), from which the following relation between their output fidelities can be observed:
\begin{align}
\rm DEJMPS(no\ twirl) &> \rm DEJMPS(twirled)  \notag \\
                      &= \rm BBPSSW(twirled)  \notag \\ 
                      &> \rm no\ coding  \notag \\ 
                      &> \rm BBPSSW(no\ twirl).
\end{align} 
}

Therefore, unless stated otherwise, for the rest of this paper “DEJMPS” refers to the protocol without twirling between rounds, whereas “BBPSSW” refers to the protocol with twirling applied after each round.

\newtextnew{
In Appendix~\ref{app:1G Simulator}, we provide a detailed description of the structure of our 1G simulator, which can significantly accelerate the simulation.
}

\subsection{Limitations of 1G Protocols}

\newtext{%
There are many works that study 1G purification. 
The recent paper on Adaptive Purification Controller (APC)~\cite{APC} explicitly shows a ``noise cliff'' around a gate error probability of \(10^{-3}\), beyond which additional purification rounds can become counterproductive because local gate errors, especially CNOT-induced errors, dominate the distillation gain.
Although gate noise is not considered in our work, this represents an important limitation, since 1G protocols cannot effectively address the problem of high gate noise.
Moreover, the probabilistic nature of 1G purification leads to the discarding of many raw Bell pairs, which is wasteful.
Therefore, an effective method for addressing both gate noise and non-deterministic purification is important, which motivates the use of 2G distillation based on quantum error correction.
}

\section{2G Adaptive Error Correction Protocol}
\label{sec:protocol}

\subsection{Standard 2G Protocol}

\newtext{The connection between quantum error correction and entanglement distillation was established in the pivotal work of Bennett \emph{et al.}~\cite{Bennett-pra96}.}
The protocol begins with $n$ Bell States shared between Alice and Bob. 
Instead of sending them directly, Alice projects these states to the subspace of a stabilizer code. 
Since we use Bell states, the encoding can be implemented via stabilizer measurements on the initial Bell pairs.
After encoding, Alice obtains $k$ encoded Bell pairs along with a set of stabilizers. 
The process is as follows~\cite{Wilde-isit10,Rengaswamy-quantum24}:
\begin{enumerate}
    \item \textbf{Encoding}: Alice performs stabilizer measurements on the Bell states to obtain coded Bell pairs.
    \item \textbf{Classical Communication}: Alice sends the stabilizer generators and measurement outcomes (syndromes) to Bob through a classical channel.
    \item \textbf{Noisy Transmission}: Alice transmits one half of the coded Bell pairs to Bob through a noisy quantum channel while keeping the other half.
    \item \textbf{Error Detection}: Upon receiving the transmitted part, Bob applies stabilizer-based syndrome measurements using the same stabilizer generators to detect errors introduced by the channel and obtain the detected syndromes.
    \item \textbf{Error Correction}: Using the stabilizer generators, the syndrome received from Alice, and the detected syndrome, Bob applies decoding to estimate the error and correct it using quantum gates.
\end{enumerate}

If the estimated error matches the actual error, up to stabilizers, then Bob successfully corrects it, restoring the encoded Bell pairs to a perfect state. 
After successful error correction, Alice and Bob share a set of perfectly encoded Bell pairs, enabling robust long-distance quantum communication.

\newtext{%
However, when the distance between Alice and Bob becomes very long, the input fidelity decreases rapidly. Therefore, repeaters are needed to improve the fidelity along the communication path. Since the operation of a repeater is based on teleportation, it can also introduce additional fidelity loss. 
As a result, if the same code is applied after each quantum swap in the repeater chain, the protocol rate can become very low.
Therefore, it is important to develop an adaptive strategy to determine how to apply 2G distillation when multiple repeaters are introduced along the path. 
Based on this insight, we develop an adaptive 2G distillation protocol.
}

\subsection{Stabilizer Codes}

In this work, we obtain stabilizer codes from the \texttt{CodeTables} database~\cite{Grassl:codetables} and then transform them into standard form as outlined in~\cite[Chapter 10]{NielsenChuang2010}.
We describe the codes in the form of matrices whose entries are Pauli operators. 
Each row of the stabilizer matrix $H$ (resp. logical operator matrix $\overline{X}$ or $\overline{Z}$) corresponds to an independent stabilizer (resp. logical $X$ or logical $Z$) of the code.
To illustrate the main ideas, we focus on the \(\llbracket 9,1,3\rrbracket\), \(\llbracket 9,2,3\rrbracket\) and \(\llbracket 9,3,3\rrbracket\) codes shown in Table~\ref{tab:9-qubit_codes}. 
As an initial test for uncovering structural properties relevant to distillation, we also examine the well-known \(\llbracket 5,1,3\rrbracket\) and \(\llbracket 7,1,3\rrbracket\) codes.
However, we emphasize that our method can be suitably adapted to other \newtext{appropriate} sequences of stabilizer codes.
\newtext{While the quantitative results will change for other code sequences, we expect the qualitative results for the above codes to extend to other sequences.
The optimal design of sequences of codes for achieving a target metric under practical constraints is a challenging problem, which we will address in future work.}

\begin{table}[]
    \centering
    \hspace*{-5pt}
    \begin{tabular}{c|c|c}
    \toprule
       $\llbracket 9,1,3 \rrbracket$ & $\llbracket 9,2,3 \rrbracket$ & $\llbracket 9,3,3 \rrbracket$ \\
    \midrule
       $H=\begin{bmatrix}
\texttt{YIZIIIIXY} \\
\texttt{ZYZIIIIIX} \\
\texttt{ZZYIIIIXI} \\
\texttt{IIIXIIIII} \\
\texttt{IIIIXIIII} \\
\texttt{IIIIIXIII} \\
\texttt{IIIIIIXII} \\
\texttt{IZZIIIIZZ}
\end{bmatrix}$ &
       $H=\begin{bmatrix}
\texttt{YZZZIIXII} \\
\texttt{ZYZIZIXYY} \\
\texttt{ZIYZIIXYX} \\
\texttt{ZIIXIIIIY} \\
\texttt{IIZIYIIXI} \\
\texttt{IIIIIXIII} \\
\texttt{ZZZZZIZZZ} 
\end{bmatrix}$ & 
       $H=\begin{bmatrix}
\texttt{YZIZIIYXX} \\
\texttt{IXZZIXYIY} \\
\texttt{ZIYZIXIYX} \\
\texttt{IZIYIXXYZ} \\
\texttt{IIIIXIIII} \\
\texttt{ZZZZIZZZZ} 
\end{bmatrix}$ \\
       $\overline{X}=
\begin{bmatrix}
\texttt{ZIIIIIIXX}
\end{bmatrix}$ & 
       $\overline{X}=
\begin{bmatrix}
\texttt{IZZIIIXXI} \\
\texttt{IZIZIIXIX} 
\end{bmatrix}$ & 
       $\overline{X}=
\begin{bmatrix}
\texttt{ZZIIIXXII} \\
\texttt{IIZZIXIXI} \\
\texttt{IZIZIXIIX} 
\end{bmatrix}$ \\
       $\overline{Z}=
\begin{bmatrix}
\texttt{ZZIIIIIIZ} 
\end{bmatrix}$ & 
       $\overline{Z}=
\begin{bmatrix}
\texttt{IZZIZIIZI} \\
\texttt{IZZZIIIIZ} 
\end{bmatrix}$ & 
       $\overline{Z}=
\begin{bmatrix}
\texttt{ZZIZIIZII} \\
\texttt{ZIZZIIIZI} \\
\texttt{ZZZIIIIIZ} 
\end{bmatrix}$ \\
  \bottomrule
    \end{tabular}
    \caption{The three $9$-qubit codes that we use for simulations.}
    \label{tab:9-qubit_codes}
\end{table}

\subsection{Decoder Based on Lookup Table}
\label{sec:lut_decoder}

The lookup table decoder is a method for quantum error correction that maps syndromes to error patterns using a precomputed table. 
It operates with the following steps:
\begin{enumerate}
    \item \textbf{Syndrome Measurement}:  
    The stabilizer measurements yield an error syndrome, $\mathbf{s} = (\lambda_1, \lambda_2, \dots, \lambda_m) \in \{ \pm 1 \}^m$.

    \item \textbf{Error Identification via Lookup Table}:  
    The syndrome is matched with a stored minimum-weight error in the lookup table as the estimated error. This is the maximum likelihood decoding strategy for the depolarizing channel (i.e., Werner states).

    \item \textbf{Error Correction}:  
    If the estimated error exactly matches the actual error, up to a stabilizer, then decoding is successful.
    Otherwise, logical operators are used to check which logical qubits are in error.
\end{enumerate}

For $k=1$ logical qubit, the logical error rate is the same as the per-logical-qubit error rate. 
However, for $k>1$ logical qubits, each logical qubit can suffer from different logical error rates. 
In our results, a lower-bound strategy is used (for output fidelity): the logical error rate for each qubit is taken to be the overall logical error rate of the code.
This method provides an efficient way to handle errors in small-distance quantum codes and can be used for practical quantum networking applications.

\subsection{Simulation Setup}

\begin{figure}[t]  
    \centering  
    \includegraphics[width=0.3\textwidth]{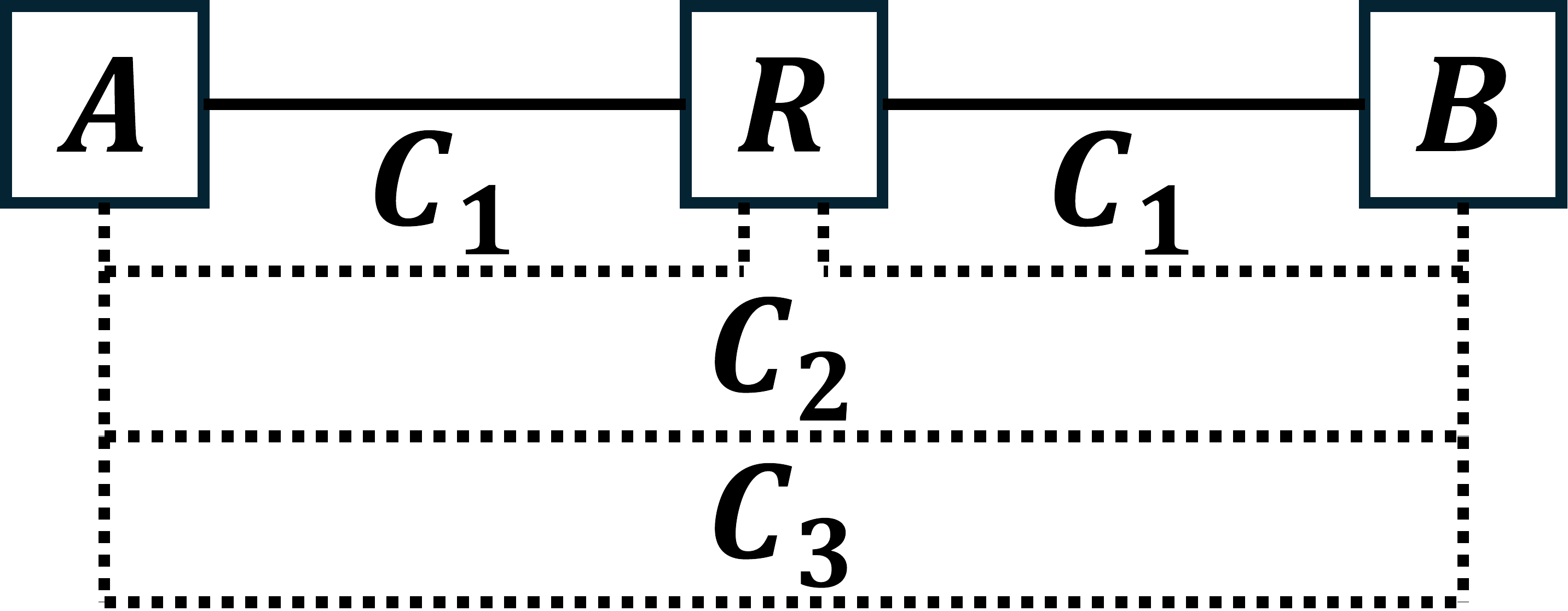}
    \caption{Protocol diagram for a single repeater, where we still perform three rounds of distillation.} 
    \label{fig:1R}  
\end{figure}

\begin{figure}[t]  
    \centering  
    \includegraphics[width=0.48\textwidth]{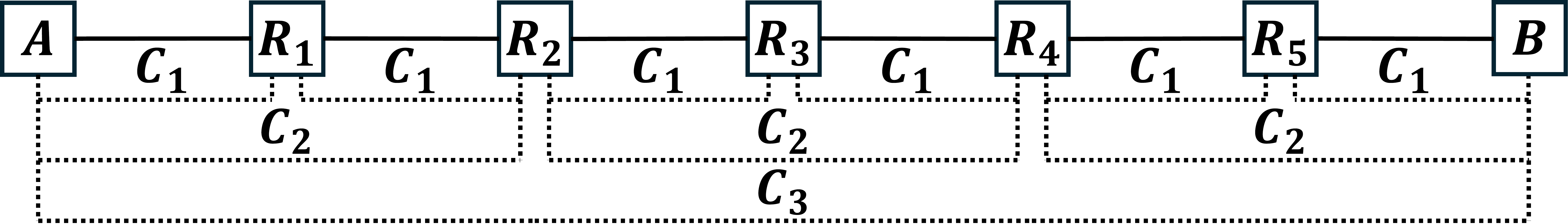}
    \caption{Protocol diagram for five repeaters, where we perform only three rounds of distillation.} 
    \label{fig:5R}  
    \vspace*{-15pt}
\end{figure}

\newtext{
In Appendix~\ref{app:3 rounds}, we investigate the effect of different number of rounds of distillation in a linear repeater chain.
Using the well-known $\llbracket 5,1,3 \rrbracket$ and $\llbracket 7,1,3 \rrbracket$ codes as illustrative examples, we conclude that three rounds of distillation suffice for our purposes, independent of the number of repeaters in the chain.
}
To clarify, let's start with a single repeater, with the structure shown in Fig. \ref{fig:1R}. This means we have Alice, the repeater, and Bob. First, Alice and the repeater, as well as the repeater and Bob, will each generate an entanglement link. Next, they will perform distillation by using code $C_1$ separately.  Then, we will perform a quantum swap to generate the entanglement link between Alice and Bob. Afterward, we will conduct distillation by using code $C_2$ between Alice and Bob. 
In this process, we aim to perform distillation between adjacent nodes in the updated effective chain after each quantum swap.
Finally, an end-to-end distillation by using code $C_3$ will be performed between Alice and Bob. In this specific situation, we will perform distillation between Alice and Bob twice to align with our three-round distillation strategy.
A more general case is shown in Fig.~\ref{fig:5R} for the setting of five repeaters, still with three rounds of distillation.
\newtext{The proposed approach can be extended to larger repeater chains by increasing the number of repeaters while maintaining the same three-round distillation strategy. 
This is the minimum number of rounds to demonstrate the non-trivial benefits of our adaptive approach for any number of repeaters.}

\subsection{Assumptions}

For our error correction-based distillation model, we use distillable entanglement as part of the end-to-end metric. 
To simplify the model, we assume that the only source of noise is the depolarizing channel, while all other components are ideal. 
This means that whenever we attempt to generate entanglement between two nodes, it will always succeed, and there is no loss in the channel.
Our primary goal here is to demonstrate the effectiveness of our adaptive distillation protocol in the simplest nontrivial setting.

\newtext{Note that there are physical scenarios where even the current depolarizing model captures distance-dependence.
For example, if the raw Bell pairs are encoded in the GKP basis with photon loss accounted for, then the input Bell pairs for our discrete variable protocols are actually logical Bell pairs of the GKP code.
Hence, the resulting logical Bell pair fidelity (i.e., the input fidelity for our protocols) indirectly depends on the effect of photon loss on GKP error correction.}

We assume that the distance and circumstances between each pair of nodes are identical, meaning they will have the same input fidelity. Therefore, we can use a single input fidelity value to represent the fidelity between all adjacent nodes. Thus, we define this input fidelity as our $F_{\rm in}$.

Quantum memories are assumed to hold qubits indefinitely without any decoherence or loss. 
Additionally, distillation and swapping operations are treated as ideal, with no errors. If we consider circuit-level errors, then the entire scheme must be made fault-tolerant for our sequence of codes, which warrants an entire paper for itself that considers all the subtle details. 
We do intend to perform that analysis as a natural follow-up to this work.
These are the minimal assumptions that we need to demonstrate our adaptive error correction protocol.

\newtextnew{
We further note that our assumption of deterministic entanglement swapping is physically motivated by repeater architectures based on stationary matter qubits, such as trapped ions, atoms, or solid-state qubits. In such architectures, once the elementary Bell pairs have been successfully established and stored in local quantum memories, entanglement swapping can be implemented through a local Bell-state measurement using quantum gates and measurements, rather than through a probabilistic linear-optical Bell-state measurement. Under our assumption of ideal local operations, we therefore take the swapping success probability to be unity.

The assumption of deterministic elementary-link generation is instead a deliberate simplification of the network layer in the present work. Consequently, the end-to-end metric reported here does not account for stochastic link-generation waiting times or the reduction in physical throughput caused by a finite link-generation success probability. If elementary-link generation were probabilistic but heralded, the actual wall-clock entanglement-generation rate would be reduced accordingly. However, conditioned on successfully generated Bell pairs, the adaptive distillation decision studied in this work would still be determined primarily by the fidelity of the available pairs. A complete treatment of probabilistic link generation, waiting-time effects, and finite-memory decoherence is beyond the scope of the present work and is left for future study.

}

\section{Simulation Results}
\label{sec:results}

\begin{figure}[t]
    \centering
    \vspace*{-7pt}
    \hspace*{-20pt}
    \includegraphics[scale=0.5,keepaspectratio]{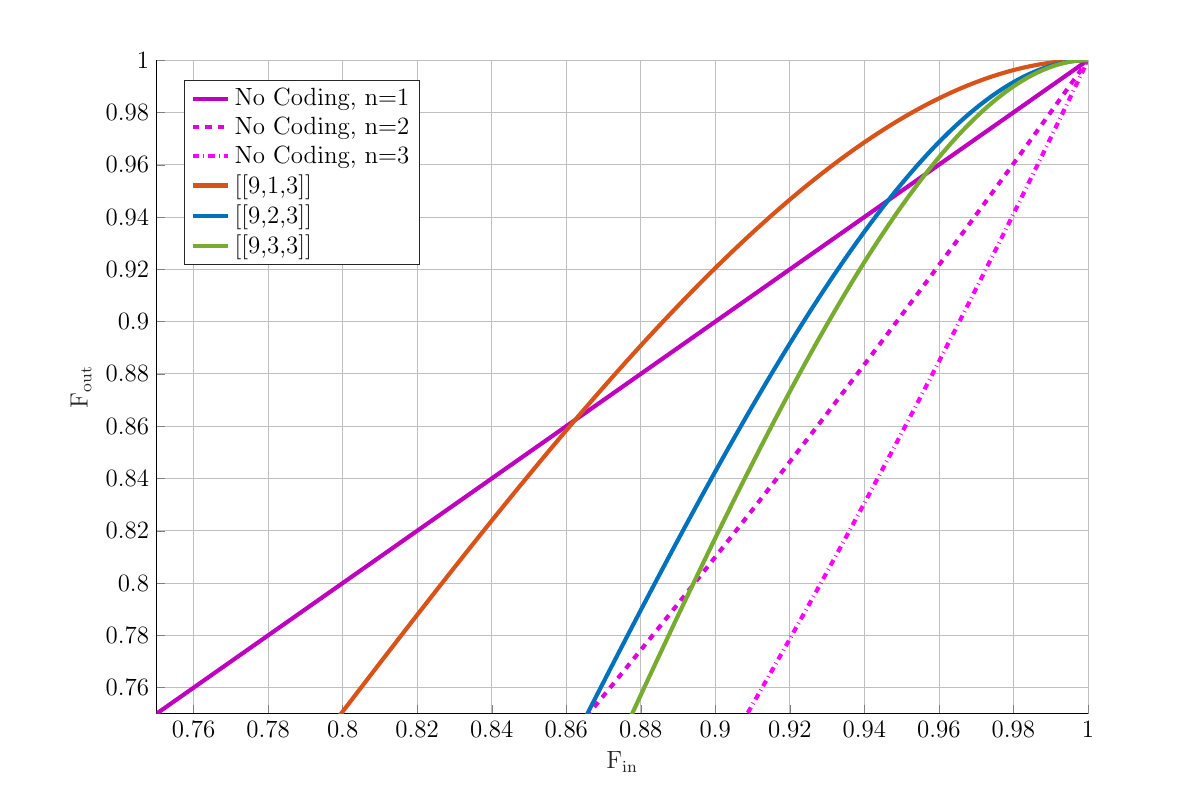}
    \caption{The $F_{\rm in}$-$F_{\rm out}$ map for the 0-repeater case. As an example, the $F_{\rm in}$-$F_{\rm out}$ map is shown for $\llbracket 9,1,3\rrbracket$, $\llbracket 9,2,3\rrbracket$ and $\llbracket 9,3,3\rrbracket$. Additionally, the no-coding error rates are provided as references, where $F_{\rm out}^{\text{No Coding}} = F_{\rm in}^n$.}
    \label{fig:0R_F} 
\end{figure}

\newtext{%
Our simulation setup is explained in detail in Appendix~\ref{app:sim_setup}.
The simulation of various protocols below builds on the $F_{\rm in}$-$F_{\rm out}$ map for the 0-repeater case (labeled ``0R''), i.e., the logical error rates of the three codes that we consider.
This plot is shown in Fig.~\ref{fig:0R_F}, where $F_{\rm out}^{\text{No Coding}} = F_{\rm in}^n$, i.e., without error correction, the output fidelity is simply the probability that all $n$ input qubits are noiseless.
}

\subsection{Goal}

Based on the single repeater \newtext{(``1R'')} situation, we can use various coding combinations. 
To describe this clearly, we will use the notation $\llbracket n_1,k_1,d_1\rrbracket-\llbracket n_2,k_2,d_2\rrbracket-\llbracket n_3,k_3,d_3\rrbracket$ to represent the codes used at each round. 
For example, $\llbracket 9,1,3\rrbracket-\llbracket 9,2,3\rrbracket-\llbracket 9,3,3\rrbracket$ means we will use $\llbracket 9,1,3\rrbracket$ for the first round. 
To maintain system stability, we will use the same code for each link within the same round. 
Then, we will use $\llbracket 9,2,3\rrbracket$ for the second round and $\llbracket 9,3,3\rrbracket$ for the final end-to-end round. 
Here, we will compare these protocols:
\begin{itemize}
    \item ${\rm Protocol}_1$: $\llbracket 9,1,3\rrbracket-\llbracket 9,1,3\rrbracket-\llbracket 9,1,3\rrbracket$
    \item ${\rm Protocol}_2$: $\llbracket 9,1,3\rrbracket-\llbracket 9,2,3\rrbracket-\llbracket 9,2,3\rrbracket$
    \item ${\rm Protocol}_3$: $\llbracket 9,1,3\rrbracket-\llbracket 9,2,3\rrbracket-\llbracket 9,3,3\rrbracket$
    \item ${\rm Protocol}_4$: $\llbracket 9,2,3\rrbracket-\llbracket 9,2,3\rrbracket-\llbracket 9,2,3\rrbracket$
\end{itemize}

Because the number of outputs (i.e., logical qubits) is different across these codes, we will use the fidelity lower bound (see Section~\ref{sec:lut_decoder}) to describe the performance of each. 
This means that all output qubits being correct simultaneously is considered successful, while any single output qubit failure is considered a total failure. 
In practice, we wouldn't know which logical qubits are in error, so this is reasonable.
Our goal is to develop an adaptive strategy to switch between protocols at relatively optimal thresholds, balancing the trade-off between rate, fidelity, and distillable entanglement.

\subsection{Efficiency Function}

To demonstrate this multi-way tradeoff, we define the \emph{efficiency} function as follows:
\begin{align}
E(F_{\rm in}) \coloneqq \frac{n_{out} D(F_{\rm out}(F_{\rm in}))}{n_{in} D(F_{\rm in})} = \frac{R_{\rm out} D_{\rm out}(F_{\rm in})}{D(F_{\rm in})},
\end{align}
where:
\begin{itemize}
    \item \( n_{\rm in} \) is the number of entangled states consumed,
    \item \( n_{\rm out} \) is the number of output entangled states,
    \item \( R_{\rm out} = \frac{n_{\rm out}}{n_{\rm in}} \) is the protocol rate,
    \item \( F_{\rm in} \) is the input fidelity,
    \item \( F_{\rm out}(F_{\rm in}) \) is the output fidelity corresponding to \( F_{\rm in} \),
    \item \( D(F_{\rm in}) \) is the input distillable entanglement,
    \item \( D_{\rm out}(F_{\rm in}) \) is the output distillable entanglement.
\end{itemize}

To calculate the protocol rate, we must ensure that the number of output Bell pairs from the previous round matches the input required by the code used in the next round. Therefore, the rate must be calculated iteratively on a round-by-round basis.

\subsubsection{First Round Using the $\llbracket n_1,k_1,d_1\rrbracket$ Code}
When discussing the first round, we need to consider the number of repeaters ($n_R$) to be used. The corresponding input and output qubits are shown below:
\begin{align}
N^{(1)}_{\rm in} &= (n_R+1) \cdot n_1,\\
K^{(1)}_{\rm out}&=k_1.
\end{align}

\subsubsection{Second Round Using the $\llbracket n_2,k_2,d_2\rrbracket$ Code}
The key point in the second round is to match $K^{(1)}_{\rm out}$ with $n_2$. Therefore, we use $L^{(1)}=\mathrm{lcm}(K^{(1)}_{\rm out}, n_2)$.
Thus, the corresponding input and output qubits are shown below:
\begin{align}
N^{(2)}_{\rm in} &= \frac{L^{(1)}}{K^{(1)}_{\rm out}} \cdot N^{(1)}_{\rm in},\\
K^{(2)}_{\rm out}&=\frac{L^{(1)}}{n_2} \cdot k_2.
\end{align}

\subsubsection{Third Round Using the $\llbracket n_3,k_3,d_3\rrbracket$ Code}
Similar to the second round, we need to use $L^{(2)}=\mathrm{lcm}(K^{(2)}_{\rm out}, n_3)$ to ensure proper matching between them.
Thus, the final input and output qubits are shown below:
\begin{align}
N^{(3)}_{\rm in} &= \frac{L^{(2)}}{K^{(2)}_{\rm out}} \cdot N^{(2)}_{\rm in},\\
K^{(3)}_{\rm out}&=\frac{L^{(2)}}{n_3} \cdot k_3.
\end{align}
Here, $N^{(3)}_{\rm in}$ and $K^{(3)}_{\rm out}$ are the total number of Bell pairs required for the input and the total number of output Bell pairs, respectively. Thus, we can set $n_{in}=N^{(3)}_{\rm in}$ and $n_{out}=K^{(3)}_{\rm out}$.

\newtext{
\subsection{Comparison with Established Metrics}

\subsubsection{Protocol Yield}

In established metrics, yield is commonly used to describe the efficiency of a distillation protocol, and it is usually defined as
\[
Y = \frac{n_{\rm out}}{n_{\rm in}} .
\]
This definition is similar to the protocol rate in our metric. However, yield mainly captures the resource conversion rate and does not directly include the amount of distillable entanglement.
This distinction becomes important when the output fidelity is low. In such cases, yield may still be high because it only counts the number of output pairs relative to the number of input pairs. However, in practice, the generated link may not be useful if the distillable entanglement is too low.

Therefore, in our proposed metric, we combine the protocol rate with the distillable entanglement. This ensures that when the output fidelity is low, the metric is dominated by the distillable entanglement contribution, reflecting the practical usefulness of the generated link. At the same time, when the output fidelity is high, the rate becomes the dominant factor, making the behavior of our metric consistent with the conventional yield interpretation.

\subsubsection{Secret Key Rate}

Our metric is application-independent and is not motivated by QKD or other applications. Instead, it is a general network-level metric that combines the protocol rate and the distillable entanglement. Most importantly, it combines the practical cost of the number of consumed Bell pairs with the asymptotic cost of distilling perfect Bell pairs from the higher fidelity Bell pairs produced by our protocols.
A similar structure that combines rate and useful entanglement has been used in the context of secret key generation~\cite{Rozp_dek_2018}. 
There, the authors do not have our denominator component, which specifies the (lower bound on) distillable entanglement of the initial raw Bell pairs. 
We think that our metric provides a better direct comparison between what one can do with the initial Bell pairs and the final distilled Bell pairs.
In addition, the hashing bound expression for distillable entanglement used in our metric also gives the asymptotic secret key rate of the six-state protocol for Bell-diagonal states~\cite{Lo-qic01,Kraus-prl05,Murta_2020}.
Hence, the new efficiency metric combines some important components of the secret key rate while remaining application-independent. 
We will continue to think of better ways to arrive at a metric that provides interpretable measures of protocol performance independent of the application.
}

\subsection{Results}

\subsubsection{Single Repeater}

\begin{figure}[t]
    \centering
    \hspace*{-20pt}
    \begin{tikzpicture}
        \node[anchor=south west,inner sep=0] (image) at (0,0) 
        {\includegraphics[scale=0.5,keepaspectratio]{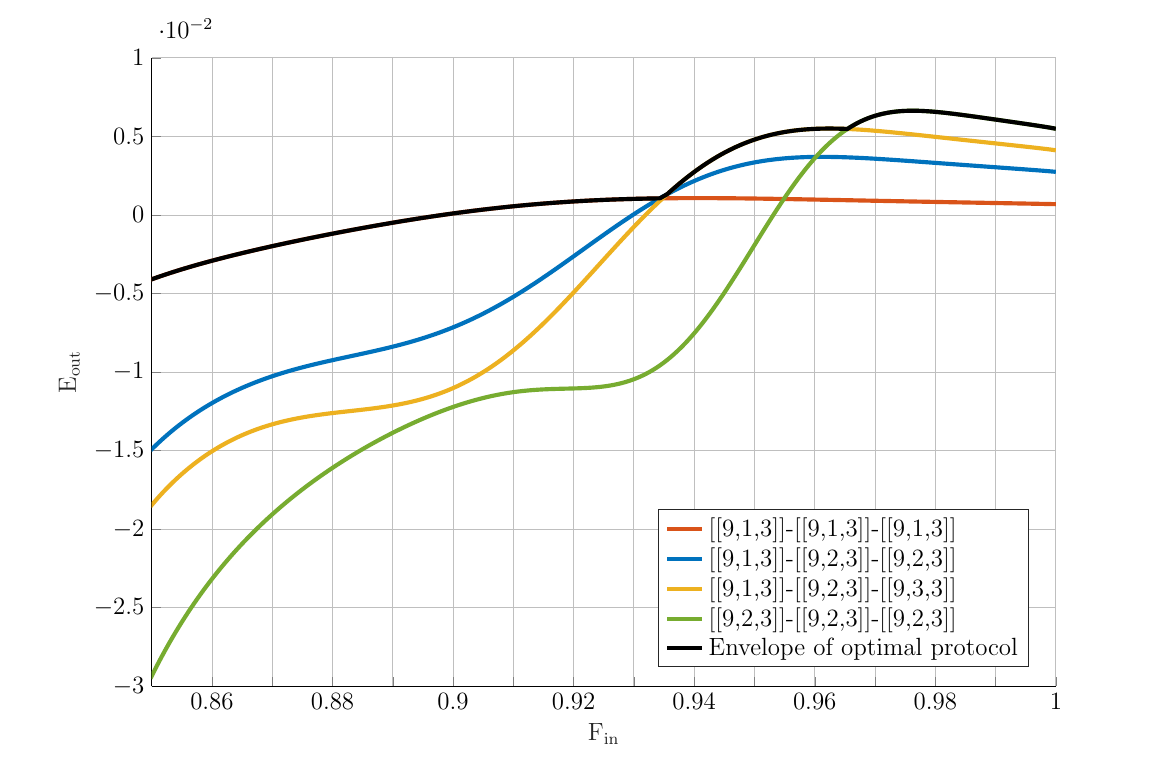}};
        \begin{scope}[x={(image.south east)},y={(image.north west)}]
            \node[anchor=south west] at (0.3,0.75)         {\includegraphics[scale=0.45,keepaspectratio,trim=240 190 210 130,clip]{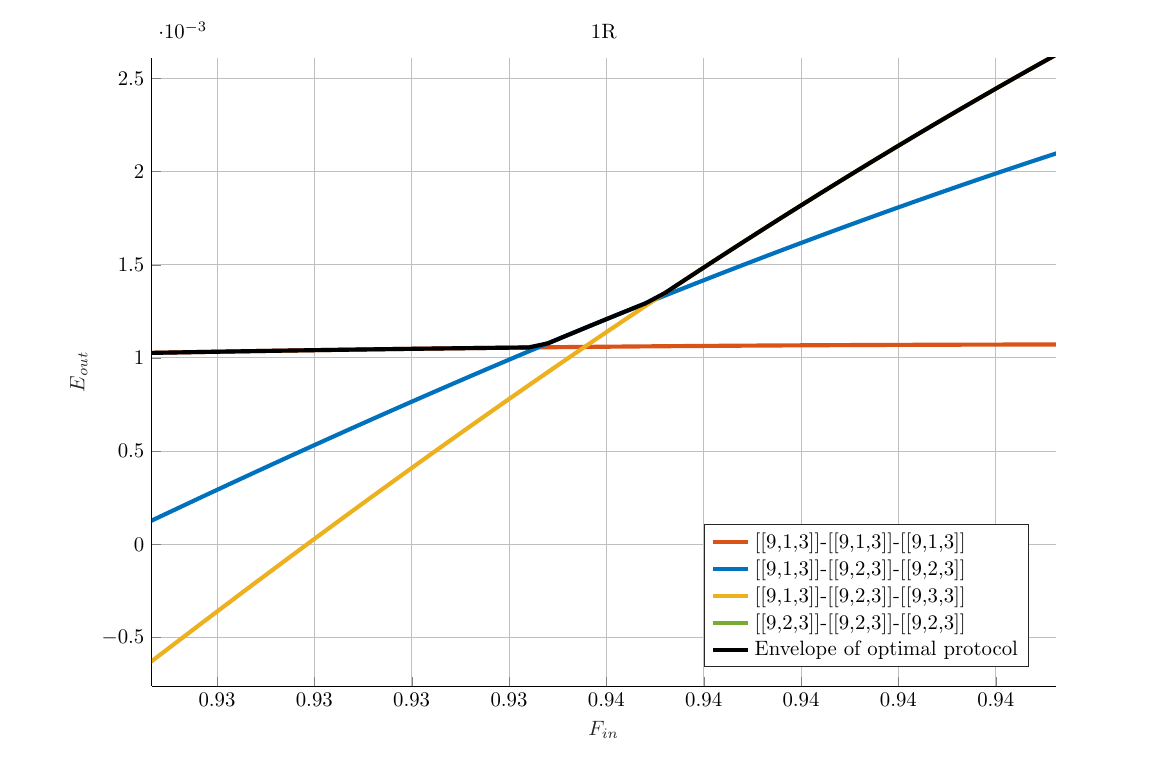}};
            \draw[draw=gray!100,thick] (0.53,0.8) rectangle (0.6,0.7);
        \end{scope}
    \end{tikzpicture}
    \caption{The $F_{\rm in}$-$E_{out}$ map for the 1-repeater case with the envelope highlighting the optimal protocol. 
    The inset highlights the switch from ${\rm Protocol}_1$ to \newtext{${\rm Protocol}_2$ to} ${\rm Protocol}_3$. \newtextnew{The key takeaway is that the optimal protocol is dependent on the range of the input fidelity and there is no global optimality.}
    } 
    \label{fig:1R_E_OL} 
    \vspace{0.5em}
    \justifying
\end{figure}

\newtext{In Fig. \ref{fig:1R_E_OL}, we plot the efficiency of the four protocols as a function of the input fidelity.}
Based on our efficiency function, we choose the protocol with the maximum efficiency as the actual protocol to be used, depending on the input fidelity, \newtext{highlighted as the envelope}. 
Finally, we obtain the switching points as follows:
\begin{itemize}
    \item From ${\rm Protocol}_1$ to ${\rm Protocol}_2$: $F_{\rm in}=0.9343$;
    \item From ${\rm Protocol}_2$ to ${\rm Protocol}_3$: $F_{\rm in}=0.9356$;
    \item From ${\rm Protocol}_3$ to ${\rm Protocol}_4$: $F_{\rm in}=0.9655$.
\end{itemize}
\newtext{
The reason we include ${\rm Protocol}_2$, even though it is optimal only for a very short input fidelity range, is that we aim to minimize changes in $k$ during each round. When switching between two closely spaced adaptive protocols, keeping $k$ as stable as possible ensures smoother system performance, especially when noise varies continuously over time.
The corresponding results for $R_{\rm out}, F_{\rm out}$ and $D_{\rm out}$ are shown in Figs.~\ref{fig:1R_R_OL},~\ref{fig:1R_F_OL} and~\ref{fig:1R_D_OL}, respectively.
}
\newtextnew{The key takeaway is that the efficiency-optimal protocol is dependent on the range of the input fidelity and there is no global optimality.}

\begingroup
\makeatletter
  \renewcommand\p@subfigure{}
  \renewcommand\thesubfigure{\thefigure.\alph{subfigure}}   
\makeatother
\begin{figure}[t]
\centering
    \begin{subfigure}[t]{0.5\textwidth}
        \centering
        \hspace*{-20pt}
        \includegraphics[scale=0.5,keepaspectratio]{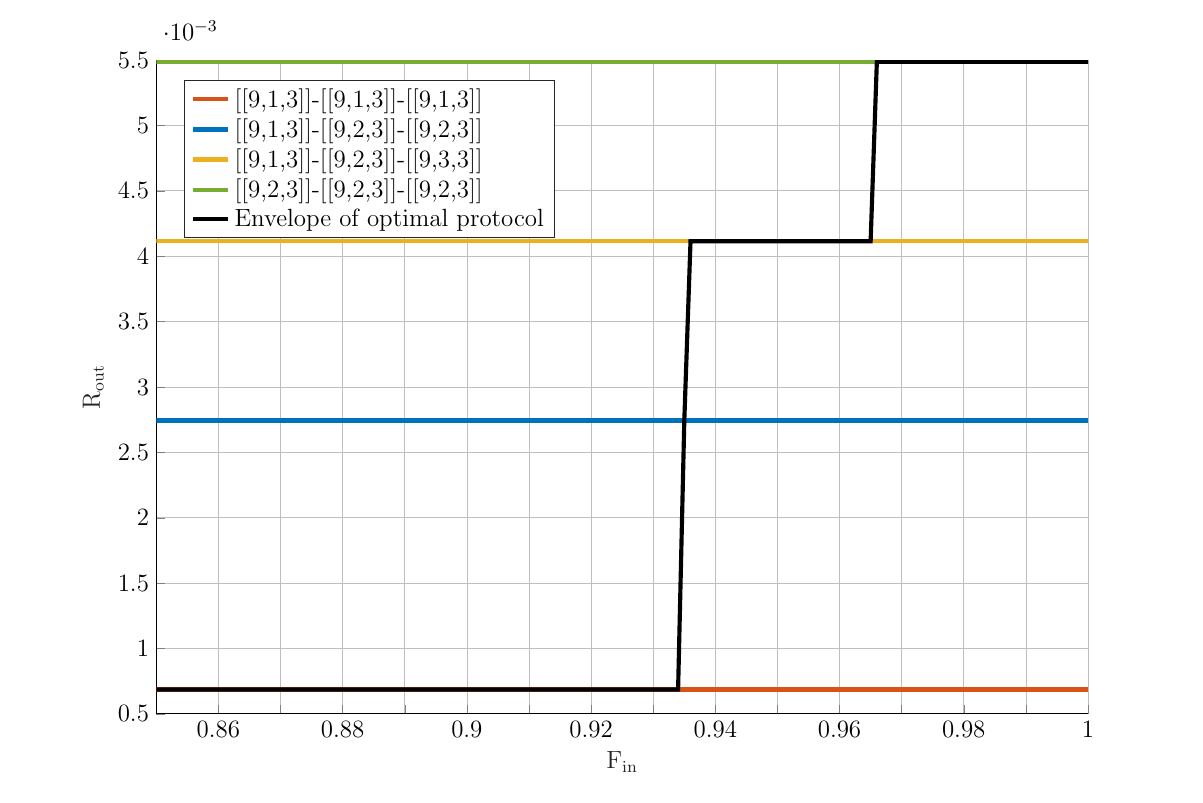} 
        \caption{The $F_{\rm in}$-$R_{out}$ map for the 1-repeater case with optimal envelope. \newtextnew{The jumps of the optimality curve correspond to the switching points of the envelope of the optimal protocol in Fig.~\ref{fig:1R_E_OL}. Hence, it is clear that the rate can be progressively increased through the adaptive switching as the input fidelity (network operating point) improves.}}
        \label{fig:1R_R_OL}
        \vspace*{-10pt}
    \end{subfigure}
\end{figure}
\begin{figure}[t]
\ContinuedFloat
\centering
    \begin{subfigure}[t]{0.5\textwidth}
        \centering
        \hspace*{-20pt}
        \includegraphics[scale=0.5,keepaspectratio]{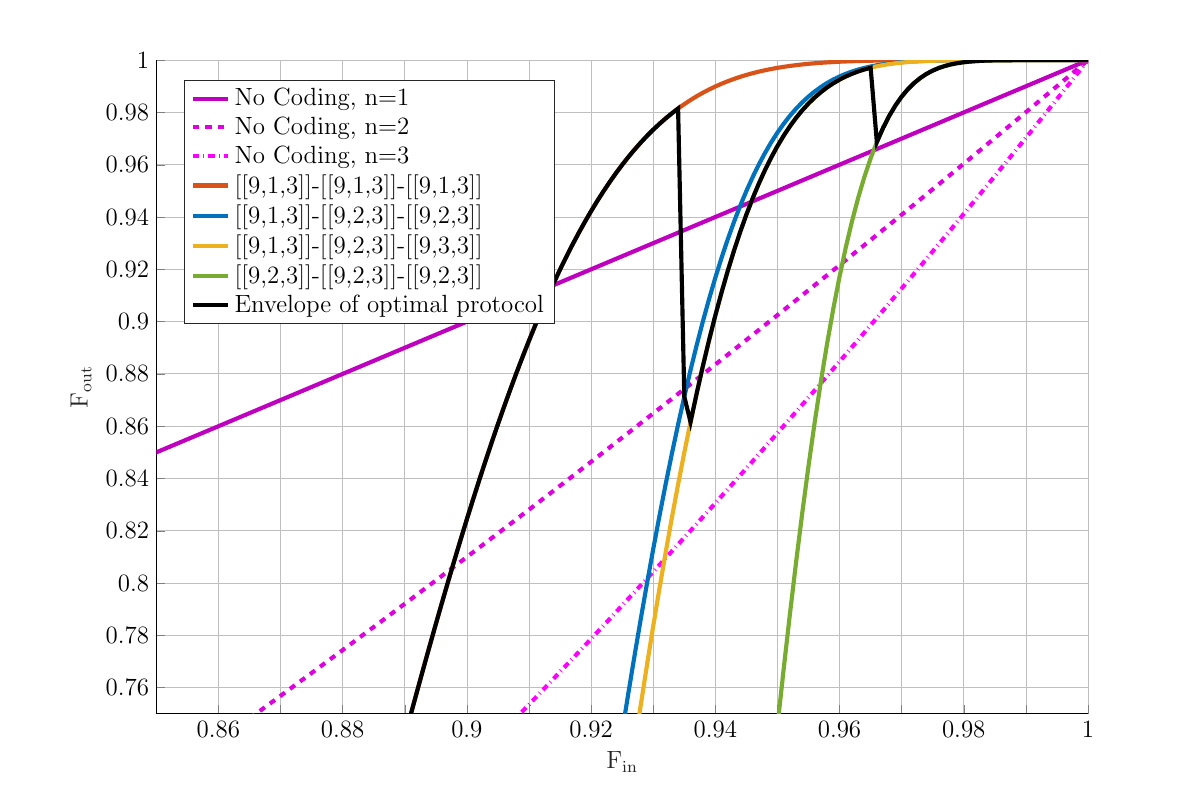} 
        \caption{The $F_{\rm in}$-$F_{\rm out}$ map for the 1-repeater case with optimal envelope. \newtextnew{The jumps of the optimality curve correspond to the switching points of the envelope of the optimal protocol in Fig.~\ref{fig:1R_E_OL}. Hence, it is interesting to note that the efficiency-improving switching points do not necessarily correspond to picking the protocol with the best fidelity. This highlights the importance of the rate-fidelity trade-off through the efficiency function, which can also be replaced with other such metrics.}}
        \label{fig:1R_F_OL}
        \vspace*{-10pt}
    \end{subfigure}
\end{figure}
\begin{figure}[t]
\ContinuedFloat
\centering
    \begin{subfigure}[t]{0.5\textwidth}
        \centering
        \hspace*{-20pt}
        \includegraphics[scale=0.5,keepaspectratio]{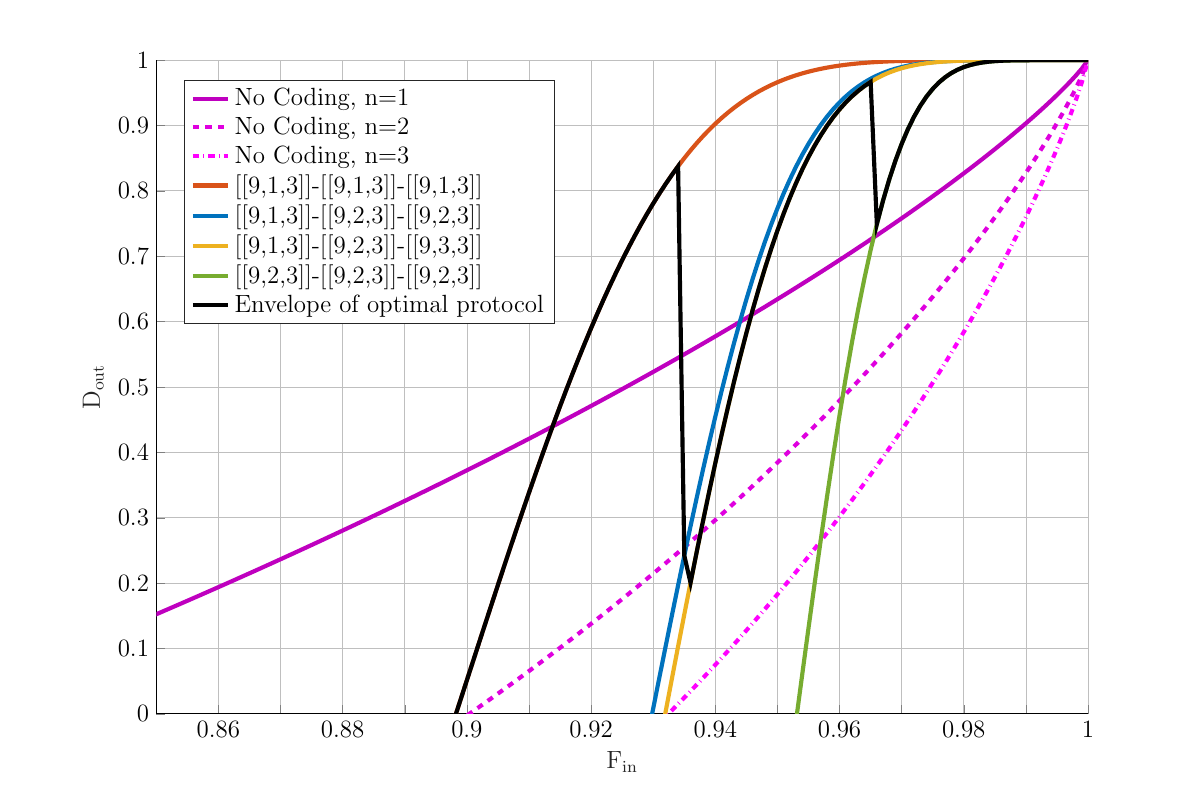} 
        \caption{The $F_{\rm in}$-$D_{out}$ map for the 1-repeater case with optimal envelope. \newtextnew{The jumps of the optimality curve correspond to the switching points of the envelope of the optimal protocol in Fig.~\ref{fig:1R_E_OL}. Evidently, the trend mirrors that of fidelity but not efficiency.}}
        \label{fig:1R_D_OL}
    \end{subfigure}
    \caption{\newtextnew{The advantage of adaptivity demonstrated in the 1-repeater case with the envelope of the optimal protocol chosen using the efficiency function. While the rate and fidelity individually differ in their choices of the optimal protocol for a given input fidelity, the efficiency function highlights the rate-fidelity trade-off through its choice of optimality.}
    }
    \label{fig:1R_OL}
\end{figure}
\endgroup

\subsubsection{Multiple Repeaters}

Based on the discussion in the single-repeater scenario, we found that the critical part is identifying the switching point based on the efficiency function.
According to our experiments, we found that for multiple repeaters, they behave similarly to the single-repeater scenario. 
However, as the number of repeaters increases, all optimal switching points move to the right, meaning that higher input fidelity is required as the trigger to switch between protocols and achieve optimal efficiency.
\newtext{See Appendix~\ref{app:multiple_repeaters} for details.}

\subsubsection{General end-to-end metrics}

Our approach aims to show that by adaptively changing the code used for distillation in each round, we can achieve a higher value for the objective function compared to using a fixed code. 
We demonstrated this using the efficiency metric \( E(F_{\rm in}) \). 
However, our key insight is that as long as a metric/objective function depends on both fidelity and rate, the adaptive approach should be advantageous. 
Even in the presence of realistic noise, an advantage should still be observed when dynamically adjusting the code/protocol depending on the network operating point. 
\newtext{Since obtaining more conclusive evidence would require building a full circuit noise model with noisy gates, developing fault tolerant methods for the involved codes, and accounting for practical constraints, which are beyond the scope of the present paper, we plan to investigate these in future work.}

\newtext{
\subsection{Limitations}

The adaptive error-correction protocol discussed above is based on quantum error correction, and in general, we refer to this as the 2G protocol~\cite{Muralidharan-scirep16}. 
From the previous discussion, we can clearly see that the pseudo-threshold is relatively high. If the input fidelity is lower than this value, then we may need to apply a 1G protocol~\cite{Muralidharan-scirep16} first to increase the input fidelity to the required threshold.
There are almost no studies in the literature regarding the transition from 1G to 2G protocols.
In the next section, we initiate this study by extending the adaptive philosophy to 1G protocols.
}

\section{Extending Adaptivity from 2G to 1G Entanglement Purification}
\label{sec:extension_to_1G}

In~\cite{10528897_Reliable}, entanglement \newtext{purification} based on DEJMPS is used to generate high-fidelity Bell pairs whose Pauli error probabilities can be strongly asymmetric. The \newtext{purified} pairs are then interpreted as an effective Pauli teleportation channel, and this asymmetry is directly exploited by asymmetric quantum error correction codes. As a result, the required number of \newtext{purification} rounds becomes a design parameter that must be optimized jointly with the chosen code, depending on the target reliability and the resource and latency constraints.

In our setting, we instead apply Werner twirling after DEJMPS (1G) to symmetrize the biased noise into an effective depolarizing model for 2G. This allows us to use a threshold-based criterion to specify the target fidelity or physical error rate required from DEJMPS, thereby providing a more transparent rule to select the number of distillation rounds. This also enables a cleaner efficiency metric that jointly accounts for distillable entanglement and the number of output qubits. In particular, when the input Bell pairs are affected by non-depolarizing noise, multiple rounds of DEJMPS do not necessarily preserve a consistent $Z$-biased structure. This makes it difficult to select or design a single bias-tailored quantum error correction code that is robust across general operating conditions. By contrast, our twirling based approach produces a uniform depolarizing description after distillation, which makes the overall design procedure easier to apply to a wide range of scenarios. In this paper, we use the depolarizing model as the main illustrative example for the input.

\subsection{1G Protocol}

In 2G protocols, all Bell pairs are retained, and QEC is applied to improve their fidelity. While it is advantageous that all Bell pairs are preserved, the threshold for effective operation is quite high.
The 1G protocol is fundamentally different. In general, it begins by generating two Bell pairs, then measuring one of them and using the result to decide whether to keep the other. This means that at least around 45\% of the Bell pairs are discarded in order to forcibly boost the output fidelity when the input fidelity is close to 0.5 in the first round. However, as the input fidelity increases, the discard rate decreases. As a result, the output fidelity can be significantly higher than that of 2G protocols, but this comes at the cost of losing many Bell pairs.

During the NISQ era, source Bell pairs are extremely precious, so using 1G protocols is not always the most practical choice. However, they serve as an effective approach to raise the input fidelity to the threshold required for 2G operation.

\subsection{BBPSSW and DEJMPS}

\begin{figure*}[t]
\centering
    \begin{subfigure}[b]{0.45\textwidth}
        \centering
        \hspace*{-30pt}
        \vspace*{-10pt}
        \includegraphics[width=1.2\textwidth]{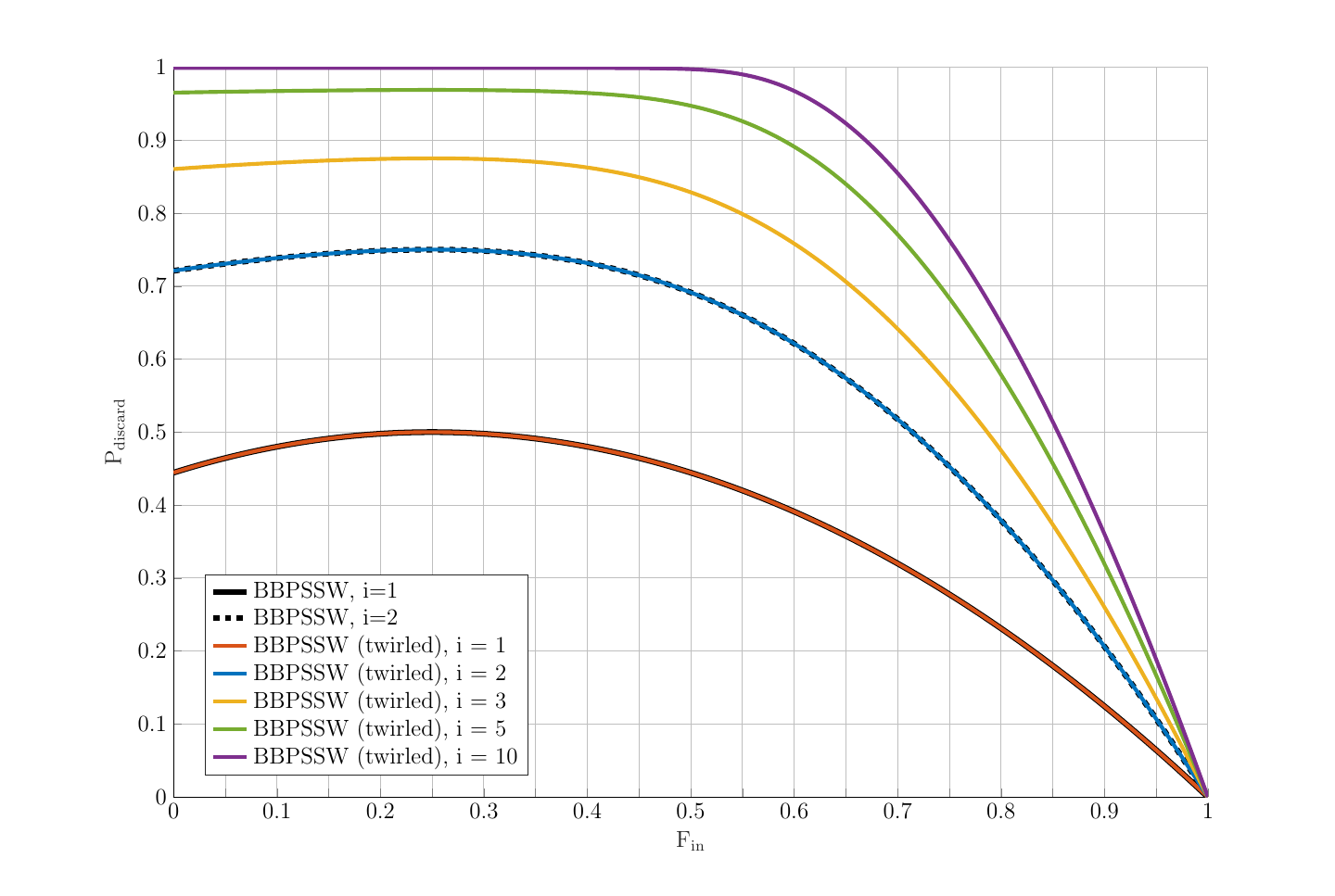} 
        \caption{BBPSSW (twirled)}
        \label{fig:L1}
    \end{subfigure}
    \hspace{0.03\textwidth}
    \begin{subfigure}[b]{0.45\textwidth}
        \centering
        \hspace*{-30pt}
        \vspace*{-10pt}
        \includegraphics[width=1.2\textwidth]{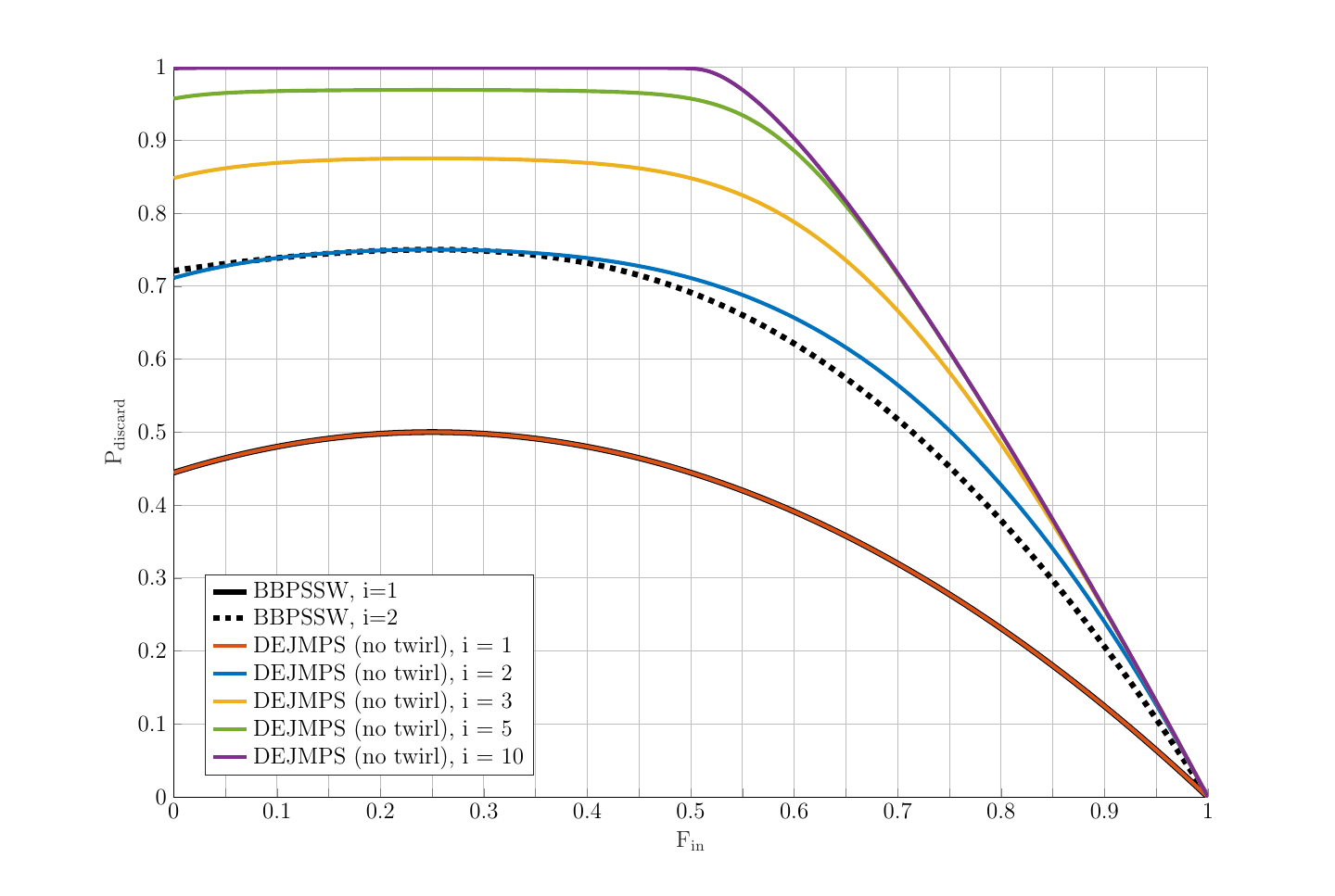} 
        \caption{DEJMPS (no twirl)}
        \label{fig:L2}
    \end{subfigure}
    \caption{Comparison of discard probabilities and fidelities between DEJMPS in Fig.~\ref{fig:DEJMPS_nt} and BBPSSW in Fig.~\ref{fig:BBPSSW_t} for different number of rounds. As an example, for $F_{\rm in}=0.6$, when $i=2$ (rounds), DEJMPS discards 0.65661 of the Bell pairs while BBPSSW discards 0.621285, which is 0.035325 (5.69\%) more. But the output fidelity of DEJMPS is 0.688616, compared to 0.644639 for BBPSSW, an improvement of 0.043977 (6.82\%). When $i=3$, DEJMPS discards 0.78774 of the Bell pairs while BBPSSW discards 0.758215, which is 0.029525 (3.89\%) more. But again, the output fidelity of DEJMPS is 0.77193, compared to 0.67288 for BBPSSW, an improvement of 0.09905 (14.72\%).}
    \label{fig:L}
    \vspace{0.5em}
    \justifying
    
\end{figure*}

The goal of using a 1G protocol is to solve the problem that when the input fidelity is lower than the pseudo-threshold of 2G protocols, the 2G protocol will reduce the output fidelity instead of improving it. Therefore, for low input fidelity, we need to select a 1G protocol that uses fewer discarded Bell pairs while still providing a fidelity improvement. Since we use a depolarizing channel in the first round, the bias begins to provide a benefit starting from the second round.

For the discard probability in BBPSSW, we can easily calculate the value for the first round based on the denominator of Eq.~\eqref{eq:BBPSSW}:
\begin{align}
    P_{discard}=1-(F_{\rm in}^2 + \frac{2}{3}F_{\rm in}(1-F_{\rm in}) + \frac{5}{9} (1-F_{\rm in})^2).
\end{align}
However, starting from the second round, we need to use Eq.~\eqref{eq:total discard} to obtain the total discard probability.
%
In Fig. \ref{fig:L}, we compare the total discard probability as a function of the input fidelity over different numbers of rounds for BBPSSW (twirled) and DEJMPS (no twirl), respectively.
%
Comparing Fig. \ref{fig:L} with Fig. \ref{fig:BBPSSW_t} and Fig. \ref{fig:DEJMPS_nt}, we can obtain some numerical results. Since we focus on the low input-fidelity region, it is appropriate to consider the case where $F_{\rm in} = 0.6$.

When $i=2$, we find that DEJMPS discards 5.69\% more Bell pairs than BBPSSW, but the fidelity is improved by $6.82\%$. As the number of rounds increases, this benefit becomes more and more significant. When $i=3$, we find that DEJMPS discards $3.89\%$ more Bell pairs than BBPSSW, but the fidelity is improved by $14.72\%$.
Therefore, we can see that as the number of rounds increases, DEJMPS continues to improve rapidly in the low input-fidelity region. In almost all performance comparisons, DEJMPS outperforms other 1G protocols. For this reason, we select DEJMPS as the representative 1G protocol for our analysis and use it as the benchmark for comparison with 2G protocols.

\subsection{Comparison of 1G and 2G Protocols without Repeaters}

For the 2G protocols, we considered the quantum error-correcting codes $\llbracket9,1,3\rrbracket$, $\llbracket9,2,3\rrbracket$, and $\llbracket9,3,3\rrbracket$. These codes exhibit relatively high input fidelity thresholds (or pseudo-thresholds). Therefore, before reaching these thresholds, it is necessary to apply 1G protocols to boost the input fidelity to a level where 2G distillation becomes effective.

We consider the $\llbracket9,3,3\rrbracket$ code as an example because its effective rate is approximately $\frac{1}{3}$, compared with $\frac{1}{9}$ for the $\llbracket9,1,3\rrbracket$ code, while maintaining the same communication distance. The threshold input fidelity for the $\llbracket9,3,3\rrbracket$ code is around $0.9563$, meaning that 1G purification must first raise the fidelity from the low input-fidelity region (e.g., between $0.5$ and $0.8$) to this threshold before any 2G correction can be meaningfully applied.

\begin{figure}
    \centering
    \hspace*{-30pt}
    \includegraphics[scale=0.43,keepaspectratio]{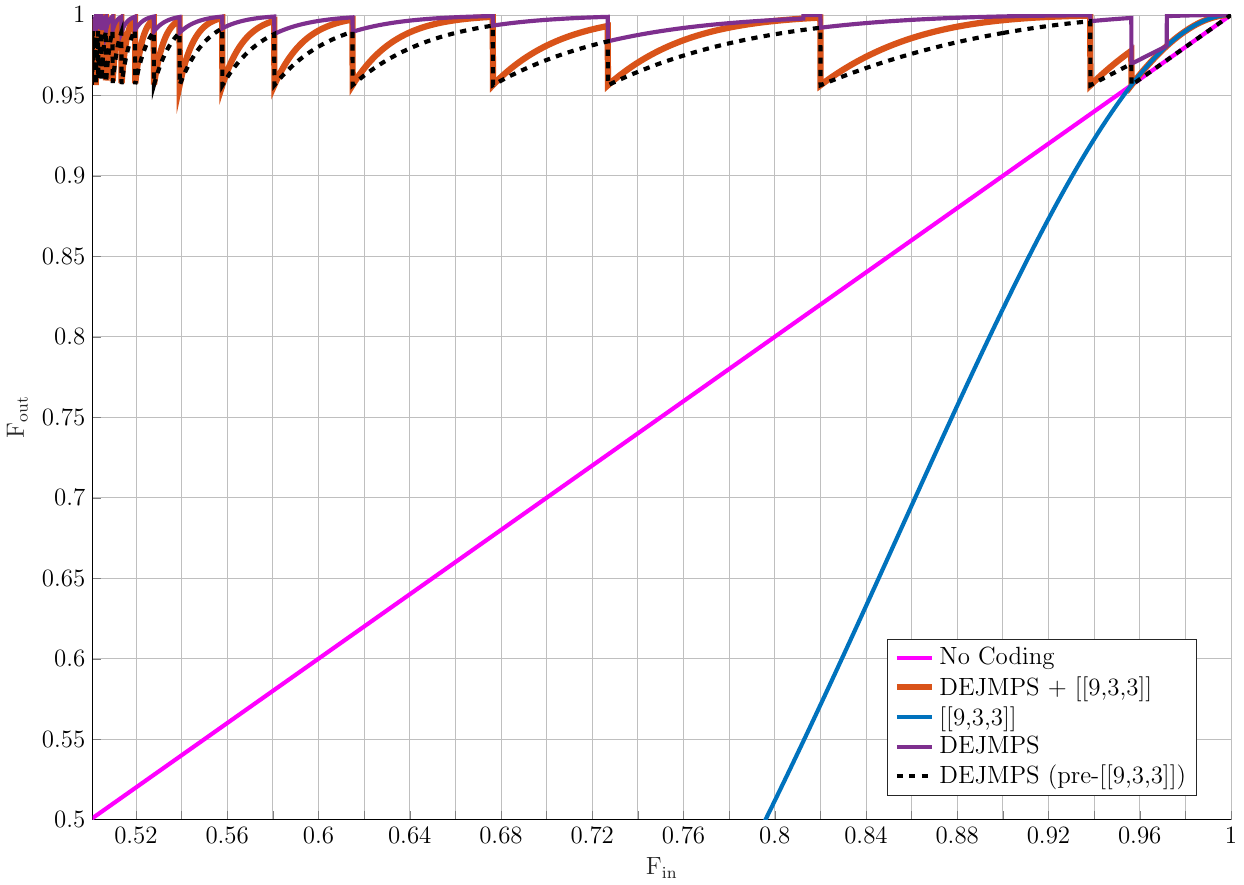}
    \caption{The $F_{\rm in}$–$F_{\rm out}$ mapping for the case without repeaters. DEJMPS (pre-$\llbracket 9,3,3\rrbracket$) refers to applying several rounds of DEJMPS until the fidelity reaches or exceeds the pseudo-threshold of $\llbracket 9,3,3\rrbracket$. Then, one additional round of $\llbracket 9,3,3\rrbracket$ coding is performed to obtain DEJMPS+$\llbracket9,3,3\rrbracket$. Finally, we set the fidelity achieved by DEJMPS+$\llbracket 9,3,3\rrbracket$ as the target and determine the minimum number of DEJMPS rounds required to match or surpass it. 
    }
    \label{fig:G1_F}
\end{figure}
In Fig. \ref{fig:G1_F}, we compare the output fidelity as a function of the input fidelity for the 1G, 2G, and hybrid cases. In particular, we can clearly see that the curve is not continuous when 1G is included in the protocol. Whenever the curve has a discontinuous jump, we call the turning point a “checkpoint”. This checkpoint is important because it marks where the protocol jumps to a different number of rounds.
When $F_{\rm in} = 0.5$, no matter how many rounds we execute, the output fidelity always remains $0.5$. Therefore, in Fig. \ref{fig:G1_F}, the x-axis range for the input fidelity is set from $0.501$ to $1$.

After the fidelity is boosted above the code threshold by 1G purification, a single round of 2G \newtext{distillation} is performed on that purified state. The output fidelity obtained after this 2G step serves as the target benchmark. \newtext{To compare this hybrid 1G+2G strategy with purely 1G}, we then examine the minimum number of 1G DEJMPS rounds required to directly achieve an output fidelity that is not lower than the benchmark. 
When the input fidelity is higher, each round still improves the fidelity, but the improvement in each new round is smaller than in the previous round.
Thus, a single round is sufficient to achieve the target fidelity over a wider range when the input fidelity is already high. As a result, only a few checkpoints appear in the high-fidelity region, while more checkpoints are observed when the input fidelity is close to $0.5$.

\begin{figure}
    \centering
    \vspace*{-5pt}
    \hspace*{-30pt}
    \includegraphics[scale=0.43,keepaspectratio]{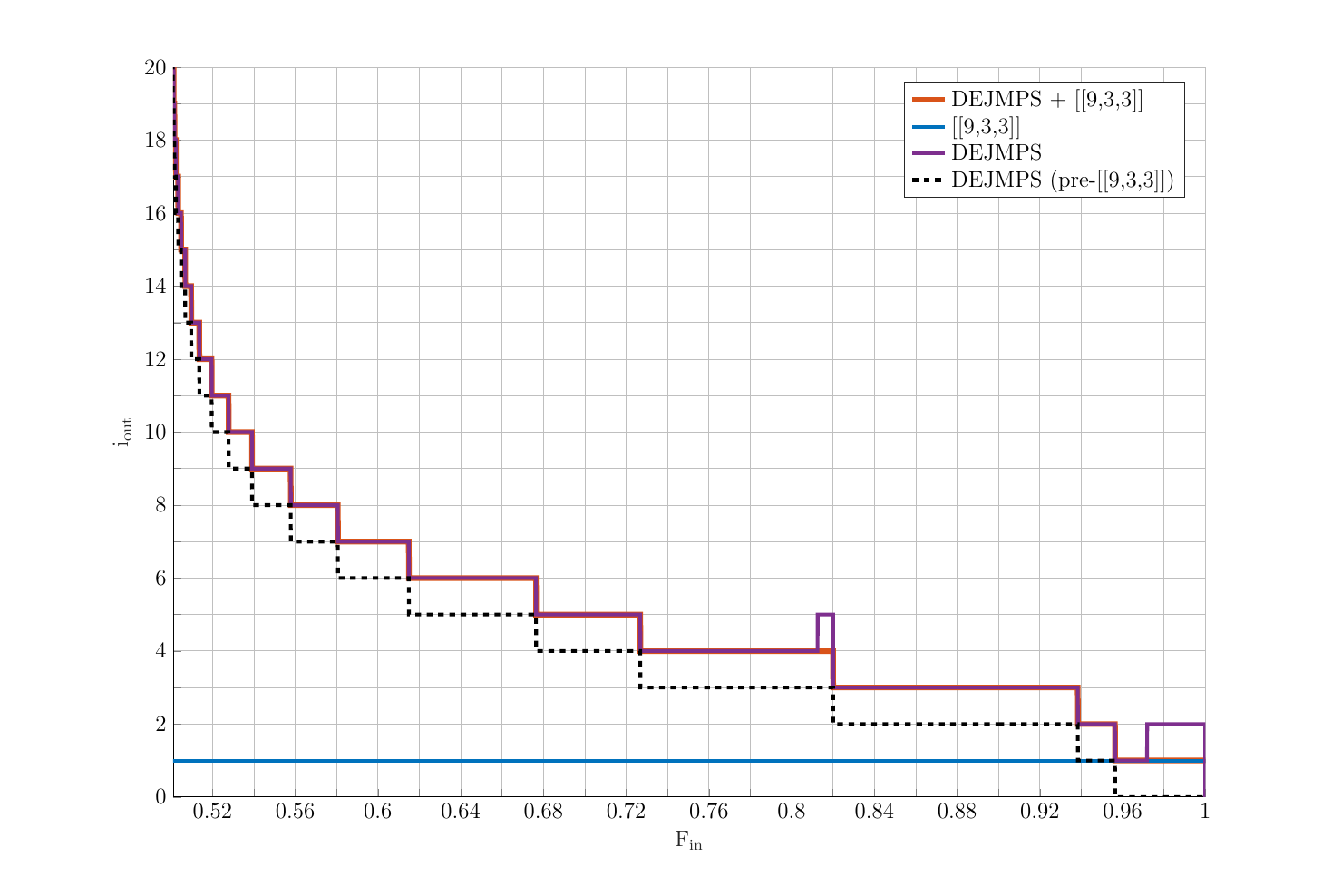}
    \vspace*{-23pt}
    \caption{The $F_{\rm in}$–$i_{\rm out}$ mapping for the case without repeaters. This plot shows the relationship between the input fidelity and the number of required protocol rounds. Several rounds of DEJMPS are applied to increase the input fidelity until it reaches the pseudo-threshold of $\llbracket 9,3,3\rrbracket$. Once the input fidelity is at or above this pseudo-threshold, DEJMPS rounds are no longer used and a single round of 2G QEC is applied instead, minimizing the total number of rounds required and avoiding unnecessary Bell-pair discards. We can observe that the difference between DEJMPS (pre-$\llbracket 9,3,3\rrbracket$) and DEJMPS+$\llbracket 9,3,3\rrbracket$ is typically only one additional round, and only a few cases require two rounds. This means that in most cases, the number of rounds required for DEJMPS and DEJMPS+$\llbracket 9,3,3\rrbracket$ is the same.}

    \label{fig:G1_i}  
\end{figure}
In Fig. \ref{fig:G1_i}, we show, as a function of the input fidelity, the number of rounds needed to reach the target output fidelity for protocols that involve 1G.
Interestingly, the results show that only one or two additional DEJMPS rounds are sufficient to reach the target benchmark after several DEJMPS rounds have raised the fidelity to the pseudo-threshold. Because only a single round of 2G QEC is performed, the total number of rounds required for purely 1G purification and the hybrid 1G+2G approach becomes the same in most cases.

\begin{figure}
    \centering
    \hspace*{-30pt}
    \includegraphics[scale=0.43,keepaspectratio]{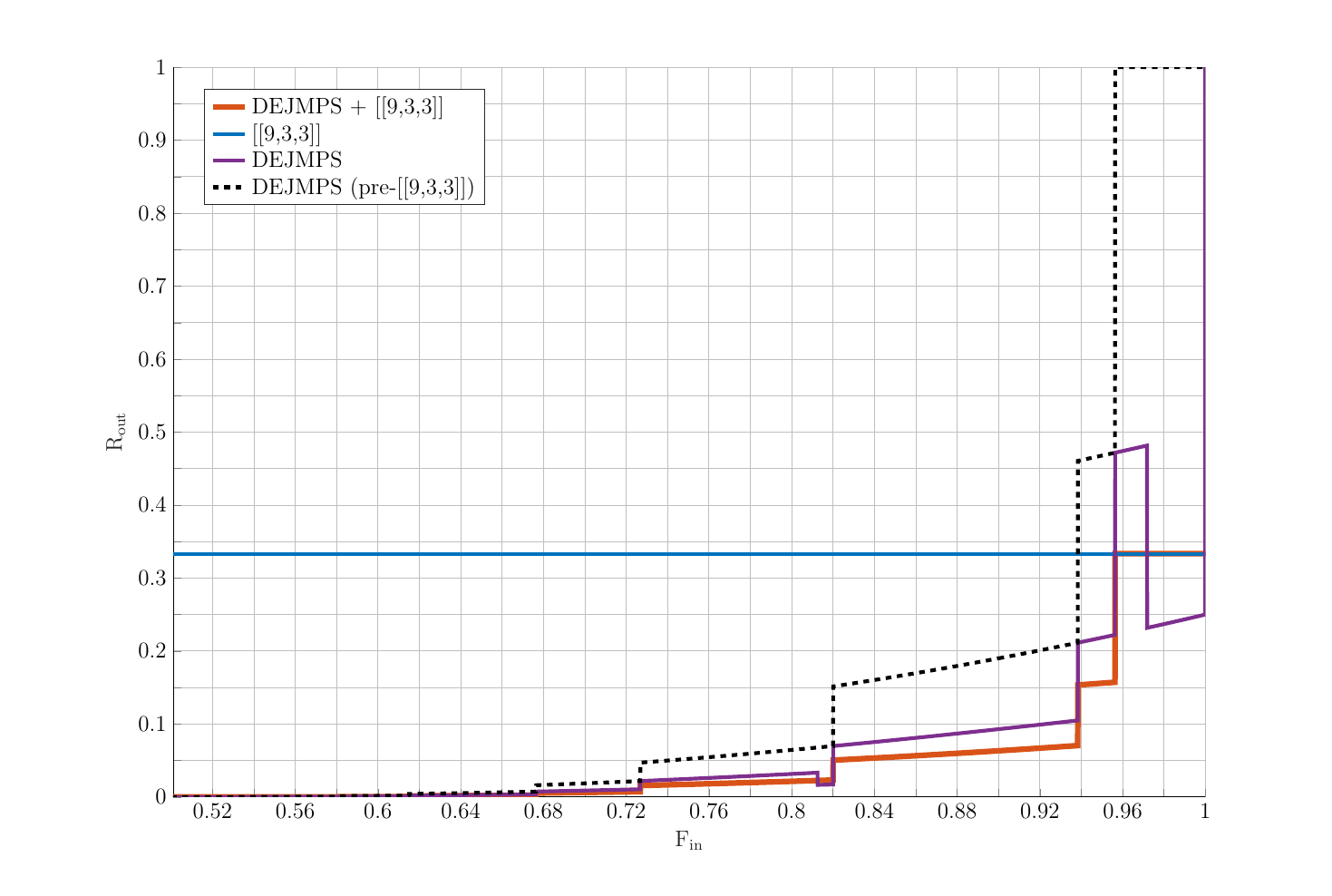}
    \caption{The $F_{\rm in}$–$R_{\rm out}$ mapping for the case without repeaters. In this plot, we consider the discard probability $P_{\rm total\ discard}$ as part of the 1G rate, such that $R_{\rm out} = \frac{n_{\rm out}}{n_{\rm in}} (1 - P_{\rm total\ discard})$.}

    \label{fig:G1_R}  
\end{figure}

Once we determine the required number of DEJMPS rounds, we can calculate the corresponding rate, where the discard probability is included as part of the efficiency, such that $R_{\rm out} = \frac{n_{\rm out}}{n_{\rm in}} (1 - P_{\rm total\ discard})$, where $n_{in}$ and $n_{out}$ represent the total number of input and output Bell pairs, respectively. Therefore, these quantities must be evaluated separately for the 1G, 2G, and hybrid 1G+2G cases.

\subsubsection{Purely 1G}
When discussing the purely 1G case, the rate calculation is straightforward. In DEJMPS, each round consumes two input Bell pairs to produce one output Bell pair or a discard. This means that once the number of DEJMPS rounds $i$ is known, the rate can be directly obtained as follows:
\begin{align}
    R_{\rm out} = \frac{1}{2^i}(1-P_{\rm total\ discard}). \label{eq:R1G}
\end{align}

\subsubsection{Purely 2G}
Since this section considers only a single 2G round and there is no discard in the purely 2G case, the rate can be directly determined by the code used. If we use a $\llbracket n,k,d\rrbracket$ code, the rate can be obtained as follows:
\begin{align}
    R_{\rm out} = \frac{k}{n}. \label{eq:R2G}
\end{align}

\subsubsection{Hybrid 1G+2G}
In our case, when using the hybrid strategy, several rounds of DEJMPS are performed first, followed by a single round of 2G QEC. Therefore, the rate can be obtained by combining Eq.~\eqref{eq:R1G} and Eq.~\eqref{eq:R2G}, when a $\llbracket n,k,d\rrbracket$ code is used for the 2G stage, as follows:
\begin{align}
    R_{\rm out} = \frac{k}{2^i n}(1-P_{\rm total\ discard}). 
\end{align}

In Fig. \ref{fig:G1_R}, we compare the rates for the 1G, 2G, and hybrid cases. We observe that the discard probability $P_{discard}$ does not significantly affect the rate when DEJMPS is involved. Moreover, since the rate of the $\llbracket 9,3,3\rrbracket$ code is always $\frac{1}{3}$, while the rate for strategies involving DEJMPS depends heavily on the number of rounds, the $\llbracket 9,3,3\rrbracket$ code maintains an advantage in most cases.

However, when examining the fidelity results in Fig. \ref{fig:G1_F}, and the zoomed-in version in Fig. \ref{fig:G2_F} that will be discussed in the next section, we observe that the $\llbracket 9,3,3\rrbracket$ code does not provide any fidelity advantage. This indicates that we need to \newtext{adapt the efficiency metric for strategies involving 1G purification}.

\subsection{Exploring an Efficiency Function for Hybrid 1G+2G Protocols Without Repeaters}

A new efficiency metric is required, because it not only allows us to determine which strategy involving 1G purification performs better, but also indicates when to switch from one protocol to another. In the previous section, we introduced an efficiency function to describe the 2G QEC strategy under different repeater scenarios and used it to determine the optimal switching point between protocols:
\begin{align}
E(F_{\rm in}) = \frac{n_{out} D(F_{\rm out}(F_{\rm in}))}{n_{in} D(F_{\rm in})}.
\end{align}
However, when 1G purification is involved, the discard probability $P_{discard}$ must be included in the formula. Although it is not the primary factor affecting the overall trend of the curve, it reflects the intrinsic behavior of the 1G protocol. Therefore, we update the efficiency function as follows:
\begin{align}
E(F_{\rm in}) = \frac{n_{out} D(F_{\rm out}(F_{\rm in}))}{n_{in} D(F_{\rm in})}(1-P_{\rm total\ discard}).
\end{align}

\begin{figure*}
\centering
    \begin{subfigure}[b]{0.45\textwidth}
        \centering
        \hspace*{-30pt}
        \vspace*{-10pt}
        \includegraphics[width=1.2\textwidth]{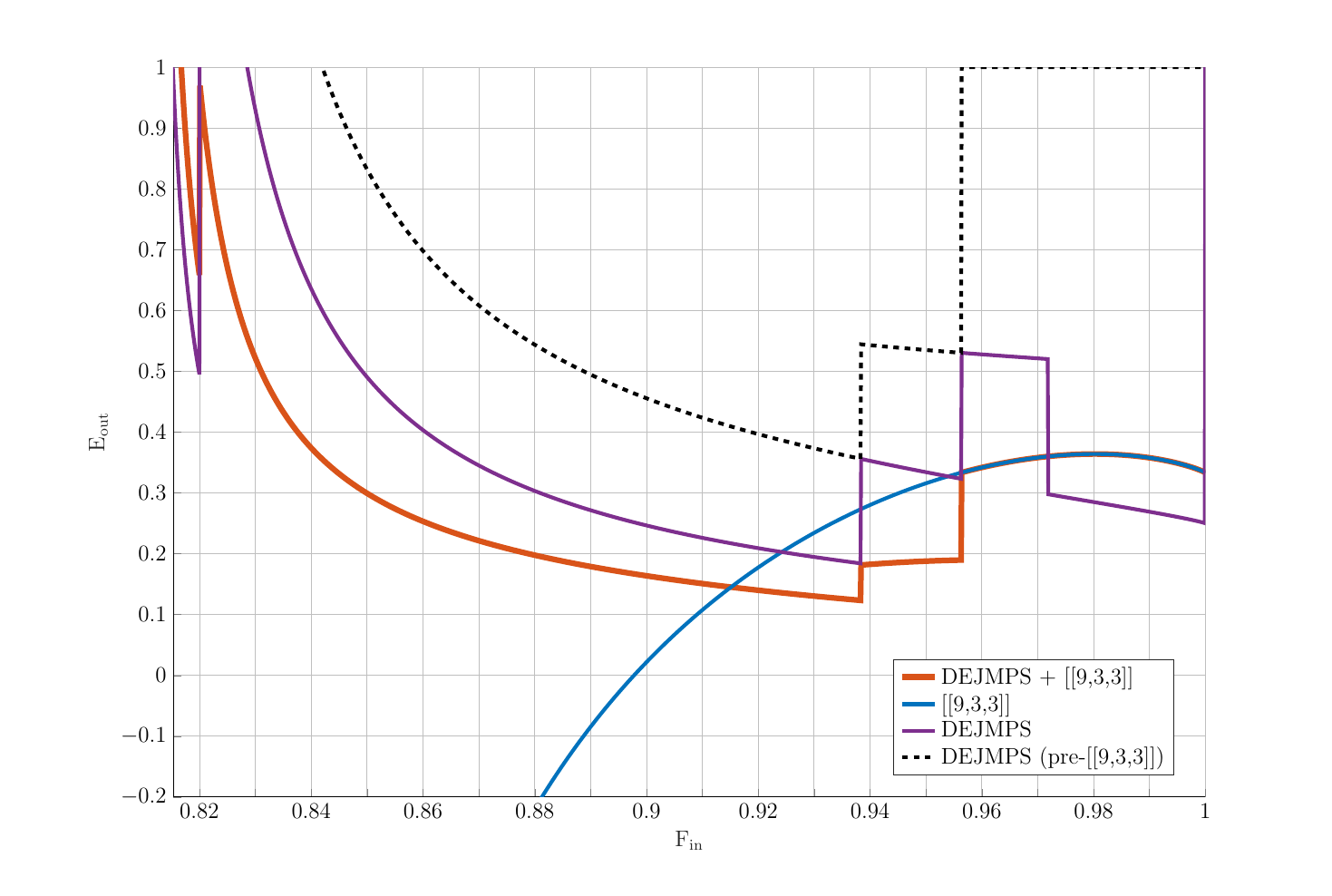} 
        \caption{Efficiency}
        \label{fig:G2_E}
    \end{subfigure}
    \hspace{0.03\textwidth}
    \begin{subfigure}[b]{0.45\textwidth}
        \centering
        \hspace*{-30pt}
        \vspace*{-10pt}
        \includegraphics[width=1.2\textwidth]{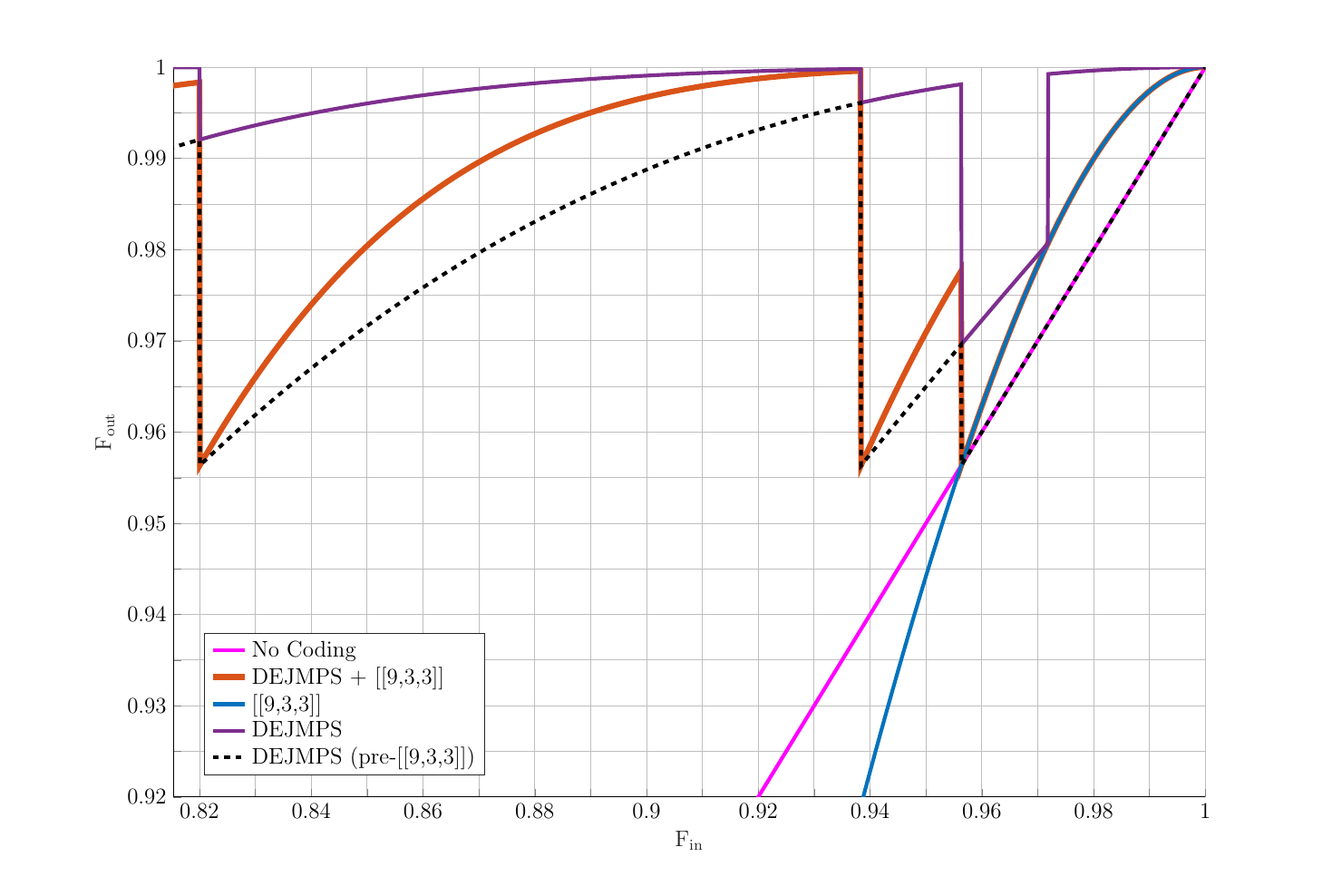} 
        \caption{Fidelity}
        \label{fig:G2_F}
    \end{subfigure}
    \caption{Using the previously defined formula to generate the efficiency function, we observe that as the input fidelity approaches 0.81071, the efficiency function becomes increasingly unstable until it diverges to infinity. In contrast, when the input fidelity is greater than 0.81071, the efficiency function accurately captures all critical points corresponding to the output fidelity.} 
    \label{fig:G2_eff}
    \vspace{0.5em}
    \justifying
\end{figure*}
In Fig. \ref{fig:G2_eff}, we show the efficiency and the corresponding output fidelity as functions of the input fidelity. We compute the efficiency using the function introduced in the previous section for 2G protocols.
We can see that this metric still effectively characterizes 1G purification. The efficiency curve behaves consistently as expected when $F_{in}>0.81071$. However, when $F_{in}<0.81071$, we have $D(F_{\rm in})<0$, which results in $E(F_{\rm in})<0$. In particular, at $F_{in}=0.81071$, $D(F_{\rm in})=0$, causing $E(F_{\rm in})$ to diverge to infinity.

In fact, when examining Fig. \ref{fig:G1_F}, there are several pieces of information that we aim to extract from the efficiency function, which become difficult to obtain when 1G purification is involved:
\begin{enumerate} 
\item We aim to identify the precise checkpoints of the output fidelity from the efficiency function. However, this becomes difficult when the input fidelity is close to or below $0.81071$, because $D(F_{\rm in}) \leq 0$ in that region.

\item We aim to determine which strategy provides a better benefit for the same input fidelity. A better strategy should exhibit higher efficiency than the others. However, once $D(F_{\rm in})<0$, the preferred strategy may instead show a lower efficiency than the alternatives.

\item Although this is not essential, to ensure that the efficiency behaves like a proper efficiency metric, we may apply normalization or a similar scaling process to keep the results within the range $[0, 1]$.
\end{enumerate}

The first issue we need to address is that $D(F_{\rm in}) \leq 0$. Since $D(F_{\rm in})$ appears in the denominator of our efficiency function, this prevents us from using $F_{\rm in}$ directly as the reference variable. Moreover, because 1G protocols generate multiple checkpoints, an appropriate way to redefine the efficiency function is to incorporate DEJMPS performance into the metric.
After several attempts, we found that using the minimum number of DEJMPS rounds required to reach a distillable entanglement of $0.12$ is a reasonable \newtext{baseline} to include in our efficiency function. Thus, we update the efficiency function as follows:
\begin{align}
E(F_{\rm in}) = \frac{n_{out} D(F_{\rm out}(F_{\rm in}))}{n_{in} D(F_{\rm in})_{\rm DEJMPS>=0.12}}(1-P_{\rm total\ discard}).
\end{align}
Here, $D(F_{\rm in})_{\rm DEJMPS>=0.12}$ indicates that if the input distillable entanglement is below $0.12$, then a minimum number of DEJMPS rounds must be applied to raise it to at least $0.12$. If the input distillable entanglement is already greater than $0.12$, no purification is needed for the \newtext{denominator baseline}. 
Since our goal here is only to address the issues arising from $D(F_{\rm in})$, we continue to use the total numbers for $n_{in}$ and $n_{out}$ without excluding the DEJMPS rounds in the baseline.
Since we want the efficiency to reflect a balance between the communication rate and the distillable entanglement, we do not modify the other components of the metric. Meanwhile, we expect the efficiency to remain below one. 
Through extensive checks, we found that $0.12$ serves as an effective threshold for the input distillable entanglement and keeps the maximum value of the efficiency function below 1. Currently, we do not have a good theoretical understanding of the significance of this value, but we plan to study this in future work.

To address the issue that the efficiency may become negative when the output fidelity is lower than 0.81071, we simply set any negative efficiency value to zero. Therefore, the final efficiency function is given below:
\begin{align}
& E(F_{\rm in}) \notag \\
&= \max\left\{ \frac{n_{out} D(F_{\rm out}(F_{\rm in}))}{n_{in} D(F_{\rm in})_{\rm DEJMPS>=0.12}}(1-P_{\rm total\ discard}), 0 \right\}.
\end{align}

\begin{figure*}
\centering
    \begin{subfigure}[b]{0.45\textwidth}
        \centering
        \hspace*{-30pt}
        \vspace*{-10pt}
        \includegraphics[width=1.2\textwidth]{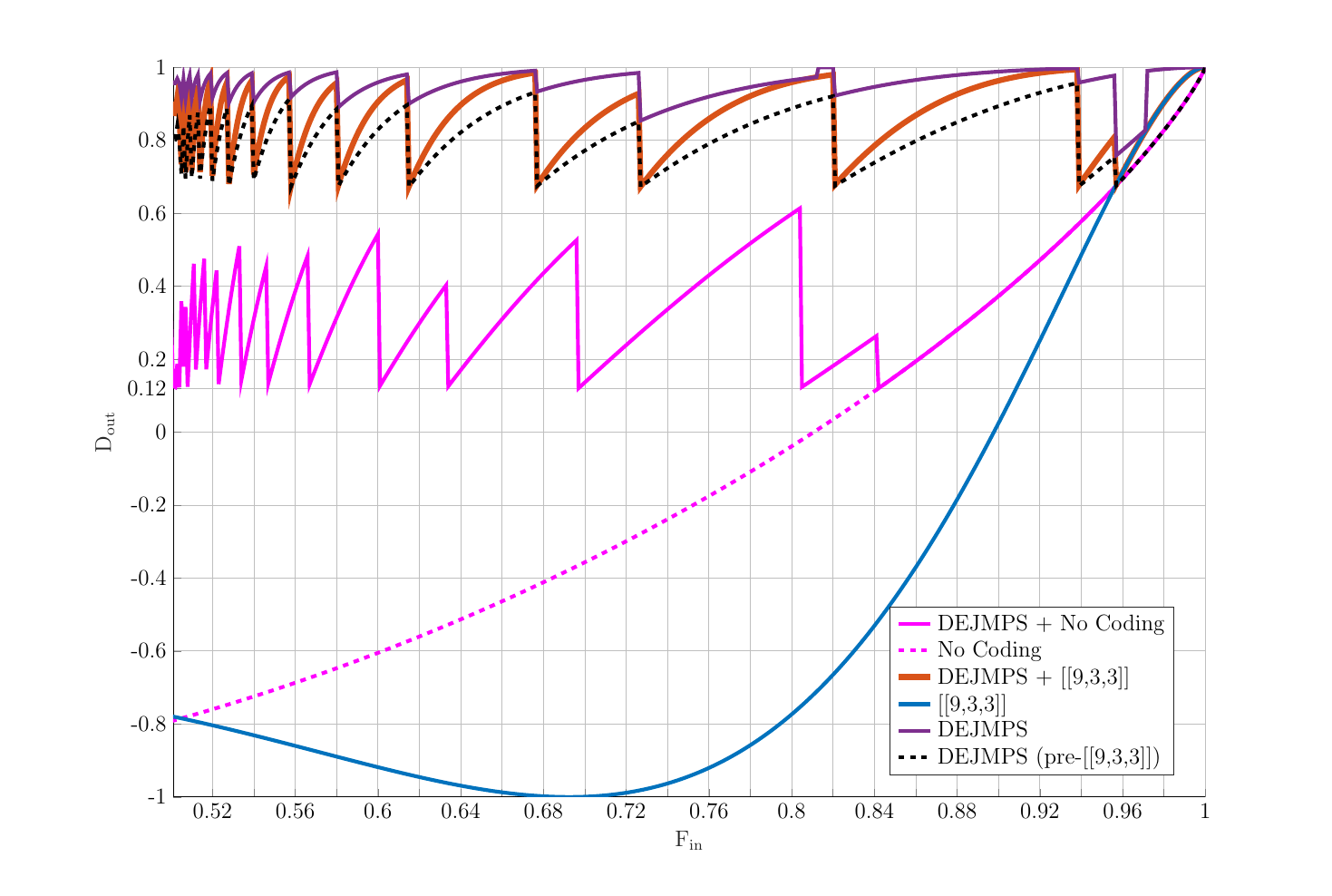} 
        \caption{Distillable Entanglement}
        \label{fig:G3_D}
    \end{subfigure}
    \hspace{0.03\textwidth}
    \begin{subfigure}[b]{0.45\textwidth}
        \centering
        \hspace*{-30pt}
        \vspace*{-10pt}
        \includegraphics[width=1.2\textwidth]{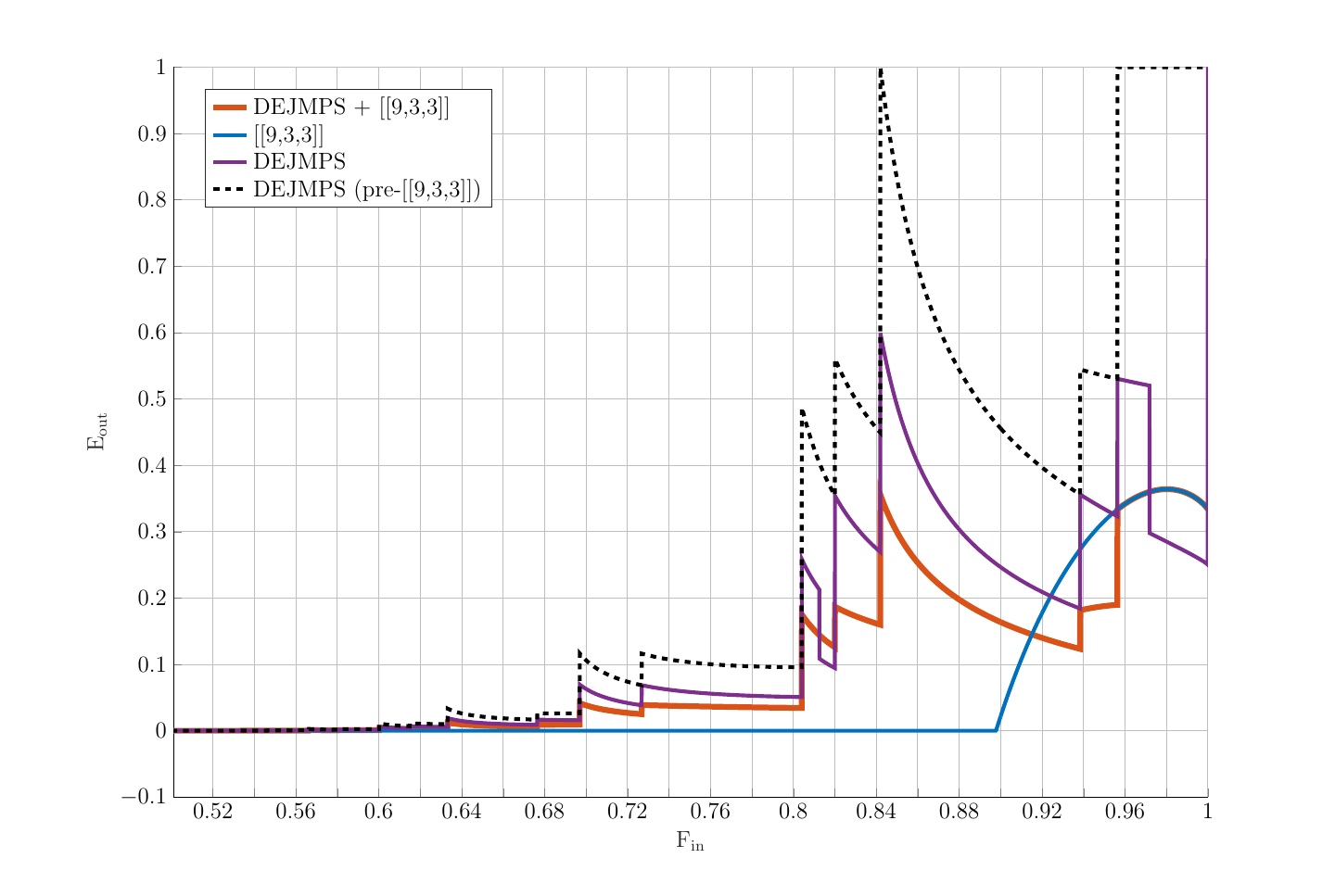} 
        \caption{New Efficiency Strategy}
        \label{fig:G3_E}
    \end{subfigure}
    \caption{The distillable entanglement and new efficiency metric by using the DEJMPS + No-Coding strategy as the baseline for a single link.
    } 
    \label{fig:G3_ED}
    \vspace{0.5em}
    \justifying
    
\end{figure*}

In Fig. \ref{fig:G3_ED}, we plot the efficiency computed using the updated efficiency function, along with the corresponding distillable entanglement.
We can see that the new efficiency function can clearly reveal the checkpoints corresponding to the output fidelity. The better strategy consistently exhibits a higher efficiency, and the resulting values remain within the range between 0 and 1.

The only remaining issue is that when the DEJMPS baseline reaches a checkpoint but the protocol under evaluation does not, the efficiency function still displays a checkpoint artifact. For example, when the input fidelity is around 0.84, there should clearly be no checkpoint for any protocol. However, the efficiency function still displays one, simply because the DEJMPS baseline reaches a checkpoint at that point. 

Since we aim to develop a general version of the efficiency function that can be applied to any type of protocol without modifying the baseline, the strategy we propose provides a good balance among all relevant factors.

\newtext{
\subsection{Practical Considerations}

While the protocols are dynamically chosen based on the input fidelity, they only depend on the current snapshot of the network, i.e, the link parameters and local gate noise (currently captured through a simple depolarizing model).
For a small network, a central controller can frequently update its network snapshot and use that to decide the sequence of codes for a given number of rounds of purification and distillation to achieve the target efficiency metric.
Note that pre-generated input-output fidelity plots are sufficient for this purpose.
Once it decides, it can broadcast the policy to all repeaters, which is then implemented at each node.
For a larger network, there can be a hierarchy of controllers that govern smaller and local sub-networks.
We don't expect the network parameters to change drastically until the entire protocol is completed.
Hence, besides the classical communication needed for DEJMPS purification as necessary, there is no further communication needed to choose the codes; each node implements the policy by keeping track of the current round of distillation.
The only other classical communication is one-way communication of the syndromes for each distillation between repeaters.
It would be interesting to consider a database of codes and a decentralized approach where each node dynamically changes the codes as per its current estimate of fidelities.

Raw Bell pairs are generated at each link and the success is heralded to the repeaters. The repeaters can then perform either DEJMPS purification or code-based distillation as approproate. Each repeater knows when the distillation is complete on both sides with the two neighboring repeaters. Then, it will perform a swap on the distilled qubits and communicate the completion to the neighboring repeaters.
Those repeaters wait till enough distilled pairs have been generated and then perform distillation with the next level code on the longer hop.
The process repeats until the end-to-end distillation between the end nodes is complete.

We have only considered depolarizing errors and do not explicitly model photon loss or memory loss. Therefore, the current fidelity update curves should be understood as conditional fidelity curves, conditioned on successful elementary entanglement generation and successful availability of the required qubits. Under this conditional model, the repeater spacing does not directly enter the input-output fidelity map. Instead, distance would mainly affect the entanglement generation rate, the waiting time, and the amount of memory decoherence through the link success probability.
A more complete distance-dependent model should include link transmissivity, heralded entanglement generation, finite memory lifetime, and possibly loss-aware or erasure-aware error correction. In such a model, the optimal adaptive switching points may become distance-dependent because longer elementary links would reduce the success probability, increase waiting times, and introduce additional memory noise. We view this as an important direction for future work.
}

\section{Conclusion}
\label{sec:conclusion}

In this work, we considered a linear chain of repeaters that create entanglement with neighboring repeaters, perform quantum error correction-based entanglement distillation, execute entanglement swaps, and repeat the process until the end nodes have sufficient number of high-fidelity Bell pairs.
We introduced a novel adaptive strategy to switch between different codes for each round of distillation, as a function of the network operating point.
Using three $9$-qubit codes with increasing rates, we conducted simulations to compare four different protocols that perform three rounds of distillation, independent of the number of repeaters.
As our end-to-end metric, we defined a function called the \emph{efficiency}, which depends on the protocol rate as well as input and output distillable entanglement, which in turn depends on the input and output fidelities.
We conclusively demonstrated the advantages of the adaptive strategy to maximize the efficiency, where each of the four protocols become optimal for a certain region of input fidelity.

Subsequently, we combined our protocols with non-error correction-based entanglement purification protocols that have significantly smaller input fidelity thresholds.
We first showed that DEJMPS comprehensively outperforms BBPSSW aided by its asymmetric output error profile.
Then, for input fidelities below the code threshold, we began with DEJMPS rounds to cross the threshold and compared two strategies: adaptively switching to one round of code-based distillation or continuing with DEJMPS to achieve the same performance.
We highlighted the emergence of ``checkpoints'' in the fidelity plot due to the periodic decrease of required number of DEJMPS rounds with increase in input fidelity.
We observed that the rapid improvement of fidelity with each DEJMPS round makes it superior to the code when considering only the output fidelity.
But with a refined efficiency metric that combines rate and fidelity, the code emerges at the top when the input fidelity is very high.
This illuminates the importance of choosing the code and purification protocol as a function of the network operating point, thereby extending the results from the earlier adaptive coding strategy.
We expect that such a streamlined strategy will have several practical advantages in real networks.

In future work, we will extend this work to realistic noise scenarios and consider fault-tolerant schemes for performing distillation and swaps.
Specifically, we will examine the best approaches to pick adaptive code and protocol sequences, the fault-tolerance requirements, the associated resource estimates, and practical considerations.
Such studies will clearly establish the evolution from first to second generation repeaters and beyond.
In the past~\cite{Patil2024EntanglementRouting}, we considered optimal schedules for distillation and swaps at each repeater to maximize end-to-end distillable entanglement and also calculated memory requirements in the repeaters.
These schedules and resource requirements can be adapted to the adaptive distillation setting introduced in this work.
We have also considered schemes for Bell pair distillation in trapped-ion repeaters~\cite{Kang-qce23}, so our results can be contextualized for such hardware technologies.
We have showed state-of-the-art fidelity thresholds for distilling GHZ states using quantum low-density parity-check (LDPC) codes~\cite{Rengaswamy-quantum24}.
This enables a path to extend this adaptive distillation approach to GHZ states, which will play an important role in distributed quantum computing and sensing applications.
Overall, our ultimate goal is to find the best adaptive strategies as a function of the desired type of entanglement, the hardware technology, the end-to-end metric, and the network topology.

\section*{Data Availability}
The repository to reproduce the data in our figures and tables is provided at the GitHub/Zenodo link, 
\href{https://doi.org/10.5281/zenodo.20551766}{10.5281/zenodo.20551766}. Although it does not show how the data were generated, it provides additional details for reproducing the figures and tables in our paper.

\section*{Acknowledgments}

We would like to thank Kaushik Seshadreesan, Filip Rozp{\k e}dek, and Stefan Krastanov for helpful discussions.
\newtext{We also thank our reviewers for their careful reading and thoughtful feedback that made the paper significantly better.}
This work was supported by the U.S. National Science Foundation under the Engineering Research Centers program, Grant Nos. 1941583 and 2550064 (``Center for Quantum Networks'').

\bibliography{Reference.bib}
\bibliographystyle{IEEEtran}

\appendices

\section{Stabilizer Formalism, Error Syndromes and Werner States}
\label{app:SFESWS}

\subsection{Stabilizer Formalism}
\label{app:SF}

A \emph{stabilizer group}, $\mathcal{S}$, is a set of commuting Hermitian Pauli operators that does not contain $-I$.
The commutativity implies that these operators can be simultaneously diagonalized.
Since their eigenvalues are $\pm 1$, there exist a subspace of quantum states that are $+1$-eigenvectors of all stabilizers.
This subspace is referred to as a \emph{stabilizer code}, $C(\mathcal{S})$.
For example, consider a $3$-qubit stabilizer group generated by two operators: $S_1 = ZZI, S_2 = IZZ$.
A quantum state $\ket{\psi}$ is in the stabilizer code if $S_1 \ket{\psi} = \ket{\psi}, S_2 \ket{\psi} = \ket{\psi}$.
In this case it can be checked that the state takes the form $\ket{\psi} = \alpha \ket{000} + \beta \ket{111}$, where $\alpha, \beta \in \mathbb{C}$ satisfy $|\alpha|^2 + |\beta|^2 = 1$.
The logical Pauli operators for the code are those that commute with every stabilizer but do not belong to $\mathcal{S}$ themselves.
The weight of a Pauli operator is given by the number of qubits on which it is not the identity, e.g., $ZZI$ has weight $2$.
If there are $m = n-k$ independent stabilizers on $n$ qubits, and the minimum weight of a logical operator is $d$, then the code has parameters $\llbracket n,k,d \rrbracket$.
The above example gives a $\llbracket 3,1,1 \rrbracket$ code since $ZII \equiv IZI \equiv IIZ$ is a logical operator.
This formalism simplifies the analysis of quantum error correction and is widely used for code design.

For the purposes of this paper, we will consider the distillation of the standard Bell state $\ket{\Phi^+} = \frac{1}{\sqrt{2}}(\ket{00} + \ket{11})$.
This state forms the one-dimensional $+1$-eigenspace of the stabilizer group $\mathcal{S}_{++} = \langle +XX, +ZZ \rangle$.
The other Bell states $\ket{\Phi^-} = \frac{1}{\sqrt{2}}(\ket{00} - \ket{11}), \ket{\Psi^{\pm}} = \frac{1}{\sqrt{2}}(\ket{01} \pm \ket{10})$ correspond to the stabilizer groups $S_{-+} = \langle -XX, ZZ \rangle, S_{\pm -} = \langle \pm XX, -ZZ \rangle$.
These states can be thought of as the standard Bell state with a Pauli error $XI, ZI, YI$.

\subsection{Error Syndromes}
\label{app:ES}

Errors in a quantum system can be detected using the stabilizer formalism by measuring error syndromes. 
When an error \( E \) acts on a code state \(\ket{\psi}\), it may commute or anti-commute with stabilizer operators \( S_i \):
\begin{align}
E \ket{\psi} = E S_i \ket{\psi} = \pm S_i E \ket{\psi} = \lambda_i S_i ( E \ket{\psi} ).    
\end{align}
This means that measuring the stabilizer will give an eigenvalue of \( \lambda_i = \pm 1 \), where $-1$s indicate the presence of an error.
The error syndrome is a binary vector that records these eigenvalues: $\mathbf{s} = (\lambda_1, \lambda_2, \dots, \lambda_m)$.
For example, consider an error \( E = XII \) acting on the $\llbracket 3,1,1 \rrbracket$ code defined earlier.
We calculate the syndrome as follows:
\begin{align*}
E S_1 \ket{\psi} &= XII \cdot ZZI \ket{\psi} = - ZZI\cdot XII \ket{\psi} = -S_1 E \ket{\psi}, \\
E S_2 \ket{\psi} &= XII \cdot IZZ \ket{\psi} = IZZ \cdot XII \ket{\psi} = +S_2 E \ket{\psi}.
\end{align*}
Thus, the syndrome for this error is: $\mathbf{s} = (-1, +1)$.
By measuring the stabilizers, we obtain the syndrome \(\mathbf{s}\), which helps identify which error has occurred. 
Decoding algorithms use the syndrome to determine and correct the error, restoring the original quantum state if there are not too many errors.

\subsection{Depolarizing Noise and Werner States}
\label{app:WS}

The depolarizing noise channel is a fundamental noise model in quantum information theory. It describes the effect of a qubit undergoing a random Pauli error with a certain probability. 
For a single-qubit state \( \rho \), it is defined as:
\begin{align}
\mathcal{E}_p(\rho) = (1 - p) \rho + \frac{p}{3} (X \rho X + Y \rho Y + Z \rho Z),
\end{align}
where \( p \) is the depolarizing probability. This means that with probability \( 1 - p \), the qubit remains unchanged, and with probability \( \frac{p}{3} \), one of the three Pauli errors \( X, Y, Z \) is applied.

In quantum communication, depolarizing noise reduces entanglement fidelity by introducing errors in transmitted qubits. Specifically, when applied to Bell states, depolarizing noise transforms a perfect Bell state \( |\Phi^+\rangle \) into a Werner state, a mixed state that retains some entanglement but requires purification for reliable quantum applications:
\begin{align}
\rho_W = W |\Phi^+\rangle \langle\Phi^+| + (1-W) \frac{I}{4}.
\end{align}
Since the identity can be decomposed into Bell states, i.e.,
\begin{align}
I = |\Phi^+\rangle \langle\Phi^+| + |\Phi^-\rangle \langle\Phi^-| + |\Psi^+\rangle \langle\Psi^+| + |\Psi^-\rangle \langle\Psi^-|,
\end{align}
we rewrite \( \rho_W \) as:
\begin{align}
    \rho_W &= \left(W + \frac{1}{4}(1-W)\right)|\Phi^+\rangle \langle\Phi^+| \notag \\
    &+ \frac{1}{4}(1-W) \left(|\Phi^-\rangle \langle\Phi^-| + |\Psi^+\rangle \langle\Psi^+| + |\Psi^-\rangle \langle\Psi^-|\right) \nonumber \\
    &= \frac{3W+1}{4} |\Phi^+\rangle \langle\Phi^+| \notag \\
    &+ \frac{1}{4}(1-W) \left(|\Phi^-\rangle \langle\Phi^-| + |\Psi^+\rangle \langle\Psi^+| + |\Psi^-\rangle \langle\Psi^-|\right).
\end{align}
By defining Fidelity \( F \) as $F = \frac{3W+1}{4} \Rightarrow W = \frac{4F-1}{3}$, we express \( \rho_W \) in terms of fidelity:
\begin{align}
    \rho_W &= F |\Phi^+\rangle \langle\Phi^+| \notag \\
    &+ \frac{1-F}{3} \left(|\Phi^-\rangle \langle\Phi^-| + |\Psi^+\rangle \langle\Psi^+| + |\Psi^-\rangle \langle\Psi^-|\right).
\end{align}

\section{A Detailed Comparison of BBPSSW and DEJMPS}
\label{app:BBPSSW_DEJMPS_comparison}

\begin{figure*}[h]
\centering
    \begin{subfigure}[b]{0.45\textwidth}
        \centering
        \hspace*{-30pt}
        \includegraphics[width=1.2\textwidth]{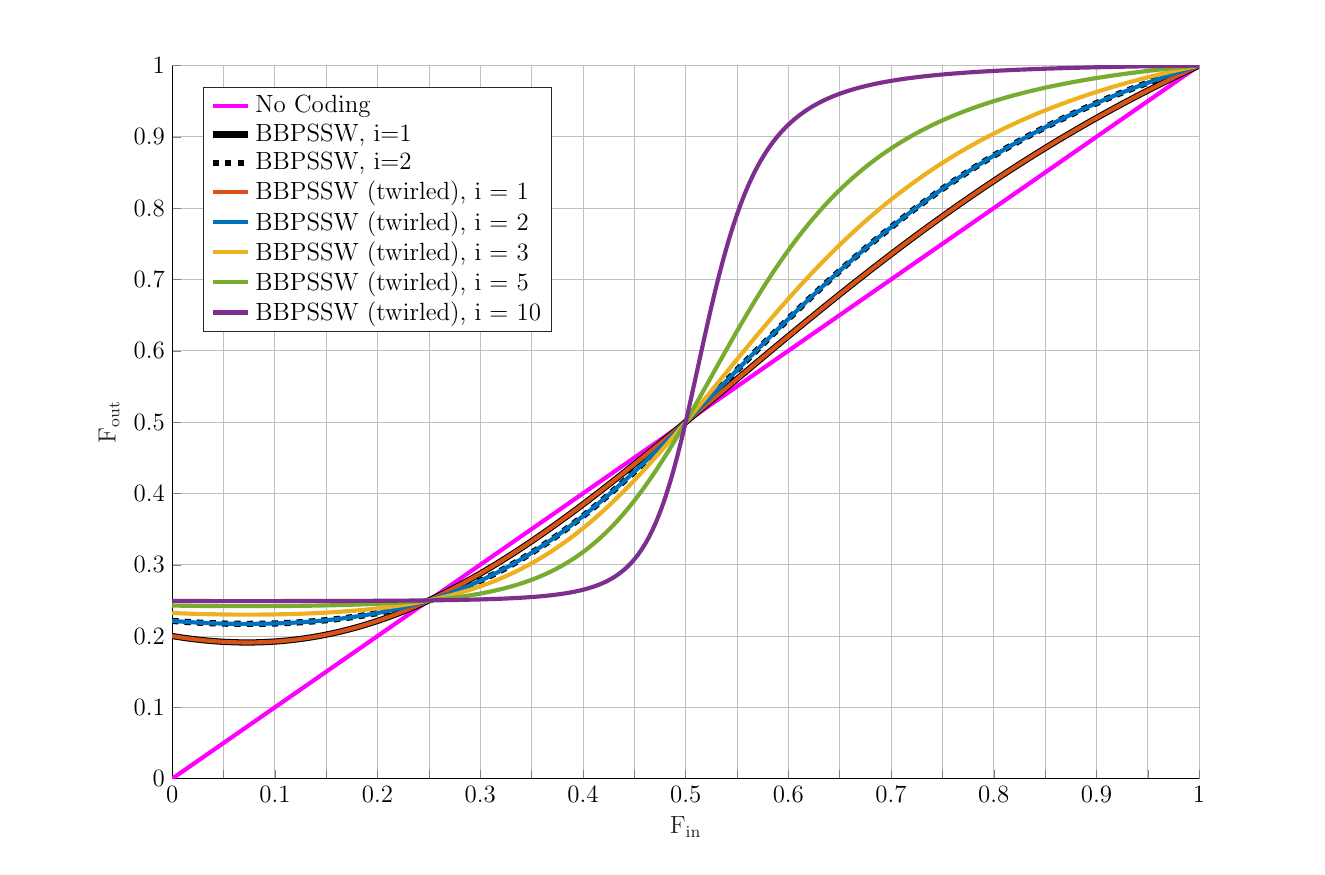}
        \caption{BBPSSW (twirled)}
        \label{fig:BBPSSW_t}
    \end{subfigure}
    \hspace{0.03\textwidth}
    \begin{subfigure}[b]{0.45\textwidth}
        \centering
        \hspace*{-30pt}
        \includegraphics[width=1.2\textwidth]{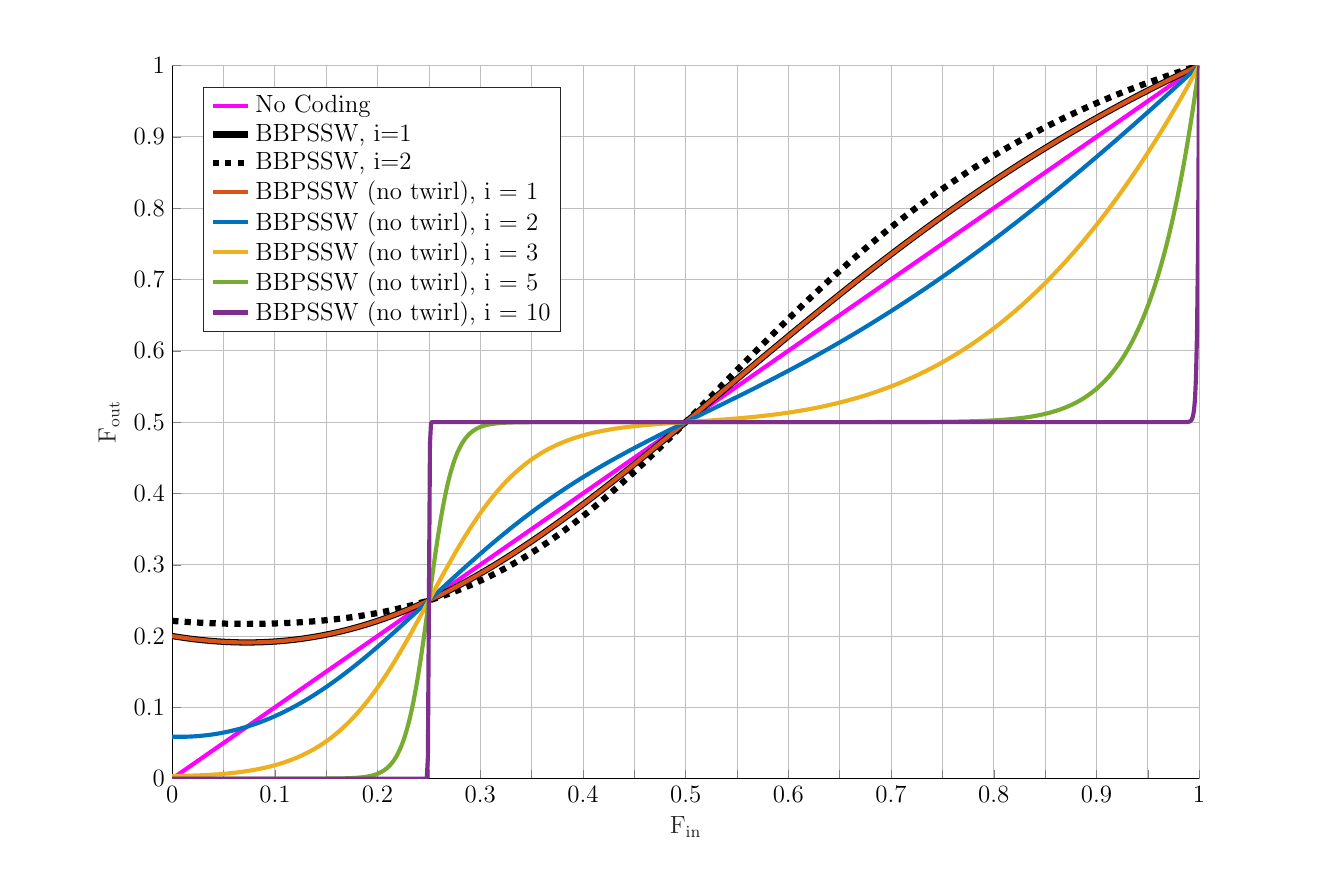} 
        \caption{BBPSSW (no twirl)}
        \label{fig:BBPSSW_nt}
    \end{subfigure}
\end{figure*}
\begin{figure*}[h]
\ContinuedFloat
\centering
    \begin{subfigure}[b]{0.45\textwidth}
        \centering
        \hspace*{-30pt}
        \includegraphics[width=1.2\textwidth]{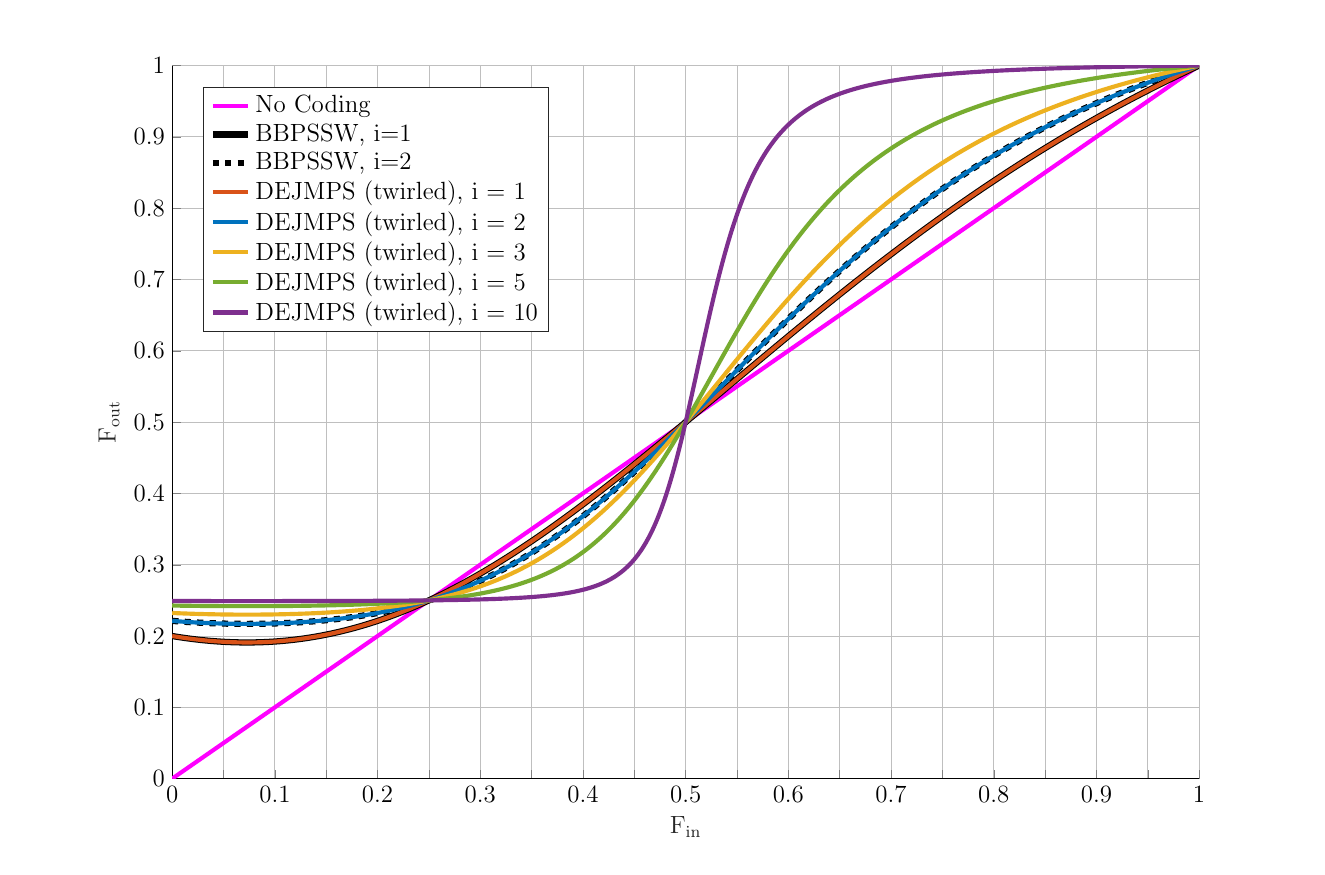} 
        \caption{DEJMPS (twirled)}
        \label{fig:DEJMPS_t}
    \end{subfigure}
    \hspace{0.03\textwidth}
    \begin{subfigure}[b]{0.45\textwidth}
        \centering
        \hspace*{-30pt}
        \includegraphics[width=1.2\textwidth]{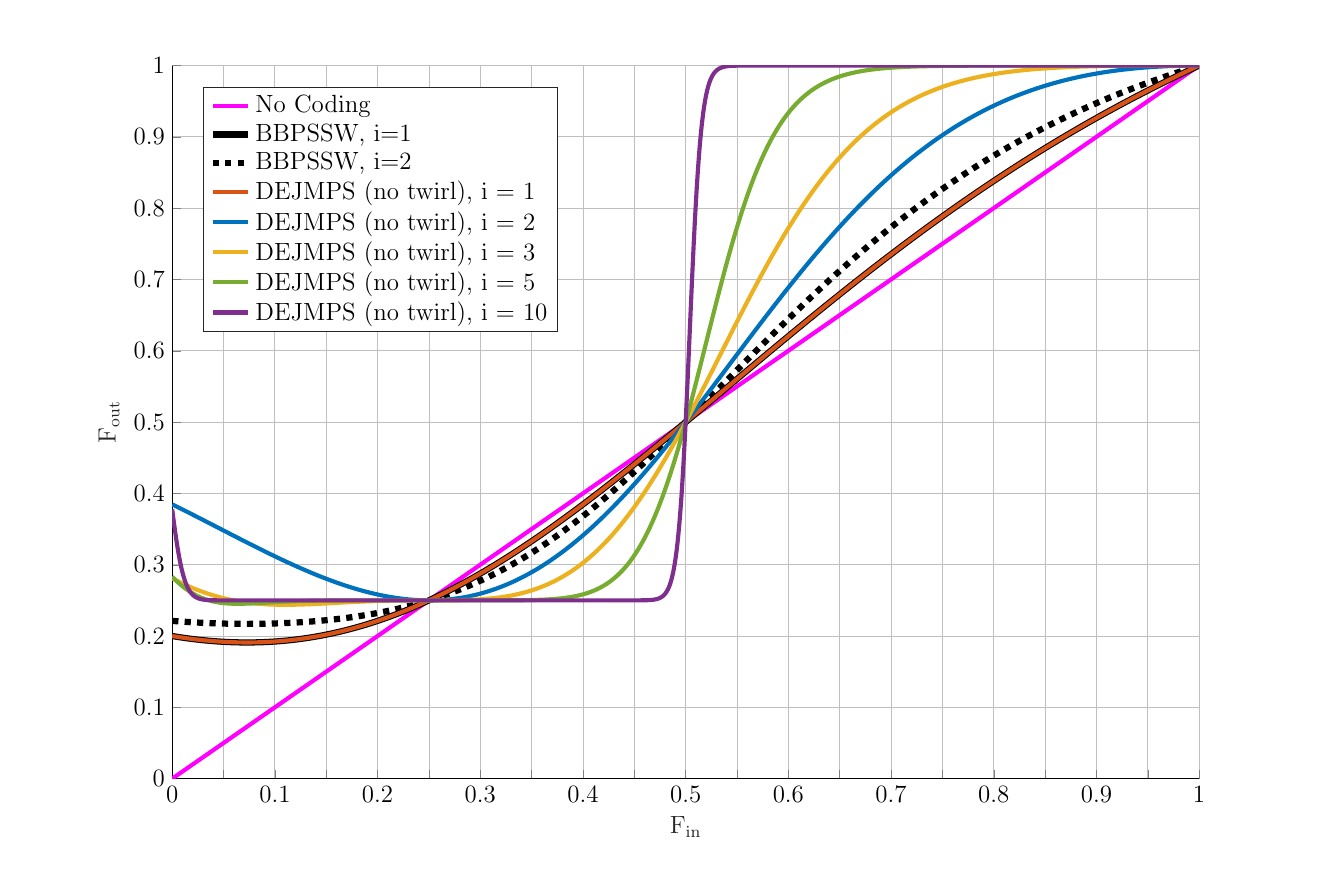} 
        \caption{DEJMPS (no twirl)}
        \label{fig:DEJMPS_nt}
    \end{subfigure} 
    \caption{Comparison of the fidelities obtained using BBPSSW and DEJMPS, including both the twirled and no-twirl versions for a single link. 
    “No Coding” indicates that $F_{\rm in} = F_{\rm out}$, and $i$ denotes the number of rounds. The black solid and black dashed lines are calculated based on Eq.~\ref{eq:BBPSSW}, while the other results are obtained from our circuit simulator.}

    \label{fig:1G_0R}
    \vspace{0.5em}
    \justifying
    
\end{figure*}

In Fig. \ref{fig:1G_0R}, we compare the output fidelity as a function of the input fidelity over multiple rounds in a single-link scenario for BBPSSW (twirled/no twirl) and DEJMPS (twirled/no twirl) respectively.
Although fidelity values below $0.5$ are generally not useful for practical entanglement distillation or purification, we include the full input fidelity range from 0 to 1 in plots to illustrate the mathematical structure of the input-output fidelity map. This helps show how the protocols behave both inside and outside the operationally useful regime.

In the idealized setting considered in the main text, the only noise source is the input depolarizing noise. Under this simplified model, the region below fidelity $0.5$ is not expected to provide a useful operating regime for distillation, and we do not use this region to support the main conclusions of the paper. Nevertheless, showing the full range is useful for visualizing the global structure of the protocol map.

We also note that, in more realistic noise models that include additional gate, measurement, or memory noise, the low fidelity region may exhibit nontrivial behavior when different 1G purification variants are combined. For example, one possible strategy is to apply BBPSSW (no twirl) when the input fidelity under depolarizing noise lies between approximately $0.25$ and $0.5$, then perform twirling, and subsequently switch to DEJMPS (no twirl) once the fidelity approaches $0.5$. In some cases, this hybrid strategy can produce a rare threshold-crossing effect, where an output state starting from below $0.5$ fidelity may, with very small probability, be pushed to a fidelity above $0.5$. This behavior can be observed in the corresponding low fidelity portion of the input-output map. However, this effect is not the main focus of the present work, and its probability is too small to affect the operating regimes studied here. Nevertheless, it motivates showing the full input fidelity range in order to make the mathematical structure of the protocol map visible.

We can clearly see that all protocols exhibit the same output fidelity in the first round. However, starting from the second round, DEJMPS achieves a higher fidelity than BBPSSW for the same number of rounds. In particular, when focusing on input fidelities greater than $0.5$ and beginning from the second round, the following relation can be observed:
\begin{align}
\rm DEJMPS(no\ twirl) &> \rm DEJMPS(twirled)  \notag \\
                      &= \rm BBPSSW(twirled)  \notag \\ 
                      &> \rm no\ coding  \notag \\ 
                      &> \rm BBPSSW(no\ twirl).
\end{align}

Therefore, unless stated otherwise, for the rest of this paper “DEJMPS” refers to the protocol without twirling between rounds, whereas “BBPSSW” refers to the protocol with twirling applied after each round.

In the literature, the BBPSSW protocol~\cite{PhysRevLett.76.722_BBPSSW} is often summarized by the closed-form recurrence for the Werner-state fidelity derived in the original paper. Since the protocol first symmetrizes an arbitrary two-qubit mixed state into a Werner state that is fully characterized by the single parameter $F$, this fidelity-only description does not track the evolution of individual Pauli-error components and therefore cannot explain how a bias in the Pauli error distribution arises during the process. The output fidelity for twirled BBPSSW is
\begin{align}
F_{\rm out} = \frac{F_{\rm in}^2 + \frac{1}{9} (1-F_{\rm in})^2}{F_{\rm in}^2 + \frac{2}{3}F_{\rm in}(1-F_{\rm in}) + \frac{5}{9} (1-F_{\rm in})^2}. \label{eq:BBPSSW}
\end{align}

The $F_{\rm out}$ from the previous round serves as the input for the next round. This approach works well when we consider only the twirling case. However, when we want to compare it with the no twirling case in DEJMPS, the formula is not sufficient to capture the underlying details. 
In the original DEJMPS paper~\cite{PhysRevLett.77.2818_DEJMPS}, the authors introduce a quantum privacy amplification procedure that tracks the Bell-basis populations. The map is formulated in terms of the Bell-diagonal parameters $A$, $B$, $C$, and $D$, but it does not directly expose the associated $I/X/Y/Z$ Pauli-error components, which makes the error-level interpretation less transparent. To make this connection explicit, we provide the following formula, which is derived based on the contribution of each error combination used to generate the final result.


For BBPSSW, 
\begin{align}
P_I&=P_{I_{\rm in}}^2+P_{Z_{\rm in}}^2,\\
P_X&=P_{X_{\rm in}}^2+P_{Y_{\rm in}}^2,\\
P_Y&=2P_{X_{\rm in}}P_{Y_{\rm in}},\\
P_Z&=2P_{I_{\rm in}}P_{Z_{\rm in}}.
\end{align}

For DEJMPS, 
\begin{align}
P_I&=P_{I_{\rm in}}^2+P_{Y_{\rm in}}^2,\\
P_X&=P_{X_{\rm in}}^2+P_{Z_{\rm in}}^2,\\
P_Y&=2P_{X_{\rm in}}P_{Z_{\rm in}},\\
P_Z&=2P_{I_{\rm in}}P_{Y_{\rm in}}.
\end{align}
Of course, the probabilities must be renormalized before being used as the input for the next round:
\begin{align}
P_{\rm discard}&=1-P_I-P_X-P_Y-P_Z,\\
F_{\rm out}&=P_{I_{\rm out}}=\frac{P_I}{1-P_{\rm discard}}, \\
P_{A_{\rm out}}&=\frac{P_A}{1-P_{\rm discard}}, A=X,Y,Z.
\end{align}

Using the BBPSSW protocol as an example, we can show that when the input fidelity satisfies $P_{I_{\rm in}}=P_{X_{\rm in}}+P_{Y_{\rm in}}+P_{Z_{\rm in}}=0.5$, the output fidelity always remains $P_{I_{\rm out}}=0.5$:
\begin{align}
    P_I&=P_{I_{\rm in}}^2+P_{Z_{\rm in}}^2 \notag \\
    &=(P_{X_{\rm in}}+P_{Y_{\rm in}}+P_{Z_{\rm in}})^2+P_{Z_{\rm in}}^2 \notag \\
    &= P_{X_{\rm in}}^2+P_{Y_{\rm in}}^2+2P_{Z_{\rm in}}^2\notag \\
    &+ 2P_{X_{\rm in}}P_{Y_{\rm in}}+2P_{Y_{\rm in}}P_{Z_{in}}+2P_{X_{\rm in}}P_{Z_{\rm in}} \notag \\
    &=(P_{X_{\rm in}}^2+P_{Y_{\rm in}}^2)+2P_{X_{\rm in}}P_{Y_{\rm in}} \notag \\
    &+ 2(P_{X_{\rm in}}+P_{Y_{\rm in}}+P_{Z_{\rm in}})P_{Z_{\rm in}} \notag \\
    &= P_X+P_Y+P_Z.
\end{align}
Based on the formulas from BBPSSW and DEJMPS, we can observe that the $R_X(\pi/2)$ and $R_X(-\pi/2)$ rotations simply switch $P_Z$ and $P_Y$. However, this modification makes DEJMPS a highly effective bias strategy.

\begin{figure*}[h]
\centering
    \begin{subfigure}[b]{0.45\textwidth}
        \centering
        \hspace*{-30pt}
        \includegraphics[width=\textwidth]{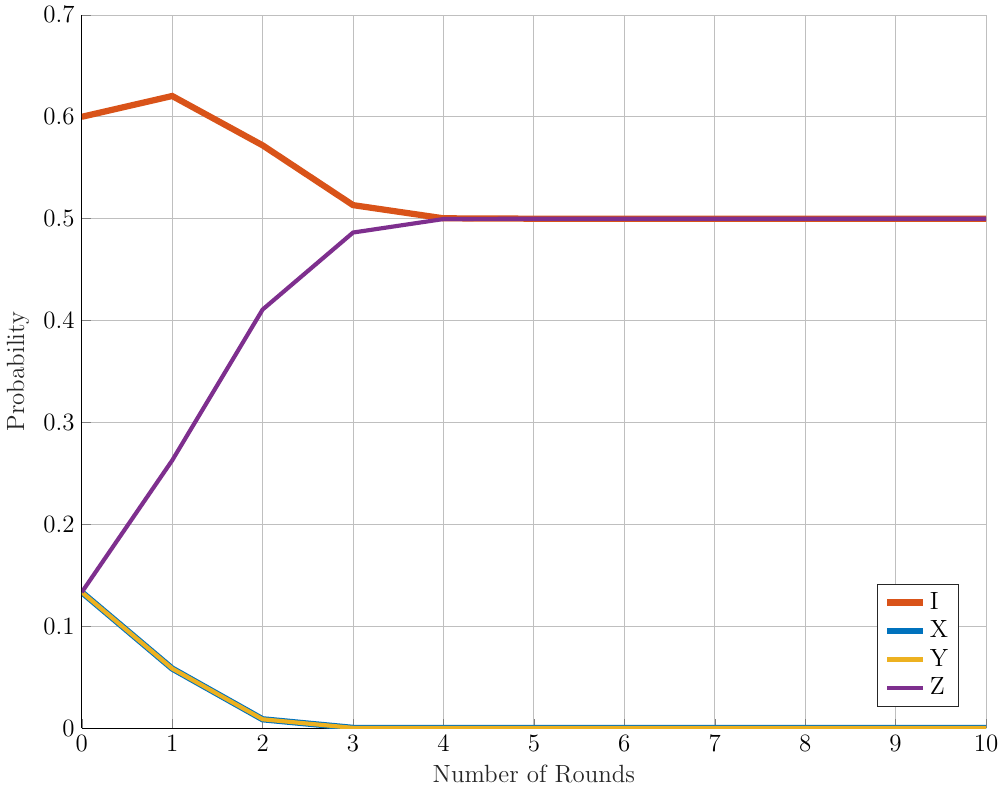} 
        \caption{BBPSSW (no twirl), $F_{\rm in}=0.6$}
        \label{fig:G_IXYZ1}
    \end{subfigure}
    \hspace{0.03\textwidth}
    \begin{subfigure}[b]{0.45\textwidth}
        \centering
        \hspace*{-30pt}
        \vspace*{-10pt}
        \includegraphics[width=1.2\textwidth]{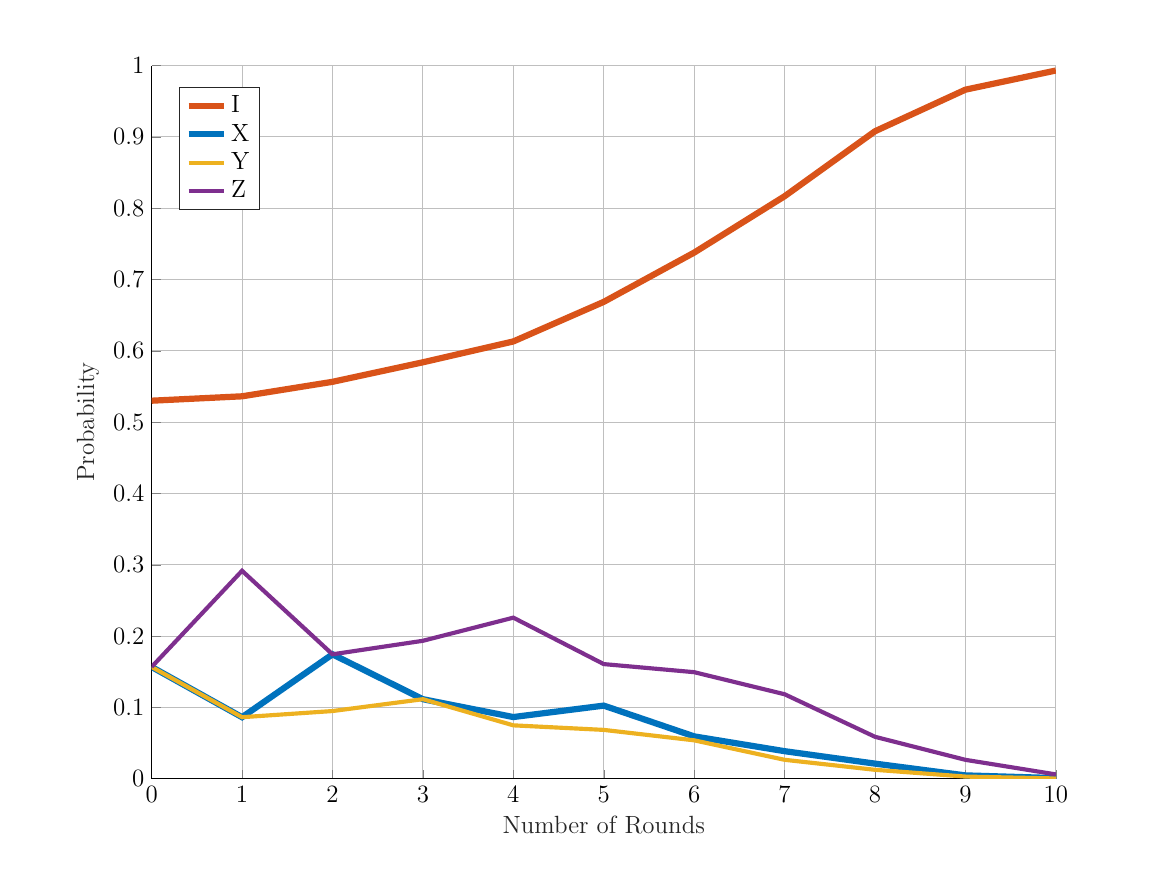} 
        \caption{DEJMPS (no twirl), $F_{\rm in}=0.53$}
        \label{fig:G_IXYZ2}
    \end{subfigure}
    \caption{As a function of the number of rounds, we show how the probabilities of $I$, $X$, $Y$, and $Z$ evolve. To a large extent, DEJMPS (no twirl) can be viewed as BBPSSW (no twirl) with an added $R_X$ rotation gate. Therefore, it is necessary to compare how the bias behaves in both protocols. To illustrate this clearly, we set $F_{\rm in} = 0.6$ for BBPSSW without twirling and $F_{\rm in} = 0.53$ for DEJMPS without twirling, so that we can show in detail how the bias evolves round by round.}

    \label{fig:1G_bias}
    \vspace{0.5em}
    \justifying
    
\end{figure*}
Furthermore, to understand how the bias provides an advantage in DEJMPS (no twirl), we first need to examine how BBPSSW behaves in the no-twirling case.

In Fig. \ref{fig:1G_bias}, we show the evolution of the I, X, Y, and Z error components over successive rounds for BBPSSW (no twirl) and DEJMPS (no twirl).
We can observe that regardless of whether the protocol is BBPSSW (no twirl) or DEJMPS (no twirl), the probability of a $Z$ error is always higher than that of the other errors. Therefore, we refer to these as “bias-on-$Z$ protocols.” In such protocols, we find that the $X$ and $Y$ errors exhibit the primary decreasing trend. 

As a result, after DEJMPS (no twirl) applies $R_X(\pi/2)$ and $R_X(-\pi/2)$, the roles of $P_Y$ and $P_Z$ are effectively swapped. 
Using this insight, we perform a detailed analysis of BBPSSW and DEJMPS without twirling in Appendix~\ref{app:BBPSSW_DEJMPS}.
The calculations show that, because $P_Y$ and $P_Z$ swap, the recursive equations for $P_I$ and $P_Z$ imply that $P_I$ goes to 1 and $P_Z$ goes to 0 for DEJMPS, whereas both go to 0.5 for BBPSSW. This agrees with the plot in Fig.~\ref{fig:1G_bias}.

\begin{theorem}
\label{thm:BBPSSW}
Consider the BBPSSW protocol.
Let $P_{I_{\rm in}}=a_0$, $P_{X_{\rm in}}=b_0$, $P_{Y_{\rm in}}=c_0$, $P_{Z_{\rm in}}=d_0$, where the subscript denotes the number of rounds. 
Assume
\begin{align}
a_0 > \frac{1}{2},b_0>0,c_0>0,d_0>0,\notag\\
a_0+b_0+c_0+d_0=1,\notag\\
P_{I_n}=a_n^2+d_n^2, P_{X_n}=b_n^2+c_n^2,\notag\\
P_{Y_n}=2b_n c_n, P_{Z_n}=2a_n d_n,\notag\\
P_{T_n}=P_{I_n}+P_{X_n}+P_{Y_n}+P_{Z_n}, \notag\\
a_{n+1}=\frac{P_{I_n}}{P_{T_n}},
b_{n+1}=\frac{P_{X_n}}{P_{T_n}}, \notag\\
c_{n+1}=\frac{P_{Y_n}}{P_{T_n}},
d_{n+1}=\frac{P_{Z_n}}{P_{T_n}}.\notag
\end{align}
Then, as $n\to\infty$, we have
\begin{align*}
a_n+d_n \to 1 &, b_n+c_n \to 0,\\
a_n \to 0.5, d_n \to 0.5 &,
b_n \to 0, c_n \to 0.
\end{align*}
\end{theorem}
\begin{IEEEproof}
See Appendix~\ref{app:BBPSSW}.
\end{IEEEproof}

\begin{observation}
\label{thm:DEJMPS}
Consider the DEJMPS protocol.
Let $P_{I_{\rm in}}=a_0$, $P_{X_{\rm in}}=b_0$, $P_{Y_{\rm in}}=c_0$, $P_{Z_{\rm in}}=d_0$, where the subscript denotes the number of rounds. 
Assume
\begin{align}
a_0>\frac{1}{2},b_0>0,c_0>0,d_0>0,\notag\\
a_0+b_0+c_0+d_0=1,\notag\\
P_{I_n}=a_n^2+c_n^2, P_{X_n}=b_n^2+d_n^2,\notag\\
P_{Y_n}=2b_n d_n, P_{Z_n}=2a_n c_n,\notag\\
P_{T_n}=P_{I_n}+P_{X_n}+P_{Y_n}+P_{Z_n}, \notag\\
a_{n+1}=\frac{P_{I_n}}{P_{T_n}}, b_{n+1}=\frac{P_{X_n}}{P_{T_n}},\notag\\
c_{n+1}=\frac{P_{Y_n}}{P_{T_n}}, d_{n+1}=\frac{P_{Z_n}}{P_{T_n}}.\notag
\end{align}
Then, as $n\to\infty$, we have
\begin{align*}
a_n+d_n \to 1 &, b_n+c_n \to 0,\\
a_{n+1} \to \frac{a_n^2+c_n^2}{(a_n+c_n)^2}=1 &,
d_{n+1} \to \frac{2a_n c_n}{(a_n+c_n)^2}=0,\\
b_n \to 0 &,
c_n \to 0.
\end{align*}
\end{observation}
\begin{IEEEproof}[\textcolor{black}{Numerical observation}]
See Appendix~\ref{app:DEJMPS}.
\end{IEEEproof}


\section{BBPSSW (no twirl) and DEJMPS (no twirl)}
\label{app:BBPSSW_DEJMPS}

In~\cite{MACCHIAVELLO1998385_QPA}, Macchiavello's main objective was to establish analytical convergence of the quantum privacy amplification (QPA) map for Bell diagonal states. The paper provides a convergence proof that is not included in the original Deutsch \emph{et al.} protocol~\cite{PhysRevLett.77.2818_DEJMPS}, where several conclusions are supported by numerical analysis. The treatment is largely framed as an iteration of Bell basis populations, and it is not intended to offer an operational description of how the underlying Pauli error components evolve through the procedure. The monotone used in the proof is introduced as a mathematical device, and its physical meaning is left unclear.

Motivated by these limitations, we develop a more intuitive description at the error level and use it to compare the BBPSSW (no twirl) and DEJMPS (no twirl) protocols. Our study still leverages numerical simulations, but we use them to validate the proposed intuition and to quantify protocol dependent differences under the noise models of interest.

\subsection{BBPSSW (no twirl)}
\label{app:BBPSSW}

For BBPSSW (no twirl), the main text gives the following relationship:
\begin{align}
P_I&=P_{I_{\rm in}}^2+P_{Z_{\rm in}}^2,\\
P_X&=P_{X_{\rm in}}^2+P_{Y_{\rm in}}^2,\\
P_Y&=2P_{X_{\rm in}}P_{Y_{\rm in}},\\
P_Z&=2P_{I_{\rm in}}P_{Z_{\rm in}},\\
P_T&=P_I+P_X+P_Y+P_Z,\\
P_{I_{\rm out}}&=\frac{P_I}{P_T},\\
P_{X_{\rm out}}&=\frac{P_X}{P_T},\\
P_{Y_{\rm out}}&=\frac{P_Y}{P_T},\\
P_{Z_{\rm out}}&=\frac{P_Z}{P_T}.
\end{align}
Using this, we want to analytically derive the performance that is observed in Fig.~\ref{fig:1G_bias}.

\begin{IEEEproof}[Proof of Theorem~\ref{thm:BBPSSW}]
Recall that
\begin{align}
a_0&>\frac{1}{2},\notag\\
b_0&>0,c_0>0,d_0>0,\notag\\
a_0&+b_0+c_0+d_0=1,\notag\\
P_{I_n}&=a_n^2+d_n^2,\notag\\
P_{X_n}&=b_n^2+c_n^2,\notag\\
P_{Y_n}&=2b_n c_n,\notag\\
P_{Z_n}&=2a_n d_n,\notag\\
P_{T_n}&=P_{I_n}+P_{X_n}+P_{Y_n}+P_{Z_n}, \notag\\
a_{n+1}&=\frac{P_{I_n}}{P_{T_n}},\notag\\
b_{n+1}&=\frac{P_{X_n}}{P_{T_n}},\notag\\
c_{n+1}&=\frac{P_{Y_n}}{P_{T_n}},\notag\\
d_{n+1}&=\frac{P_{Z_n}}{P_{T_n}}.\notag
\end{align}
To prove the result, we first derive the following relationship:
\begin{align}
P_{I_0}&=a_0^2+d_0^2\\
P_{X_0}&=b_0^2+c_0^2\\
P_{Y_0}&=2b_0 c_0\\
P_{Z_0}&=2a_0 d_0\\
P_{T_0}&=P_{I_0}+P_{X_0}+P_{Y_0}+P_{Z_0} \notag\\
&=a_0^2+d_0^2+b_0^2+c_0^2 + 2b_0 c_0 + 2a_0 d_0\notag \\
&=(a_0+d_0)^2+(b_0+c_0)^2 \notag \\
a_1&=\frac{a_0^2+d_0^2}{(a_0+d_0)^2+(b_0+c_0)^2}\\
b_1&=\frac{b_0^2+c_0^2}{(a_0+d_0)^2+(b_0+c_0)^2}\\
c_1&=\frac{2b_0 c_0}{(a_0+d_0)^2+(b_0+c_0)^2}\\
d_1&=\frac{2a_0 d_0}{(a_0+d_0)^2+(b_0+c_0)^2}
\end{align}
So,
\begin{align}
a_1+d_1&=\frac{(a_0+d_0)^2}{(a_0+d_0)^2+(b_0+c_0)^2}\\
b_1+c_1&=\frac{(b_0+c_0)^2}{(a_0+d_0)^2+(b_0+c_0)^2}
\end{align}
Let $s=a+d$, $t=b+c$,
\begin{align}
s_1&=\frac{s_0^2}{s_0^2+t_0^2}\\
t_1&=\frac{t_0^2}{s_0^2+t_0^2}
\end{align}
Let $u=\frac{s}{t}$,
\begin{align}
u_1=\frac{s_1}{t_1}=\frac{s_0^2}{t_0^2}=u_0^2
\end{align}
So,
\begin{align}
u_n=u_{n-1}^2=u_0^{2^n}
\end{align}
Therefore, when $u_0>1$ and $n\to\infty$, we have $u_n\to\infty$.

Since $u_0>1$ implies $s_0>t_0$, it follows that
$F_{I_{\rm in}}+F_{Z_{\rm in}} > F_{X_{\rm in}}+F_{Y_{\rm in}}$.
Moreover, as discussed in the main text, $F_{I_{\rm in}} > 0.5$.
Therefore, the condition $u_0>1$ is always satisfied.

Therefore, as $n\to\infty$, we have $u_n=\frac{s_n}{t_n}\to\infty$.

So,
\begin{align}
s_{n+1}&=\frac{s_n^2}{s_n^2+t_n^2}=\frac{(\frac{s_n}{t_n})^2}{(\frac{s_n}{t_n})^2+1}=\frac{(u_n)^2}{(u_n)^2+1}\to 1\\
t_{n+1}&=\frac{t_n^2}{s_n^2+t_n^2}=\frac{1}{(\frac{s_n}{t_n})^2+1}=\frac{1}{(u_n)^2+1}\to 0
\end{align}
Therefore, as $n\to\infty$, we have $t_n\to 0$, and hence $P_{X_{\rm out}}$ and $P_{Y_{\rm out}}$ converge to $0$.
Moreover, $P_{I_{\rm out}} + P_{Z_{\rm out}} \to 1$.

Therefore,
\begin{align}
a_n+d_n&=1\\
b_n&=0\\
c_n&=0
\end{align}
So, 
\begin{align}
a_{n+1}&=\frac{a_n^2+d_n^2}{(a_n+d_n)^2+(b_n+c_n)^2}=\frac{a_n^2+d_n^2}{(a_n+d_n)^2}\\
d_{n+1}&=\frac{2a_n d_n}{(a_n+d_n)^2+(b_n+c_n)^2}=\frac{2a_n d_n}{(a_n+d_n)^2}
\end{align}
Let $r=\frac{d}{a}$,
\begin{align}
r_{n+1}=\frac{2a_n d_n}{a_n^2+d_n^2}=\frac{2(\frac{d_n}{a_n})}{1+(\frac{d_n}{a_n})^2}=\frac{2r_n}{1+r_n^2}
\end{align}
Because $r_0=\frac{d_0}{a_0}$, we have $0<r_0<1$.
Therefore, $0<\frac{2r_n}{1+r_n^2}<1$,
and hence $0<r_{n+1}<1$.
Let $q=\frac{1-r}{1+r}$.
Thus, $0<q_n<1$.

Therefore,
\begin{align}
q_{n+1}=\frac{1-r_{n+1}}{1+r_{n+1}}=\frac{\frac{1+r_n^2-2r_n}{1+r_n^2}}{\frac{1+r_n^2+2r_n}{1+r_n^2}}=(\frac{1-r_n}{1+r_n})^2=q_n^2
\end{align}

Since $0<q_n<1$, we have $q_n \to 0$ as $n \to \infty$.

So,
\begin{align}
q_n&=\frac{1-r_n}{1+r_n}\notag \\
q_n(1+r_n)&=1-r_n\notag \\
q_n+q_n r_n &= 1-r_n\notag \\
r_n(q_n+1)&=1-q_n\notag\\
r_n&=\frac{1-q_n}{1+q_n}\to 1\notag
\end{align}
Therefore,
\begin{align}
\frac{d_n}{a_n}&=1\\
a_n&=d_n
\end{align}
Since $a_n + d_n = 1$, we have $a_n = d_n = 0.5$.

Therefore, for BBPSSW (no twirl), as $n\to\infty$, we have
\begin{align}
a_{n+1}+d_{n+1}&=1\\
b_n+c_n&=0\\
a_{n+1}&=\frac{a_n^2+d_n^2}{(a_n+d_n)^2}=0.5\\
d_{n+1}&=\frac{2a_n d_n}{(a_n+d_n)^2}=0.5\\
b_n&=0\\
c_n&=0
\end{align}
\end{IEEEproof}

\subsection{DEJMPS (no twirl)}
\label{app:DEJMPS}

In the DEJMPS (no twirl) setting, the analysis is slightly more involved than for BBPSSW (no twirl). As we derive below, the recursion naturally separates into two cases. In the $n$th round, the improvement does not always come from the $(n\!-\!1)$st round. In some cases, it comes from the $(n\!-\!2)$nd round. As a result, DEJMPS (no twirl) may converge to $0$ more slowly than BBPSSW (no twirl) for the $X$, $Y$, and $Z$ components.

For DEJMPS (no twirl), the main text gives the following relationship:
\begin{align}
P_I&=P_{I_{\rm in}}^2+P_{Y_{\rm in}}^2,\\
P_X&=P_{X_{\rm in}}^2+P_{Z_{\rm in}}^2,\\
P_Y&=2P_{X_{\rm in}}P_{Z_{\rm in}},\\
P_Z&=2P_{I_{\rm in}}P_{Y_{\rm in}},\\
P_T&=P_I+P_X+P_Y+P_Z,\\
P_{I_{\rm out}}&=\frac{P_I}{P_T},\\
P_{X_{\rm out}}&=\frac{P_X}{P_T},\\
P_{Y_{\rm out}}&=\frac{P_Y}{P_T},\\
P_{Z_{\rm out}}&=\frac{P_Z}{P_T}.
\end{align}
Using this, we want to analytically derive the performance that is observed in Fig.~\ref{fig:1G_bias}.

\begin{IEEEproof}[\textcolor{black}{Numerical Observation}~\ref{thm:DEJMPS}]
Recall that
\begin{align}
a_0&>\frac{1}{2},\notag\\
b_0&>0,c_0>0,d_0>0,\notag\\
a_0&+b_0+c_0+d_0=1,\notag\\
P_{I_n}&=a_n^2+c_n^2,\notag\\
P_{X_n}&=b_n^2+d_n^2,\notag\\
P_{Y_n}&=2b_n d_n,\notag\\
P_{Z_n}&=2a_n c_n,\notag\\
P_{T_n}&=P_{I_n}+P_{X_n}+P_{Y_n}+P_{Z_n}, \notag\\
a_{n+1}&=\frac{P_{I_n}}{P_{T_n}},\notag\\
b_{n+1}&=\frac{P_{X_n}}{P_{T_n}},\notag\\
c_{n+1}&=\frac{P_{Y_n}}{P_{T_n}},\notag\\
d_{n+1}&=\frac{P_{Z_n}}{P_{T_n}}.\notag
\end{align}
To prove the result, we first derive the following relationship:
\begin{align}
P_{I_0}&=a_0^2+c_0^2\\
P_{X_0}&=b_0^2+d_0^2\\
P_{Y_0}&=2b_0 d_0\\
P_{Z_0}&=2a_0 c_0\\
P_{T_0}&=P_{I_0}+P_{X_0}+P_{Y_0}+P_{Z_0} \notag\\
&=a_0^2+c_0^2+b_0^2+d_0^2 + 2b_0 d_0 + 2a_0 c_0\notag \\
&=(a_0+c_0)^2+(b_0+d_0)^2 \notag \\
a_1&=\frac{a_0^2+c_0^2}{(a_0+c_0)^2+(b_0+d_0)^2}\\
b_1&=\frac{b_0^2+d_0^2}{(a_0+c_0)^2+(b_0+d_0)^2}\\
c_1&=\frac{2b_0 d_0}{(a_0+c_0)^2+(b_0+d_0)^2}\\
d_1&=\frac{2a_0 c_0}{(a_0+c_0)^2+(b_0+d_0)^2}
\end{align}
So,
\begin{align}
a_1+d_1&=\frac{(a_0+c_0)^2}{(a_0+c_0)^2+(b_0+d_0)^2}\\
b_1+c_1&=\frac{(b_0+d_0)^2}{(a_0+c_0)^2+(b_0+d_0)^2}
\end{align}
To prepare for the case split in the proof below, we first need to show that $a_1>\frac{1}{2}$. Note that $a_0+b_0+c_0+d_0=1$.
\begin{align}
a_0&>\frac{1}{2}\notag\\
a_0&>1-a_0\notag\\
a_0-c_0&>1-a_0-c_0\notag\\
a_0-c_0&>b_0+d_0\notag\\
(a_0-c_0)^2&>(b_0+d_0)^2\notag\\
a_0^2-2a_0 c_0+c_0^2&>(b_0+d_0)^2\notag\\
2a_0^2+2c_0^2&>a_0^2+2a_0 c_0+c_0^2+(b_0+d_0)^2\notag\\
2(a_0^2+c_0^2)&>(a_0+c_0)^2+(b_0+d_0)^2\notag\\
\frac{a_0^2+c_0^2}{(a_0+c_0)^2+(b_0+d_0)^2}&>\frac{1}{2}\notag\\
a_1&>\frac{1}{2}
\end{align}
Next, we apply the same analysis as in BBPSSW (no twirl) to DEJMPS (no twirl).

Let $s=a+c$, $t=b+d$,
\begin{align}
a_1+d_1&=\frac{s_0^2}{s_0^2+t_0^2}\\
b_1+c_1&=\frac{t_0^2}{s_0^2+t_0^2}
\end{align}
So,
\begin{align}
s_1&=a_1+c_1=\frac{s_0^2}{s_0^2+t_0^2}-d_1+c_1\\
t_1&=b_1+d_1=\frac{t_0^2}{s_0^2+t_0^2}-c_1+d_1
\end{align}
Thus,
\begin{align}
s_1&=a_1+c_1=\frac{s_0^2-2a_0 c_0+2b_0 d_0}{s_0^2+t_0^2}\\
t_1&=b_1+d_1=\frac{t_0^2-2b_0 d_0+2a_0 c_0}{s_0^2+t_0^2}
\end{align}
Let $u=\frac{s}{t}$,
\begin{align}
u_1&=\frac{s_1}{t_1}=\frac{s_0^2-2a_0 c_0+2b_0 d_0}{t_0^2-2b_0 d_0+2a_0 c_0}\\
&=\frac{s_0^2}{t_0^2}+\frac{s_0^2-2a_0 c_0+2b_0 d_0}{t_0^2-2b_0 d_0+2a_0 c_0}-\frac{s_0^2}{t_0^2}\\
&=\frac{s_0^2}{t_0^2}+\frac{t_0^2(-2a_0 c_0+2b_0 d_0)-s_0^2(-2b_0 d_0+2a_0 c_0)}{t_0^2((b_0+d_0)^2-2b_0 d_0+2a_0 c_0)}\\
&=u_0^2+\frac{2(t_0^2+s_0^2)(b_0 d_0-a_0 c_0)}{t_0^2(b_0^2+d_0^2+2a_0 c_0)}
\end{align}
So,
\begin{align}
u_n&=u_{n-1}^2+\frac{2(t_{n-1}^2+s_{n-1}^2)(b_{n-1} d_{n-1}-a_{n-1} c_{n-1})}{t_{n-1}^2(b_{n-1}^2+d_{n-1}^2+2a_{n-1} c_{n-1})}\\
&=u_{n-1}^2+o(b_{n-1} d_{n-1}-a_{n-1} c_{n-1})
\end{align}
Because $2(t_{n-1}^2+s_{n-1}^2)>0$ and
$t_{n-1}^2(b_{n-1}^2+d_{n-1}^2+2a_{n-1}c_{n-1})>0$,
the term $(b_{n-1}d_{n-1}-a_{n-1}c_{n-1})$ determines the sign of the second component.

Because the first component already contains $u_{n-1}^2$, if the second component is always positive, we can conclude that, when $u_0>1$ and $n\to\infty$, we have $u_n\to\infty$.

Now, we consider the three cases separately.

In the first case, when $b_{n-1}d_{n-1}-a_{n-1}c_{n-1}>0$, we can directly conclude that, if $u_0>1$ and $n\to\infty$, then $u_n\to\infty$.

In the second case, when $b_{n-1}d_{n-1}-a_{n-1}c_{n-1}=0$, we obtain
\begin{align}
u_n=u_{n-1}^2=u_0^{2^n}
\end{align}
Therefore, when $u_0>1$ and $n\to\infty$, we have $u_n\to\infty$.

In the third case, when $b_{n-1}d_{n-1}-a_{n-1}c_{n-1}<0$, the analysis becomes slightly more subtle. In this case, $u_n$ does not necessarily improve from $u_{n-1}$, but may instead improve from $u_{n-2}$.

First, it does not necessarily improve from $u_{n-1}$. Note that $t_0+s_0=1$.
\begin{align}
u_1&=\frac{s_0}{t_0}+\frac{s_0^2-2a_0 c_0+2b_0 d_0}{t_0^2-2b_0 d_0+2a_0 c_0}-\frac{s_0}{t_0}\\
&=\frac{s_0}{t_0}+\frac{t_0(s_0^2-2a_0 c_0+2b_0 d_0)-s_0(t_0^2-2b_0 d_0+2a_0 c_0)}{t_0((b_0+d_0)^2-2b_0 d_0+2a_0 c_0)}\\
&=u_0+\frac{t_0 s_0(s_0-t_0)+2(t_0+s_0)(b_0 d_0-a_0 c_0)}{t_0^2(b_0^2+d_0^2+2a_0 c_0)}\\
&=u_0+\frac{t_0 s_0(2s_0-1)+2(b_0 d_0-a_0 c_0)}{t_0^2(b_0^2+d_0^2+2a_0 c_0)}
\end{align}
When $b_0d_0-a_0c_0<0$, the quantity $t_0s_0(2s_0-1)+2(b_0d_0-a_0c_0)$ may also be negative. For example, for $a_0=0.7$ and $b_0=c_0=d_0=0.1$, it is less than $0$.

Second, it can improve from $u_{n-2}$.
\begin{align}
u_{n-1}&=u_{n-2}^2+\frac{2(t_{n-2}^2+s_{n-2}^2)(b_{n-2} d_{n-2}-a_{n-2} c_{n-2})}{t_{n-2}^2(b_{n-2}^2+d_{n-2}^2+2a_{n-2} c_{n-2})}\\
u_{n}&=(u_{n-2}^2+o_1(b_{n-2} d_{n-2}-a_{n-2} c_{n-2}))^2\\
&+o_2(b_{n-1} d_{n-1}-a_{n-1} c_{n-1})
\end{align}

Based on numerical simulations, we found that $u_n$ may not always show an obvious improvement over $u_{n-2}$. However, we observed that $u_n$ always shows an obvious improvement over $u_{n-m}$ for some $m$.

We tried different initial conditions and consistently observed that the sequence eventually starts increasing more clearly, but the iteration index at which this increase begins depends on the initial conditions. We observe this behavior numerically, but we do not yet know how to prove it mathematically.

Therefore, when $u_0>1$ and $n\to\infty$, we have $u_n\to\infty$.

Since $u_0>1$ implies $s_0>t_0$, it follows that
$F_{I_{\rm in}}+F_{Y_{\rm in}} > F_{X_{\rm in}}+F_{Z_{\rm in}}$.
Moreover, as discussed in the main text, $F_{I_{\rm in}} > 0.5$.
Therefore, the condition $u_0>1$ is always satisfied.

Therefore, as $n\to\infty$, we have $u_n=\frac{s_n}{t_n}\to\infty$.

So,
\begin{align}
a_{n+1}+d_{n+1}&=\frac{s_n^2}{s_n^2+t_n^2}=\frac{(\frac{s_n}{t_n})^2}{(\frac{s_n}{t_n})^2+1}=\frac{(u_n)^2}{(u_n)^2+1}\to 1\\
b_{n+1}+c_{n+1}&=\frac{t_n^2}{s_n^2+t_n^2}=\frac{1}{(\frac{s_n}{t_n})^2+1}=\frac{1}{(u_n)^2+1}\to 0
\end{align}
Therefore, as $n\to\infty$, we have $b_{n+1}+c_{n+1}\to 0$, and hence $P_{X_{\rm out}}$ and $P_{Y_{\rm out}}$ converge to $0$.
Moreover, $P_{I_{\rm out}} + P_{Z_{\rm out}} \to 1$.

Therefore,
\begin{align}
a_n+d_n&=1\\
b_n&=0\\
c_n&=0
\end{align}
So, 
\begin{align}
a_{n+1}&=\frac{a_n^2+c_n^2}{(a_n+c_n)^2+(b_n+d_n)^2}=\frac{a_n^2+c_n^2}{(a_n+c_n)^2}=1\\
d_{n+1}&=\frac{2a_n c_n}{(a_n+c_n)^2+(b_n+d_n)^2}=\frac{2a_n c_n}{(a_n+c_n)^2}=0
\end{align}
Therefore, for DEJMPS (no twirl), as $n\to\infty$, we have
\begin{align}
a_{n+1}+d_{n+1}&=1\\
b_n+c_n&=0\\
a_{n+1}&=\frac{a_n^2+c_n^2}{(a_n+c_n)^2}=1\\
d_{n+1}&=\frac{2a_n c_n}{(a_n+c_n)^2}=0\\
b_n&=0\\
c_n&=0
\end{align}
\end{IEEEproof}

Recall that, for BBPSSW (no twirl), as $n\to\infty$, we have
\begin{align}
a_{n+1}+d_{n+1}&=1\\
b_n+c_n&=0\\
a_{n+1}&=\frac{a_n^2+d_n^2}{(a_n+d_n)^2}=0.5\\
d_{n+1}&=\frac{2a_n d_n}{(a_n+d_n)^2}=0.5\\
b_n&=0\\
c_n&=0
\end{align}
We observe that, after the $Y$ and $Z$ swap in DEJMPS (no twirl), the evolution of the $I$ and $Z$ components, which was previously dominated by $Z$, becomes dominated by $Y$. Moreover, since $Y$ converges to $0$ strictly, the modified dynamics in DEJMPS(no twirl) drives $Z$ to $0$, and hence $I$ converges to $1$.

\section{1G Simulator}
\label{app:1G Simulator}

In general, the 1G protocol is based on a quantum circuit. To simulate it in a clear and systematic way, we use the stabilizer formalism to calculate all results step by step, gate by gate. Similar to the previous section, we also list all possible errors and calculate the probability of each one. 
However, the 1G protocol exhibits a sharp transition at $F_{\rm in} = 0.5$, making the $F_{\rm in} - F_{\rm out}$ mapping effectively step-like (discontinuous and piecewise constant). Therefore, using \texttt{interp1()} to densify the mapping (e.g., from 1{,}000 points to 10{,}000) is not valid globally: near this sharp jump, interpolation can produce non-physical intermediate values or become numerically unstable. In practice, interpolation may only be reliable when restricted to $F_{\rm in} > 0.5 + \epsilon$ (with a small $\epsilon > 0$) and sufficiently far from the transition. For robustness across all input fidelities, we do not use \texttt{interp1()} for the 1G protocol.
Instead, we must either generate the 1G protocol results in real time according to the protocol’s sequence or precompute all required data points in advance. The process for generating the plots of the 1G purification protocols is summarized as follows:

\begin{enumerate} [label=Step \arabic*:, leftmargin=4em]

\item Set the required input: \\
Provide the complete protocol circuit and define the input state based on the stabilizer formalism. Then, specify several groups, each consisting of two qubits. Later, if the measurement results of any group do not match, its outcome will be marked as discarded. Finally, specify the number of rounds to be executed.

\item Construct the $F_{\rm in}$ Sequence: \\
In this step, we uniformly select 10000 points from $F_{\rm in}=0$ to $F_{\rm in}=1$. At the same time, we generate the probabilities $P_X$, $P_Y$, and $P_Z$ based on the characteristics of depolarizing channel or input by hand.

\item Determine whether to apply twirling: \\
If the protocol requires twirling in each round, this step redistributes the errors according to the depolarizing model. Otherwise, the probabilities of $P_I$, $P_X$, $P_Y$, and $P_Z$ are passed directly to the next step as a no-twirl case.

\item Construct a circuit table based on the protocol circuit: \\
We developed a quantum circuit simulator in MATLAB. In this model, based on the input defined in Step 1, possible errors are added at the beginning of the main circuit. Then, using the stabilizer formalism, the circuit is evaluated gate by gate to calculate the resulting states. Consequently, each outcome of the circuit corresponds to a specific combination of input errors. Finally, this process generates a table that lists all possible cases, showing the output (I, X, Y, Z, or discard) along with the corresponding input errors.

\item 
Calculate the results: \\
Based on the probabilities obtained in Step 3 and the circuit table generated in Step 4, we can summarize the overall probabilities of $P_I$, $P_X$, $P_Y$, $P_Z$, and discard corresponding to each input fidelity. Since discard events occur in this step, the probabilities of $P_I$, $P_X$, $P_Y$, and $P_Z$ must be renormalized using the formula below.
\begin{align}
P_{\rm discard}&=1-P_I-P_X-P_Y-P_Z\\
F_{\rm out}&=P_{I_{\rm out}}=\frac{P_I}{1-P_{\rm discard}} \\
P_{A_{\rm out}}&=\frac{P_A}{1-P_{\rm discard}}, A=X,Y,Z
\end{align}

\item Calculate the rate and discard probability: \\
The total discard probability up to the current round is given by
\begin{align}
P_{\rm total\ discard} &= P_{\rm last\ discard} \notag \\
&+ (1-P_{\rm last\ discard})P_{\rm discard}, \label{eq:total discard}
\end{align}
where $P_{\rm last\ discard}$ is the total discard probability accumulated up to the previous round, and $P_{\rm discard}$ is the discard probability for the current round. Therefore, the total rate for the current round is 
\begin{align}
R_{\rm total} = \frac{1}{2^n}(1-P_{\rm total\ discard}),
\end{align}
where $n$ is the number of the current round.

\item Loop or output:\\
Return to Step 3, using the output of the current round as the input for the next round. Alternatively, stop at the current round to generate and plot the results.

\end{enumerate}

\newtext{
\section{Quantitative Analysis of Three-Round Distillation}
\label{app:3 rounds}

We want to determine a scalable structure for a chain of repeaters based on several rounds of distillation and entanglement swaps to increase the output fidelity.
\begin{figure}[h]  
    \centering  
    \includegraphics[width=0.45\textwidth]{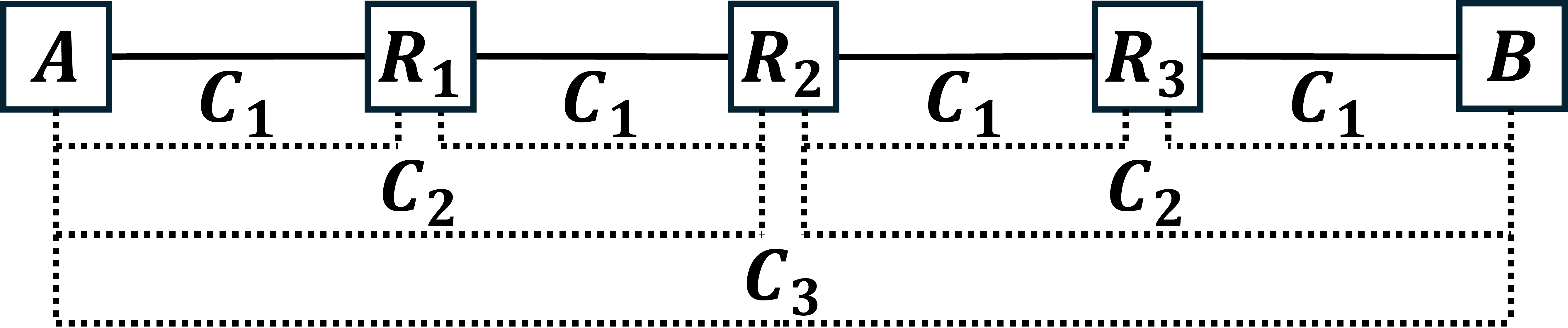}
    \caption{Protocol diagram for three repeaters.} 
    \label{fig:3R}  
    \vspace*{-5pt}
\end{figure}
To determine the optimal number of distillation rounds, the three-repeater situation provides a suitable case for analysis. To clarify, let's start with a three-repeater configuration, with the structure shown in Fig. \ref{fig:3R}. This setup includes Alice, a chain of repeaters from R1 to R3, and Bob.

First, entanglement links are established between the following adjacent pairs: Alice and R1, R1 and R2, R2 and R3, and R3 and Bob. Next, entanglement distillation is performed independently for each link using code $C_1$. Then, quantum swaps are performed to generate entanglement links between Alice and R2, and between R2 and Bob.

Subsequently, distillation is applied using code $C_2$ between Alice and R2, and between R2 and Bob. In this process, distillation is always performed between the two nodes closest to the current entanglement link. Next, a final quantum swap is performed to establish an entanglement link between Alice and Bob. Finally, end-to-end distillation is carried out between Alice and Bob using code $C_3$.

In the first test experiment, we want to start with something simple. So, we use these two codes: $\llbracket 5,1,3 \rrbracket$ and $\llbracket 7,1,3 \rrbracket$, which are well known for encoding a single logical qubit, but one has a lower rate than the other.

\begin{figure*}[h]
\centering
    \begin{subfigure}[b]{0.45\textwidth}
        \centering
        \includegraphics[width=\textwidth]{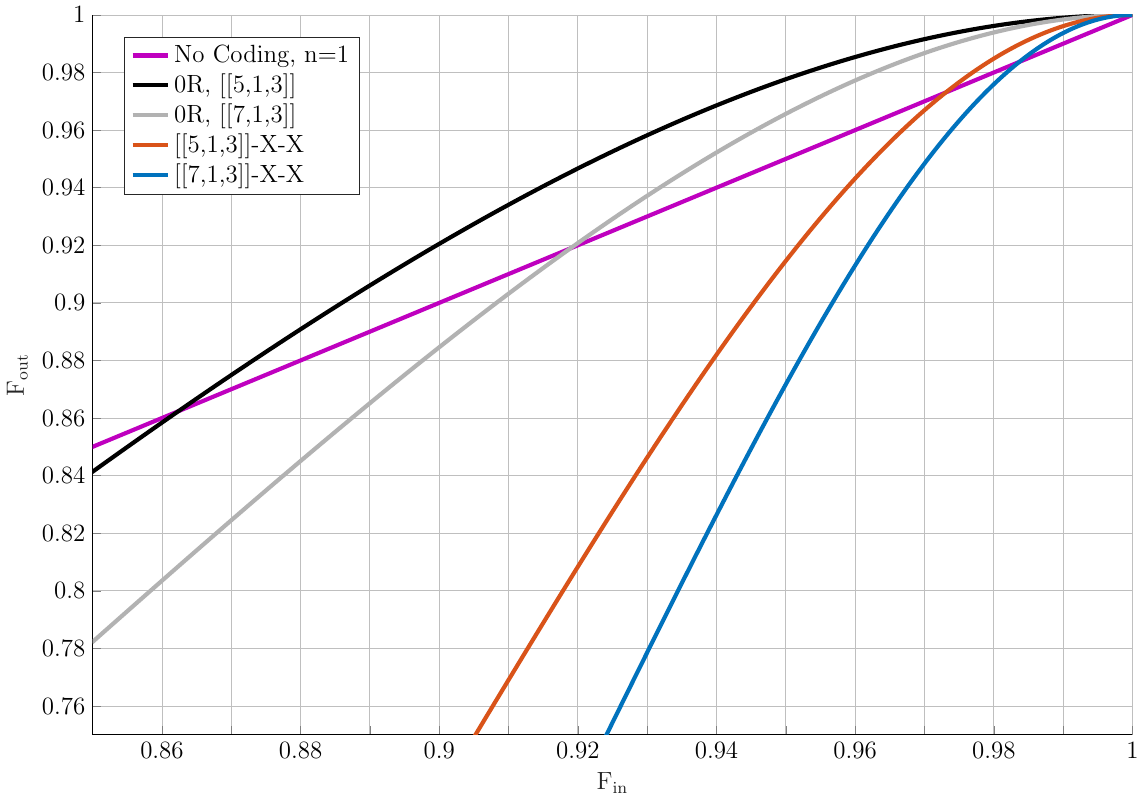}
        \caption{C-X-X}
        \label{fig:CXX}
    \end{subfigure}
    \hspace{0.03\textwidth}
    \begin{subfigure}[b]{0.45\textwidth}
        \centering
        \includegraphics[width=\textwidth]{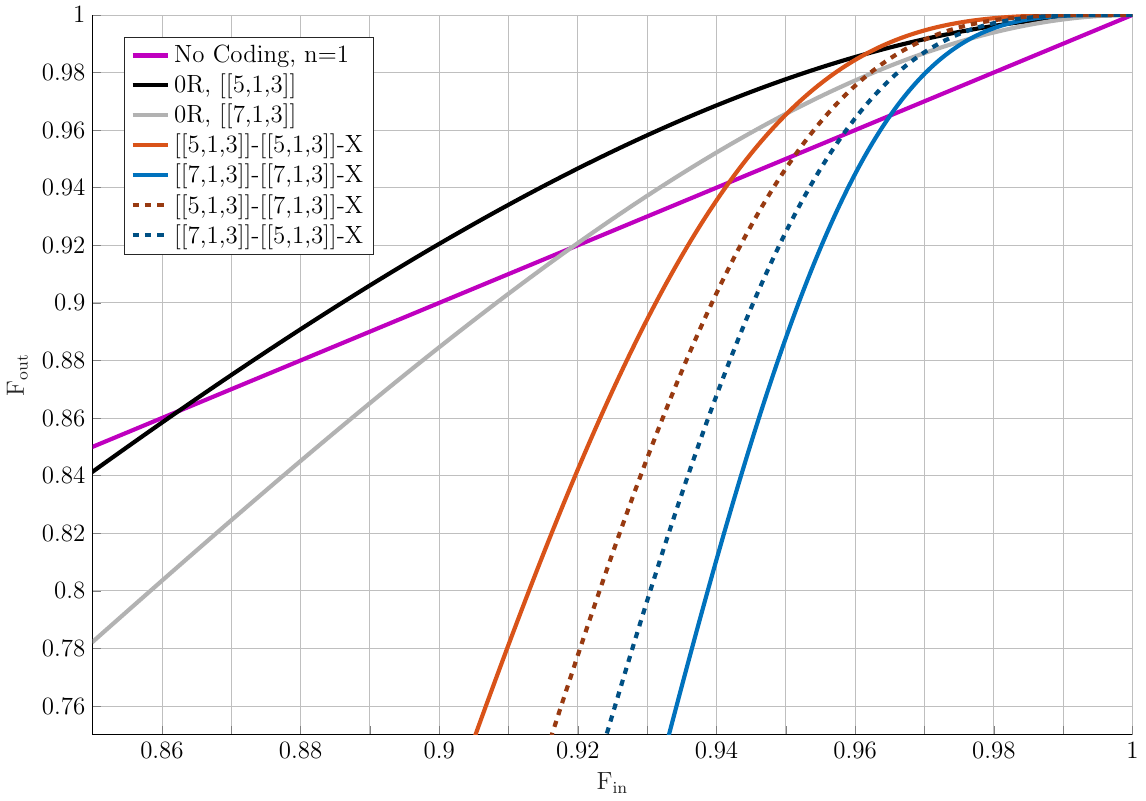} 
        \caption{C-C-X}
        \label{fig:CCX}
    \end{subfigure}
\end{figure*}
\begin{figure*}[h]
\ContinuedFloat
\centering
    \begin{subfigure}[b]{0.45\textwidth}
        \centering
        \includegraphics[width=\textwidth]{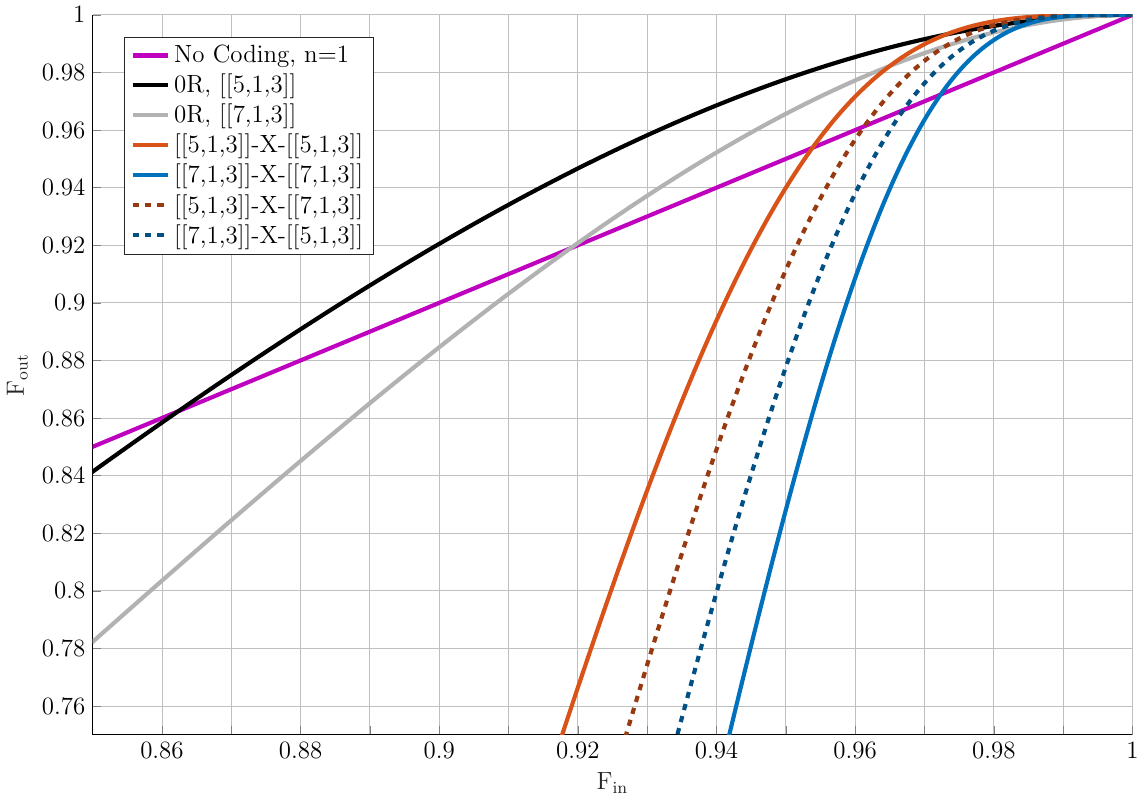} 
        \caption{C-X-C}
        \label{fig:CXC}
    \end{subfigure}
    \hspace{0.03\textwidth}
    \begin{subfigure}[b]{0.45\textwidth}
        \centering
        \includegraphics[width=\textwidth]{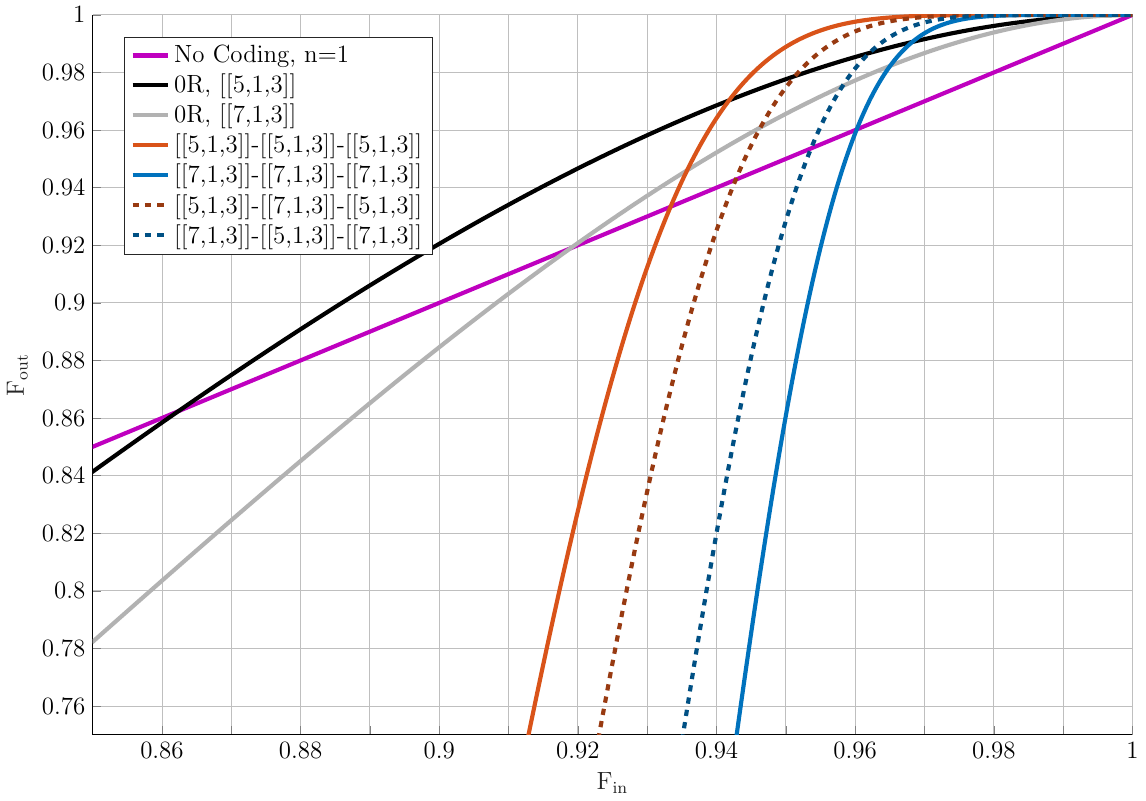} 
        \caption{C-C-C}
        \label{fig:CCC}
    \end{subfigure}
    \caption{Three Rounds of Distillation. Legend entry ``No Coding'' means $F_{\rm out}=F_{\rm in}$, \newtext{i.e., no distillation is done}.} 
    \label{fig:3R_XXX}
    \vspace{0.5em}
    \justifying
    
\end{figure*}

In Fig. \ref{fig:3R_XXX}, the title of each subfigure follows the format ``C-X-X", where ``C" indicates that distillation is performed in that round using either code, and ``X" means that distillation is not performed in the corresponding round. 
Legends beginning with ``0R" indicate cases without repeaters (i.e., these are just logical error rates for the codes). For example:

(a) represents performing only the first round of distillation. ``0R, $\llbracket 5,1,3 \rrbracket$" and ``0R, $\llbracket 7,1,3 \rrbracket$" denote directly establishing the entanglement link between Alice and Bob without repeaters, using the $\llbracket 5,1,3 \rrbracket$ and $\llbracket 7,1,3 \rrbracket$ codes, respectively. These curves are simply the logical error rates of the codes, plotted as input fidelity vs output fidelity. In this context, ``$\llbracket 5,1,3 \rrbracket$-X-X" indicates that only the first round of distillation is performed, using the $\llbracket 5,1,3 \rrbracket$ code.

(b) refers to performing the first and second rounds.

(c) refers to performing the first and third rounds.

(d) refers to performing all three rounds.

Since we assume that the depolarizing noise is applied to the initial Bell pairs before the first round of distillation, the first round plays a critical role in the overall distillation performance.
The final end-to-end round of distillation is also important because it can fix errors generated during intermediate swaps, which makes it the second most important. 
Comparing Fig. \ref{fig:CCX} and Fig. \ref{fig:CXC}, both apply the first round, but one includes the second and the other includes the final. As shown in the plots, Fig. \ref{fig:CCX} performs better, suggesting that having only the first and final rounds may not be enough.

 Comparing Fig. \ref{fig:CXX}, Fig. \ref{fig:CCX} and Fig. \ref{fig:CCC}, we find that as the number of rounds increases, the last-performed round in each configuration contributes the least benefit. So the second round ranks third in importance. This suggests that adding more rounds of distillation in the middle provides diminishing benefits as the number of rounds increases, potentially leading to a decrease in the link-building speed.  Therefore, it is reasonable to limit the number of distillation rounds to three.

To summarize this analysis,  
we use $1-F_{\rm th}$ to compare the applicable ranges of different codes, where $F_{\rm th}$ denotes the pseudo-threshold fidelity. A larger value of $1-F_{\rm th}$ indicates
a wider applicable region, since the code can be applied over a larger range of input fidelities.

\subsection{The Position of the First Round}

\begin{table}[H]
    \centering
    \begin{tabular}{|c|c|c|c|c|c|c|}
        \toprule 
        \multicolumn{2}{|c|}{$F_{\rm th}$} & 3R & \multicolumn{4}{|c|}{$1-F_{\rm th}$} \\
        \midrule
        $\llbracket 5,1,3 \rrbracket$ & $\llbracket 7,1,3 \rrbracket$ & Type & \multicolumn{2}{|c|}{$\llbracket 5,1,3 \rrbracket$} & \multicolumn{2}{|c|}{$\llbracket 7,1,3 \rrbracket$}\\
        \midrule
        0.973 & 0.984 & CXX & 0.027 & & 0.016 & \\
        0.986 & 0.992 & XCX & 0.013 & -49.55\% & 0.008 & -49.69\%\\
        0.993 & 0.996 & XXC & 0.007 & -49.78\% & 0.004 & -49.94\%\\
        \bottomrule
    \end{tabular}
    \caption{In the 3-repeater case, the table shows the relationship between the pseudo-threshold and different protocol types. In the Type column, ``C'' denotes a round with QEC-based distillation, while ``X'' denotes an idle round with no operation.}
    \label{tab:1_round}
\end{table}

As shown in Table~\ref{tab:1_round}, the CXX protocol provides the largest benefit, while the benefits of the other protocols decrease progressively by about 50\%.
Therefore, applying the code in the first round ($C_1$) provides the largest operable region, making it the most suitable choice.

\subsection{The Position of the Second Round}
Now, let us keep $C_1$ fixed and discuss the effects of adding $C_2$ and $C_3$.

\begin{table}[H]
    \centering
    \begin{tabular}{|c|c|c|c|c|c|c|}
        \toprule 
        \multicolumn{2}{|c|}{$F_{\rm th}$} & 3R & \multicolumn{4}{|c|}{$1-F_{\rm th}$} \\
        \midrule
        $\llbracket 5,1,3 \rrbracket$ & $\llbracket 7,1,3 \rrbracket$ & Type & \multicolumn{2}{|c|}{$\llbracket 5,1,3 \rrbracket$} & \multicolumn{2}{|c|}{$\llbracket 7,1,3 \rrbracket$}\\
        \midrule
        0.973 & 0.984 & CXX & 0.027 & & 0.016 & \\
        0.942 & 0.965 & CCX & 0.058 & +116.16\% & 0.035 & +114.09\%\\
        0.933 & 0.960 & CCC & 0.067 & +14.50\% & 0.040 & +13.82\%\\
        \bottomrule
    \end{tabular}
    \caption{In the 3-repeater case, the table shows the relationship between the pseudo-threshold and different protocol types. Here, the same code is used in multiple rounds depending on the protocol type. }
    \label{tab:2_round}
\end{table}

As shown in Table~\ref{tab:2_round}, we can easily see that as the number of rounds increases, the additional benefit gradually decreases. The largest benefit comes from adding $C_2$. 
Therefore, applying the code in the second round ($C_2$) provides the largest operable region, making it the second suitable choice.

\subsection{The Position of the Third Round}
Now, let us keep $C_1$ and $C_2$ fixed and discuss the effect of adding $C_3$ for different numbers of repeaters.

\begin{table}[H]
    \centering
    \begin{tabular}{|c|c|c|c|c|c|c|}
        \toprule 
        \multicolumn{2}{|c|}{$F_{\rm th}$} & CCX & \multicolumn{4}{|c|}{$1-F_{\rm th}$} \\
        \midrule
        $\llbracket 5,1,3 \rrbracket$ & $\llbracket 7,1,3 \rrbracket$ & Type & \multicolumn{2}{|c|}{$\llbracket 5,1,3 \rrbracket$} & \multicolumn{2}{|c|}{$\llbracket 7,1,3 \rrbracket$}\\
        \midrule
        0.922 & 0.953 & 1R & 0.078 & & 0.046 & \\
        0.914 & 0.949 & CCC & 0.086 & +10.05\% & 0.051 & +9.53\%\\
        \midrule
        0.942 & 0.965 & 3R & 0.058 & & 0.035 & \\
         0.933 & 0.960 & CCC & 0.067 & +14.50\% & 0.040 & +13.82\%\\
        \midrule
        0.950 & 0.970 & 5R & 0.049 &  & 0.030 & \\
        0.942 & 0.965 & CCC & 0.058 & +16.85\% & 0.035 & +16.14\%\\
        \bottomrule
    \end{tabular}
    \caption{Comparison of the CCX and CCC cases for different numbers of repeaters.}
    \label{tab:3_round}
\end{table}

As shown in Table~\ref{tab:3_round}, adding $C_3$ indeed increases the operable region. Moreover, the relative improvement becomes larger as the number of repeaters increases. However, the incremental gain becomes smaller. For example, using the upper bound, the improvement increases by \(4.45\%\) from the 1R case to the 3R case, but only by \(2.35\%\) from the 3R case to the 5R case. This indicates a diminishing marginal benefit as the number of repeaters increases.

Although the marginal benefit of adding $C_3$ decreases as the number of repeaters increases, $C_3$ remains important because its contribution is still positive and it acts directly on the final end-to-end Bell pair. Its role is to suppress the residual errors accumulated through repeater operations and entanglement swapping, thereby extending the applicable region of the protocol. Thus, $C_3$ should be viewed as a final stabilization step rather than the dominant source of improvement.
}

\newtextnew{
Under the ideal-operation assumptions considered in this work, the final end-to-end distillation round provides a comparatively modest additional improvement. Its role in the present model is therefore better interpreted as a final stabilization step rather than as the dominant source of performance enhancement.

This conclusion is specific to the ideal-noise model used here. In a circuit-level setting with imperfect local gates, the final end-to-end distillation step may become substantially more important, since it acts on errors accumulated throughout the entire repeater chain. A quantitative analysis of this regime requires a fault-tolerant treatment of the local operations and is left for future work.

}

\newtext{%
\section{Simulation Setup}
\label{app:sim_setup}

Our simulation setup is built on MATLAB. 
When we try to build our algorithm, there are two major issues that need to be solved.
The first one is the noise generation problem. 
When generating random Pauli noise, we expect it to be truly random. 
As is well-known, random numbers generated by modern classical computers are all pseudo-random. 
Thus, generating results that approximate true randomness correctly is a significant goal for our algorithm.
The second issue arises from the fact that the final result represents a probability distribution, which means we need a large number of noise simulations to obtain reliable final results. 
Based on our experiments, each mapping point between input fidelity $F_{\rm in}$ and output fidelity $F_{\rm out}$ requires at least $10^6$ simulations to obtain a relatively accurate result. As a result, accelerating this process becomes the core goal of the algorithm.

To address these issues, we propose an algorithm which uses a fixed probability to replace random error generation, \newtext{i.e., we avoid such exorbitant Monte Carlo simulations}.
\subsubsection{Part I: Constructing the Logical Error Rates (0R)}
First, as shown in Fig. \ref{fig:0R_F}, we need to construct the $F_{\rm in}$-$F_{\rm out}$ map for the 0-Repeater case, which will serve as a reference for subsequent parts.
When multiple repeaters are introduced, if the input noise is depolarizing, then after performing a quantum swap, the resulting noise remains depolarizing. 
Therefore, when performing error correction in the next round, we can directly use the $F_{\rm in}$-$F_{\rm out}$ map from the 0-Repeater case to determine $F_{\rm out}$ after the next round of error correction, thereby accelerating the simulation process.

\begin{enumerate} [label=Step \arabic*:, leftmargin=4em]

    \item Construct the $F_{\rm in}$ Sequence: \\
In this step, we uniformly select 1000 points from $F_{\rm in}=0$ to $F_{\rm in}=1$. At the same time, we generate the probabilities $P_X$, $P_Y$, and $P_Z$ based on the characteristics of depolarizing noise.

    \item Construct the All-Error List:\\
We list all possible error events (including the case where no error occurs) and sort them in ascending order based on their error weight. That is, the first error in the list is the most probable error, the second error is the second most probable one, and so on. If we describe the error-correcting code using $\llbracket n, k, d \rrbracket$, then the total number of possible error combinations based on the syndrome is $2^{(n-k)}$.

    \item Write Data into the All-Error List: \\
In this step, we need to perform three tasks. Using the points obtained from \emph{Step 1}, calculate the probability of each error combination (for each of the 1000 selected points) and write this into the all-error list. Then, construct the Lookup Table and obtain the logical error based on the logical operator, then write it into the all-error list.

    \item Construct the Lookup Table: \\
This step is performed simultaneously with \emph{Step 3}. For each processed error combination, we obtain its corresponding syndrome. If this syndrome has not been recorded in the Lookup Table, we record the error combination that triggers this syndrome. If the syndrome has already been recorded, it means that the current error combination is not the smallest-weight error and does not need to be recorded, because in \emph{Step 2}, we have already sorted the errors by weight.

    \item Obtain the Logical Error Corresponding to Each Error Combination: \\
This step is performed simultaneously with \emph{Step 3}. After \emph{Step 4}, we use the logical operator to determine the logical error caused by the current error combination. Then, we retrieve the smallest weight error from the Lookup Table corresponding to this syndrome and test whether it can accurately predict the logical error caused by the current error combination: If it can predict correctly, record it as correctable by the Lookup Table. Otherwise, record it as uncorrectable. Finally, write the results into the error list.

    \item Generate the $F_{\rm in}$-$F_{\rm out}$ Map: \\
From the all-error list, read all correctable logical error probabilities and accumulate them for each of the 1000 points (generated in \emph{Step 1}).
In the end, we obtain the $F_{\rm in}$-$F_{\rm out}$ map for 1000 points spanning $F_{\rm in}=0$ to $F_{\rm in}=1$.

\end{enumerate}

\subsubsection{Part II: Constructing the Many-Repeater Situation}

Before proceeding to this part, we need to repeatedly execute \emph{Part I} to obtain all required $F_{\rm in}$-$F_{\rm out}$ maps for the error-correcting codes being used.
\begin{enumerate} [label=Step \arabic*:, leftmargin=4em]

    \item Construct the $F_{\rm in}$ sequence: \\
In this step, we uniformly select a required number of points from $F_{\rm in}=0$ to $F_{\rm in}=1$ as needed.

    \item Perform error correction in this round: \\
Based on the $F_{\rm in}$ points constructed in \emph{Step 1} and the chosen $\llbracket n, k, d\rrbracket$ error-correcting code, we use \texttt{interp1()} to look up the corresponding $F_{\rm in}$-$F_{\rm out}$ map in 0-repeater case, where the $F_{\rm out}$ value represents the post-error-correction $F_{\rm out}$ at this round.

    \item Compute the output fidelity of entanglement swap: \\
    Based on our discussion in the background (Eqn.~\eqref{eq:quantum swap}), since in this case each round of error correction is based on the same codes, with the same input fidelity at the beginning, we can simplify the formula to the following form:\\
    $F_{\rm eff} = \frac{1}{4}+\frac{3}{4}\left(\frac{4F_{\rm in}-1}{3}\right)^{n_{\rm QS}+1}$, where $n_{\rm QS}$ represents the total number of quantum swaps required in this round.
    The result obtained from this formula becomes the $F_{\rm in}$ for the next round.
    
    \item Obtain the final $F_{\rm in}$-$F_{\rm out}$ map: \\
By repeating \emph{Step 2} and \emph{Step 3} with different codes in each round, we complete error correction across all rounds and obtain the final $F_{\rm in}$-$F_{\rm out}$ map.

\end{enumerate}
}

\begin{figure}[t]
    \centering
    \hspace*{-20pt}
    \includegraphics[scale=0.5,keepaspectratio]{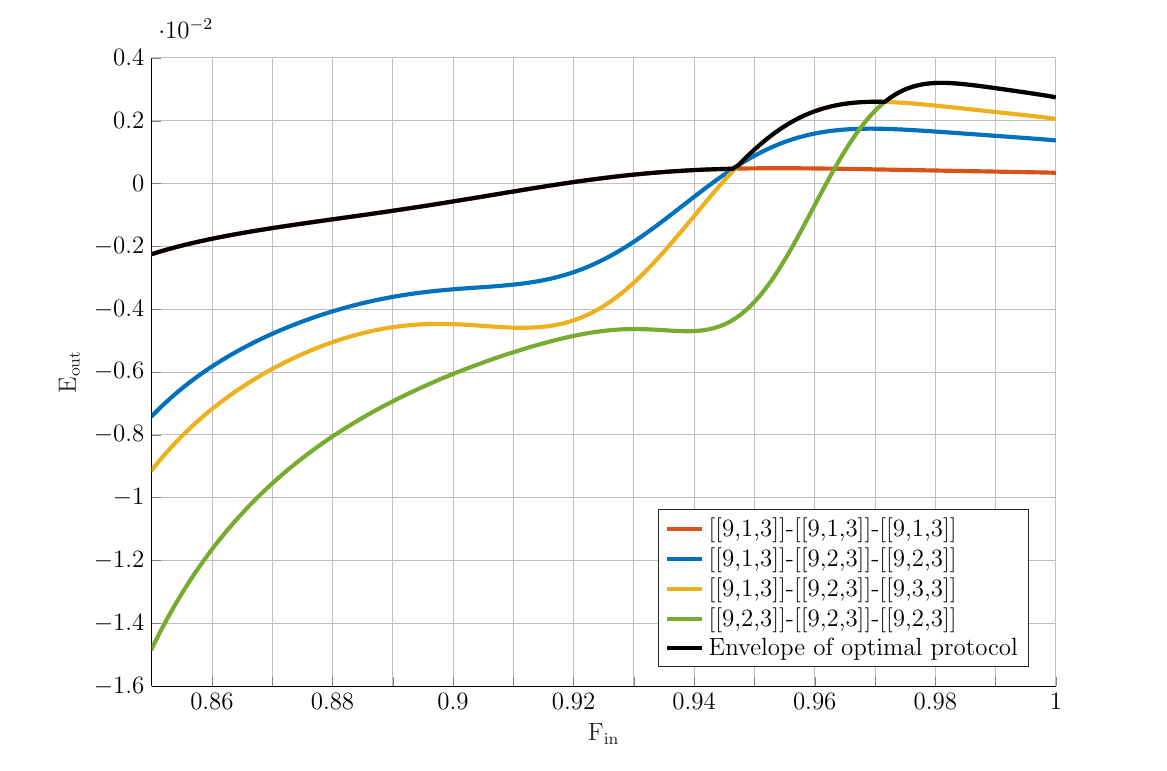}
    \caption{The $F_{\rm in}$-$E_{out}$ map for three repeaters with optimal envelope. Observe that the protocol switching points moved to the right compared to the single repeater case.} 
    \label{fig:3R_E_OL}  
\end{figure}

\begin{figure}[t]
    \centering
    \vspace*{5pt}
    \hspace*{-20pt}
    \includegraphics[scale=0.5,keepaspectratio]{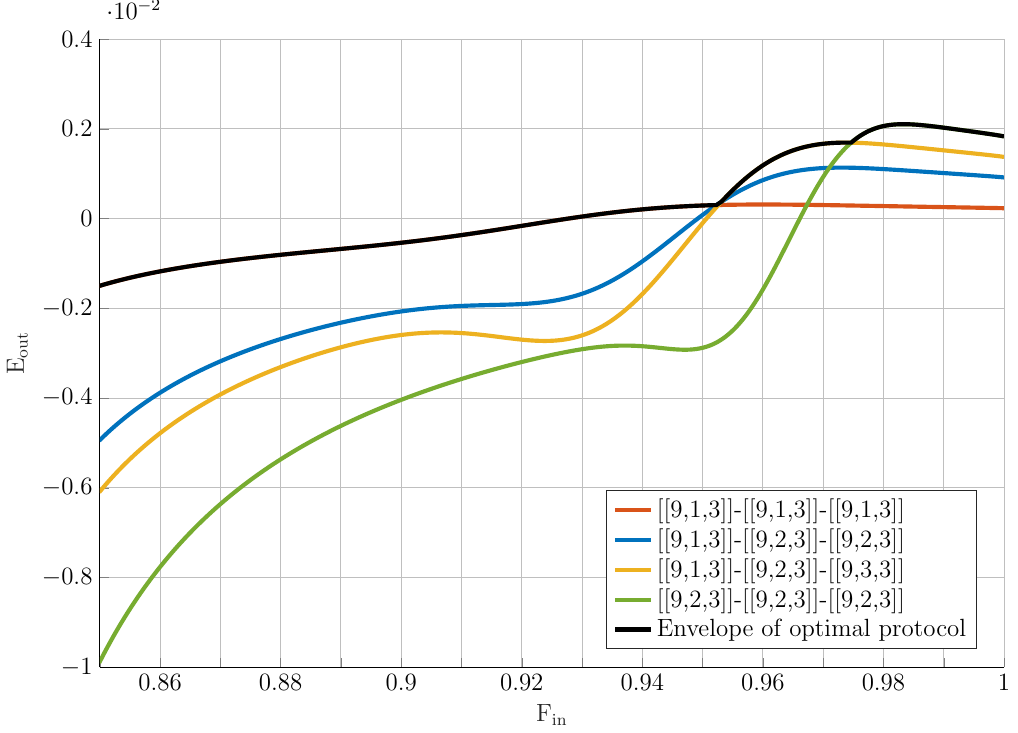}
    \caption{The $F_{\rm in}$-$E_{out}$ map for five repeaters with optimal envelope. Observe that the protocol switching points moved to the right compared to the three-repeaters case.} 
    \label{fig:5R_E_OL}  
\end{figure}

\newtext{
\section{Adaptive Protocol for Multiple Repeaters}
\label{app:multiple_repeaters}

From Fig. \ref{fig:3R_E_OL}, we can identify the switching points for the adaptive protocol on three repeaters as follows:
\begin{itemize}
    \item From ${\rm Protocol}_1$ to ${\rm Protocol}_2$: $F_{\rm in}=0.9465$;
    \item From ${\rm Protocol}_2$ to ${\rm Protocol}_3$: $F_{\rm in}=0.9474$;
    \item From ${\rm Protocol}_3$ to ${\rm Protocol}_4$: $F_{\rm in}=0.9717$.
\end{itemize}

From Fig. \ref{fig:5R_E_OL}, we can identify the switching points for the adaptive protocol on five repeaters as follows:
\begin{itemize}
    \item From ${\rm Protocol}_1$ to ${\rm Protocol}_2$: $F_{\rm in}=0.9524$;
    \item From ${\rm Protocol}_2$ to ${\rm Protocol}_3$: $F_{\rm in}=0.9532$;
    \item From ${\rm Protocol}_3$ to ${\rm Protocol}_4$: $F_{\rm in}=0.9747$.
\end{itemize}
These have progressively increased from the single repeater and three repeaters cases.
In Table~\ref{tab:nR_F}, we identify the switching points for different numbers of repeaters. 
We can easily see that the switching point increases with the number of repeaters.
\begin{table}[H]
    \centering
    \begin{tabular}{|c|c|c|c|}
        \toprule 
        \#R & $F_{SW}^{1}$ & $F_{SW}^{2}$ & $F_{SW}^{3}$ \\
        \midrule
        1  & 0.9343 & 0.9356 & 0.9655 \\
        3  & 0.9465 & 0.9474 & 0.9717 \\
        5  & 0.9524 & 0.9532 & 0.9747 \\
        7  & 0.9561 & 0.9568 & 0.9766 \\
        9  & 0.9587 & 0.9594 & 0.9780 \\
        11 & 0.9608 & 0.9614 & 0.9791 \\
        13 & 0.9624 & 0.9630 & 0.9799 \\
       101 & 0.9779 & 0.9782 & 0.9881 \\
      1001 & 0.9877 & 0.9879 & 0.9934 \\
        \bottomrule
    \end{tabular}
    \caption{The relationship between the switching points and the number of repeaters. \#R represents the number of repeaters, and $F_{SW}^{n}$ represents the switching points.}
    \label{tab:nR_F}
\end{table}
}

\end{document}